    \documentstyle[nato]{crckapb} 
\input psfig.sty

\def\D{\phi}
\def\Z{z}
\def\M{M}

\def\epsfig{\psfig}
\def\gstring{g_{\rm s}}
\def\Mstr{M_{\scriptscriptstyle\rm s}}
\def\refeq#1{(\ref{#1})}
\newcommand{\be}{\begin{equation}}
\newcommand{\ee}{\end{equation}}
\newcommand{\bea}{\begin{eqnarray}}
\newcommand{\eea}{\end{eqnarray}}
\renewcommand{\a}{\alpha}
\renewcommand{\b}{\beta}

\renewcommand{\d}{\delta}

\newcommand{\pa}{\partial}

\newcommand{\G}{\Gamma}
\newcommand{\e}{\epsilon}

\renewcommand{\l}{\lambda}
\renewcommand{\L}{\Lambda}
\newcommand{\m}{\mu}
\newcommand{\n}{\nu}
\newcommand{\x}{\chi}

\newcommand{\s}{\sigma}

\newcommand{\y}{\upsilon}

\newcommand{\ft}[2]{{\textstyle\frac{#1}{#2}}}
\newcommand{\eqn}[1]{(\ref{#1})}
\def\slash{\llap /}
\begin{opening}
\title{}
\end{opening}
\begin{document}
\vspace*{-5.5cm}
\begin{flushright}
THU-98/04
\end{flushright}
\vskip 1cm
\begin{center}
{\bf\LARGE Supersymmetry and Dualities 
in various dimensions}\footnote{Lectures given 
at the Nato Advanced Study Institute on
{\it Strings, Branes  and Dualities},
Carg\`ese, May 26 - June 14, 1997.}
\vskip .8cm
BERNARD DE WIT\\
{\it Institute for Theoretical Physics, Utrecht University \\
           Princetonplein 5, 3508 TA Utrecht, The Netherlands}\\
\vskip .5cm
JAN LOUIS\\
{\it Martin--Luther--Universit\"at Halle--Wittenberg,\\
        FB Physik, D-06099 Halle, Germany}

\end{center}
\vskip 20pt
\section{Introduction}
Since the seventies 
string theory has been discussed as a possible candidate for
a theory which unifies all known particle interactions
including gravity. Until recently, however,  
string theory has only been known in its perturbative 
regime. That is, the (particle) excitations of a 
string theory are computed in the free theory ($\gstring=0$), 
while  their scattering processes are evaluated
in a perturbative series for $\gstring\ll 1$.
The string coupling constant $\gstring$ is a free parameter of 
string theory but for $\gstring= {\cal O}(1)$ no method
of computing the spectrum or the interactions
had been known. This situation dramatically changed 
during the past
three years. For the first time it became possible to go beyond the 
purely perturbative regime and to reliably
compute some of the nonperturbative properties of string theory.
The central point of these developments rests on the idea that
the strong-coupling limit of a given string theory can be 
described in terms of another, weakly coupled, `dual theory'. 
This dual theory can take the form of either a different string 
theory, or the same string theory with a different set of 
perturbative excitations, or a new theory termed M-theory.

The precise nature of the strong-coupling limit sensitively depends 
on the number of (Minkowskian) spacetime dimensions and the amount
of supersymmetry. Supersymmetry has played a major role 
in the recent developments in two respects.
First of all, it is difficult (and it has not been satisfactorily
accomplished) to rigourously prove a string duality, since it 
necessitates a full nonperturbative formulation, which is not yet available.
Nevertheless it has been possible to perform nontrivial
checks of the conjectured dualities for quantities or couplings
whose quantum corrections are under (some) control.
It is a generic property of supersymmetry that it protects a 
subset of the couplings and implies a set of
nonrenormalization theorems. 
The recent developments heavily rely on the fact
that the mass (or tension) of BPS-multiplets
is protected and that holomorphic couplings
obey a nonrenormalization theorem. 
Thus, they can be computed in the perturbative regime of string 
theory and, under the assumption of unbroken supersymmetry, 
reliably extrapolated into the nonperturbative region.
It is precisely for these
BPS-states and holomorphic couplings that
the conjectured dualities have been successfully
verified.

Second of all, for a given spacetime dimension $D$
and a given representation of supersymmetry there can exist 
perturbatively different string theories.
For example, the heterotic SO(32) string in $D=10$
and the {type-I} string in $D=10$ share the same supersymmetry, but 
their interactions are different in perturbation theory.
However, once nonperturbative corrections are taken into account, 
it is believed that  the two theories are identical and merely 
different perturbative limits of the same 
underlying quantum theory.
A similar phenomenon is encountered with other 
string theories in different dimensions and 
the moduli space of string theory is much smaller 
than was previously known. 
In the course of these lectures we will see 
that many nontrivial relations 
exist among perturbatively distinct string vacua
and furthermore, that,  what were thought to be disconnected
components of the moduli space,  
can in fact be different perturbative 
regions of one and the same component.
Thus, in determining the properties of the underlying 
quantum theory, supersymmetry seems to play
a much more prominent role than had previously been 
appreciated.  

The purpose of these lectures is to review some of these recent developments
with particular emphasis on the role played by 
supersymmetry.\footnote{This set of lectures notes 
is an expanded version of \cite{DLtrieste}.}
In section 2 we collect the representations of supersymmetry in 
spacetime dimensions $3\le D\le 11$ from a common point of view.
Many features are only displayed in appropriate tables but we
present slightly more detail in the dimensions $D=11,10$ and 6
as representative cases. We explain a number of features of the 
dimensional reduction of supergravity, such as the emergence of hidden 
symmetries, the low-energy action in different
frames and other aspects relevant in the string context.
In section 3 we first recall the different perturbative string
theories, their (Calabi-Yau) compactifications
and the dualities which already exist at the perturbative level.
Then we discuss the various types
of possible strong-coupling limits
(S-duality, U-duality, M-theory,
F-theory) and the corresponding string vacua.
This leads to a `web' of interrelations which
we attempt to visualize at the end of section 3.
Finally, in appendix A
we review some basic properties
of the free-field representation for states of different spin, 
while in appendix B we present a more detailed  
discussion of the relation between the parameters
of string theory and those of the corresponding low-energy 
effective field theory.

\section{Supersymmetry in various dimensions}
 \subsection{The supersymmetry algebra}
An enormous amount of information about supersymmetric theories is 
contained in the structure of the underlying representations of 
the supersymmetry  
algebra \cite{general-sg}. Here we distinguish the supermultiplet 
of the fields, which transforms irreducibly under the supersymmetry 
transformations, and the supermultiplet of states described by 
the theory. The latter will depend on the dynamics encoded in a 
supersymmetric action or Hamiltonian. The generators of the 
super-Poincar\'e algebra comprise the supercharges, transforming as 
spinors under the Lorentz group, the energy and momentum 
operators, the generators of the Lorentz group, and possibly 
additional generators that commute with the supercharges. For the 
moment we ignore these additional charges, often called {\it 
central} charges\footnote{%
   The terminology adopted in the literature is  
   not always very precise. Usually, all charges that commute 
   with the supercharges, but not necessarily with all the 
   generators of the Poincar\'e algebra, are called `central 
   charges'. We will adhere to this nomenclature. }. 
There are other relevant superalgebras, such as the 
supersymmetric extension of the anti-de Sitter algebra. These 
will not be considered here, but they are unavoidable when 
considering supersymmetry in theories with a cosmological term.

Ignoring the central extensions for the moment, the most 
important anti-commutation relation is the one between the 
supercharges,
\be\label{susy-algebra}
\{Q_\a , \bar Q_\b \} = -2 i P^\m\, (\G_\m)_{\a\b}\,.
\ee
Here $\G_\m$ are the gamma matrices that generate the Clifford 
algebra with Minkowskian signature $(-,+,+,\cdots)$. 

\begin{table}
\begin{center}
\begin{tabular}{l l l l}\hline
$D$ & $Q_{\rm irr}$ &  $H_{\rm R}$   & type \\\hline
1, 3, 9, 11, mod 8   & $2^{(D-1)/2}$ & SO($N$) & M\\
5, 7, mod 8          & $2^{(D+1)/2}$ & USp($2N$) & D\\
4, 8, mod 8          & $2^{D/2}$     & U($N$) & M \\
6, mod 8             & $2^{D/2}$    & USp($2N_+)\times 
                                       $USp($2N_-$) & W\\
2, 10, mod 8         & $2^{D/2-1}$   & SO($N_+)\times $SO($N_-$) 
                                              & MW\\\hline
\end{tabular}
\end{center}
\caption{The supercharges in various spacetime dimensions $D$. In 
the second column, $Q_{\rm irr}$ specifies the real dimension of 
an irreducible spinor in a $D$-dimensional 
Minkowski spacetime. The third column specifies the group $H_{\rm 
R}$ for $N$-extended supersymmetry, defined in the text, acting 
on $N$-fold reducible spinor  
charges. The fourth column denotes the type of spinors: Majorana 
(M), Dirac (D), Weyl (W) and Majorana-Weyl (MW).}\label{susy-charges}
\end{table}

The size of a supermultiplet depends sensitively on the number of 
independent supercharge components $Q$. The first step is therefore to 
determine $Q$ for any given number of spacetime 
dimensions $D$. The result is summarized in 
Table~\ref{susy-charges}. As shown, there exist five  
different sequences of spinors, corresponding to spacetimes of 
particular dimensions. When this dimension is odd, it is 
possible in certain cases to have Majorana spinors. These cases 
constitute the first sequence.  
The second one corresponds to those odd dimensions where Majorana 
spinors do not exist. The spinors are then Dirac spinors. In even 
dimension one may distinguish three sequences. In the first one, 
where the number of dimensions is a multiple of 4, 
charge conjugation relates positive- with negative-chirality 
spinors. All spinors in this sequence can be restricted to 
Majorana spinors. For the remaining two  
sequences, charge conjugation preserves the chirality of the 
spinor. Now there are again two possibilities, depending on whether 
Majorana spinors can exist or not. The cases where we cannot have 
Majorana spinors, whenever $D=6$~mod~8, comprise the fourth sequence. 
For the last sequence Majorana   
spinors exist and we restrict the charges to Majorana-Weyl 
spinors.  

One can consider {\it extended supersymmetry}, where the spinor 
charges transform reducibly under the Lorentz group and comprise 
$N$ irreducible spinors. For Weyl charges, one can  
consider combinations of $N_+$ positive- and $N_-$ 
negative-chirality spinors. In all these cases there exists 
a group $H_{\rm R}$ that rotates the irreducible components such 
that the supersymmetry algebra is left invariant. This 
group acts exclusively on the spinor charges and commutes with 
the Lorentz transformations. The group $H_{\rm R}$ is thus the 
part of the automorphism group of the supersymmetry algebra that 
commutes with the Lorentz group. This group is often realized as 
a manifest invariance group of a supersymmetric field theory. 

Another way to present the various cases is shown in  
Table~\ref{simple-susy}. Here we list the real dimension of an 
irreducible spinor charge and its corresponding spacetime 
dimension. In addition we have included the number of states of the 
shortest\footnote{%
   By the {\it shortest} multiplet, we mean the multiplet with 
   the helicities of the states as low as possible. This is 
   usually (one of) the smallest possible supermultiplet(s). } %
supermultiplet of massless states, written as a sum of 
bosonic and fermionic states. 

\begin{table}
\begin{center}
\begin{tabular}{l l l}\hline
$Q_{\rm irr}$ & $D$  & shortest supermultiplet \\\hline
32   & $D=11$        & $128+128$\\
16   & $D= 10,9,8,7$ & $8+8$\\
8    & $D= 6,5$      & $4+4$\\
4    & $D= 4$        & $2+2$\\
2    & $D=3$         & $1+1$ \\\hline
\end{tabular}
\end{center}
\caption{Simple supersymmetry in various dimensions. We present 
the dimension of the irreducible spinor charge with $2\leq Q_{\rm 
irr}\leq 32$ and the corresponding spacetime dimensions $D$. The 
third column represents the number of bosonic + fermionic 
massless states for the shortest supermultiplet.  }\label{simple-susy}
\end{table}
%
 \subsection{Massless representations}
Because the momentum operators $P^\m$ commute with the 
supercharges, we may  
consider the states at arbitrary but fixed momentum $P^\m$, which, 
for massless representations, satisfies $P^2 =0$. The matrix 
$P^\m\G_\m$ on the right-hand side of \eqn{susy-algebra} has 
therefore zero eigenvalues. In a positive-definite Hilbert space  
some (linear combinations) of the supercharges must therefore 
vanish. To exhibit this more explicitly, let us rewrite 
\eqn{susy-algebra} as
\be
\{Q_\a , Q^\dagger_\b \} = 2 \,(P\!\slash\, \G^0)_{\a\b}\,.
\ee
For light-like $P^\m= (P^0,\vec P)$ the right-hand side is 
proportional to a  
projection operator $({\bf 1} + \G_\parallel \G^0)/2$. Here 
$\G_\parallel$ is the gamma matrix along the spatial momentum 
$\vec P$ of the states.  The supersymmetry anti-commutator can 
then be written as
\be\label{susy-algebra2}
\{Q_\a , Q^\dagger_\b \} = 2 \, P^0\Big({\bf 1} + \tilde \G_D 
\tilde \G_\perp \Big)_{\a\b}  \,.
\ee
Here $\tilde \G_D$ consists of the product of all $D$ independent 
gamma matrices, and $\tilde\G _\perp$ of the product of all $D-2$ 
gamma matrices in the transverse directions (i.e., perpendicular 
to $\vec P$), with phase factors such that
\be
(\tilde \G_D)^2 = ( \tilde \G_\perp)^2 = {\bf 1}\,,\qquad [ 
\tilde \G_D , \tilde \G_\perp] = 0\,.
\ee
This shows that the right-hand side of \eqn{susy-algebra2} is 
proportional to a projection operator, which projects out half of 
the spinor space. Consequently, half the spinors must vanish on 
physical states, whereas the other ones generate a Clifford algebra. 
Denoting the real dimension of the supercharges by $Q$, the 
representation space of the charges decomposes into the two chiral 
spinor representations of SO$(Q/2)$. When confronting these results 
with the last column in Table~\ref{simple-susy}, it turns out 
that the dimension of the shortest supermultiplet is not just 
equal to $2^{Q_{\rm irr}/4}$, as one might naively expect. For 
$D=6$, this is so because the representation is complex. For $D=3,
4$ the representation is twice  
as big because it must also accommodate fermion number (or, 
alternatively, because it must satisfy the correct CPT 
properties). The derivation for $D=4$ is presented in many 
places. For $D=3$ we refer to \cite{DWTN}.

\begin{figure}  
\epsfig{figure=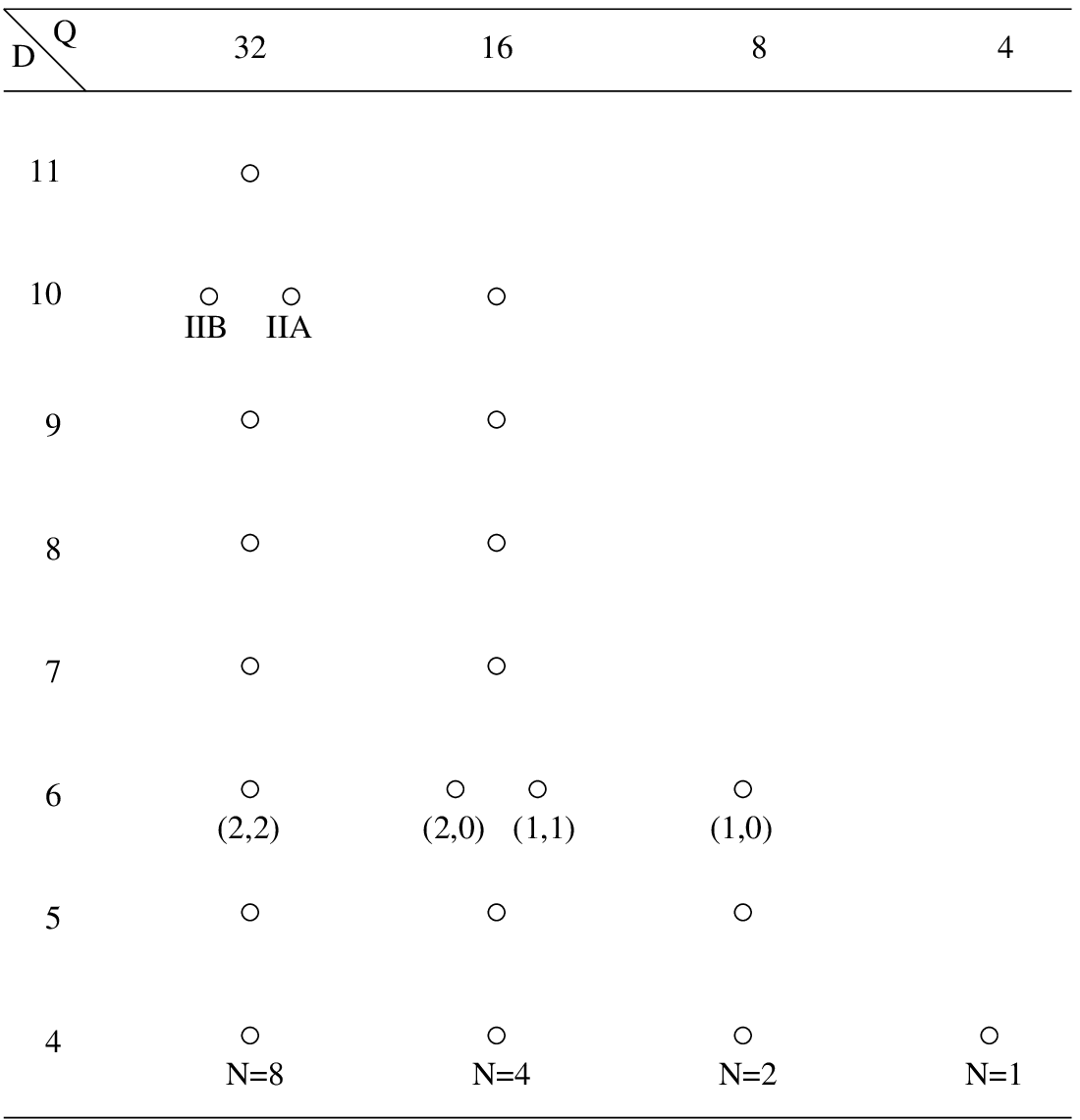}
\caption{Pure supergravity theories in dimensions $4\leq 
D\leq11$ with the number of independent 
supercharges equal to $Q=32$, 16, 8 and 4. In 3 spacetime 
dimensions, pure supergravity does not describe propagating 
degrees of freedom and is a topological theory.}
\label{fig:sugry}
\end{figure}

The two chiral spinor spaces correspond to the bosonic and 
fermionic states, respectively. For the massless 
multiplets, the dimensions are shown in Table~\ref{simple-susy}. 
Bigger supermultiplets can be obtained by combining various 
irreducible multiplets in a nontrivial way. We will demonstrate 
this below in three relevant cases, corresponding to 
$D=11, 10$ and 6 spacetime dimensions. 
For the convenience of the reader we present 
Fig.~\ref{fig:sugry}, 
which lists the  pure 
supergravity theories in dimensions $4\leq D \leq 11$ with $Q=32,
16,8,4$.\footnote{%
  In $D=4$  there exist additional theories with $Q=12,20$ and 24;  
  in $D=5$  there exists a theory with $Q=24$ \cite{cremmer2} and most likely there also exits a $Q=24$
supergravity in $D=6$. So far these 
  supergravities have played no role in string theory and hence 
  we omit them from our discussion here. } %
Some of these theories will be discussed later in more detail (in  
particular supergravity in $D=11$ and 10 dimensions). 

%
%

%
 \subsubsection{$D=11$}
In 11 dimensions we are dealing with 32 independent real 
supercharges.  
In odd-dimensional spacetimes irreducible spinors are subject to 
the eigenvalue condition $\tilde\G_D=\pm {\bf 1}$. Therefore 
\eqn{susy-algebra2} simplifies and shows that the 16 
nonvanishing spinor charges transform according to the chiral  
spinor representation of the helicity group SO(9). 

On the other hand, when regarding the 16 spinor charges as gamma 
matrices, it follows that the representation space constitutes 
the spinor representation of SO(16), which 
decomposes into two chiral subspaces, one corresponding to the 
bosons and the other one to the fermions. To determine the 
helicity content of the bosonic and fermionic states,  
one considers the embedding of the SO(9) spinor representation in 
the SO(16) vector transformation. It then turns out that one of 
the $\bf 128$ representations branches into helicity representations 
according to 
${\bf 128}\to {\bf 44} + {\bf 84}$, while the second one 
transforms irreducibly according to the $\bf 128$  
representation of the helicity group.

The above states comprise precisely the massless states 
corresponding to $D=11$ supergravity \cite{cjs}. The graviton states 
transform in the $\bf 44$, the antisymmetric tensor states in the 
$\bf 84$ and the gravitini states in the ${\bf 128}$ representations 
of SO(9). Rather than showing all this in detail, we continue 
with other cases, where the representations are smaller and the 
group theory is more transparent. The helicity representations of 
the graviton, gravitino and tensor gauge fields are discussed in 
the appendix. 
Bigger supermultiplets consist of multiples of 256 states. For 
instance, without central charges, the smallest massive 
supermultiplet comprises $32768+32768$ states. These multiplets 
will not be considered here.

 \subsubsection{$D=10$}
In 10 dimensions the supercharges are both Majorana and Weyl 
spinors. 
The latter means that they are eigenspinors of $\tilde \G_D$. 
According to \eqn{susy-algebra2}, when we have simple (i.e.,  
nonextended) supersymmetry with 16 charges, the nonvanishing charges 
transform in a chiral spinor representation of the SO(8) helicity 
group. With 8 nonvanishing supercharges we are dealing with an 
8-dimensional Clifford  
algebra, whose irreducible representation space corresponds to the 
bosonic and fermionic states, each transforming according to a 
chiral spinor representation. Hence we are dealing with three 
8-dimensional representations of SO(8), which are inequivalent. One is the    
representation to which we assign the supercharges, which we will 
denote by ${\bf 8}_s$; to the other two, denoted as the ${\bf 
8}_v$ and ${\bf 8}_c$ representations, we  
assign the bosonic and fermionic states, respectively. 
The fact that SO(8) representations  
appear in a three-fold variety is known as {\it triality}, which 
is a characteristic property of the group SO(8). With the 
exception of certain  
representations, such as the adjoint and the 
singlet representation, the three types of representation are 
inequivalent. They are traditionally distinguished by labels $s$,
$v$ and $c$ (see, for instance, \cite{slansky}). 

The smallest massless supermultiplet has now been constructed 
with 8 bosonic and 8 fermionic states and corresponds to the 
vector multiplet of supersymmetric Yang-Mills theory in 10 
dimensions \cite{10D-YM}.
Before constructing the supermultiplets that are relevant for 
$D=10$ supergravity, let us first discuss some other properties of 
SO(8) representations. One way to distinguish the inequivalent 
representations, is to investigate how they decompose into 
representations of an SO(7) subgroup. Each of the 8-dimensional 
representations leaves a different SO(7) subgroup
of SO(8) invariant. Therefore there is an 
SO(7) subgroup under which the ${\bf 8}_v$ representation 
branches into 
$$
{\bf 8}_v \longrightarrow {\bf 7} + {\bf 1} . 
$$
Under this SO(7) the other two 8-dimensional representations branch 
into
$$
{\bf 8}_s \longrightarrow {\bf 8} \, , \qquad {\bf 8}_c 
\longrightarrow {\bf 8} \,,
$$
where ${\bf 8}$ is the spinor representation of SO(7). Corresponding 
branching rules for the 28-, 35- and 56-dimensional representations are
\be
\begin{array}{rcl}
{\bf 28} & \longrightarrow  & {\bf 7}   + {\bf 21} \,, \\
{\bf 35}_v & \longrightarrow  & {\bf 1}  +  {\bf 7}  +  {\bf 27} 
\,, \\
{\bf 35}_{c,s} & \longrightarrow  & {\bf 35} \,,
\end{array}\qquad
\begin{array}{rcl}
~          &  ~               &   ~  \\
{\bf 56}_v & \longrightarrow  & {\bf 21} + {\bf 35}  \,,\\
{\bf 56}_{c,s} & \longrightarrow  & {\bf 8}   +  {\bf 48} \,.
\end{array} 
\ee

\begin{table}
\begin{center}
\begin{tabular}{l l l}\hline
supermultiplet & bosons & fermions \\\hline
vector multiplet    & ${\bf 8}_v$            & ${\bf 8}_c$\\
graviton multiplet      & ${\bf 1} +{\bf  28} + {\bf 35}_v$  & 
${\bf 8}_s +{\bf 56}_s$ \\
gravitino multiplet     &  ${\bf 1} +{\bf  28} + {\bf 35}_c$ & 
${\bf 8}_s +{\bf 56}_s$ \\
gravitino multiplet     &  ${\bf 8}_v +{\bf 56}_v$ &  ${\bf 8}_c 
+{\bf 56}_c$ \\ \hline
\end{tabular}
\end{center}
\caption{Massless $N=1$ supermultiplets in $D=10$ spacetime 
dimensions containing $8+8$ or $64 + 64$ bosonic and fermionic 
degrees of freedom.  }\label{D10N1-susy-mult}
\end{table}

In order to obtain the supersymmetry representations relevant for 
supergravity we consider tensor products of the smallest 
supermultiplet consisting of ${\bf 8}_v + {\bf 8}_c$, with one 
of the 8-dimensional
representations. There are thus three different possibilities, each
leading to a 128-dimensional supermultiplet. Using the multiplication 
rules for SO(8) representations,
\be\label{so8-multiplication}
\begin{array}{rcl}
{\bf 8}_v  \times  {\bf 8}_v & = & {\bf 1}  + {\bf 28} + {\bf 
35}_v\,, \\
{\bf 8}_s \times  {\bf 8}_s & = & {\bf 1} +  {\bf 28}   + 
{\bf 35}_s \,, \\
{\bf 8}_c  \times  {\bf 8}_c & = & {\bf 1}  +  {\bf 28}  + 
{\bf 35}_c \,,
\end{array}
\qquad
\begin{array}{rcl}
{\bf 8}_v  \times {\bf 8}_s & = & {\bf 8}_c +  {\bf 56}_c \,, \\
{\bf 8}_s  \times  {\bf 8}_c & = & {\bf 8}_v  +  {\bf 56}_v \,, \\
{\bf 8}_c  \times  {\bf 8}_v & = & {\bf 8}_s  +  {\bf 56}_s \,,
\end{array}
\ee
it is straightforward to obtain these new multiplets. Multiplying 
${\bf 8}_v$ with ${\bf 8}_v + {\bf 8}_c$ yields ${\bf 8}_v\times 
{\bf 8}_v$ bosonic and ${\bf 8}_v\times{\bf 8}_c$ fermionic 
states, and leads to the second supermultiplet shown in 
Table~\ref{D10N1-susy-mult}. This supermultiplet contains the
representation ${\bf 35}_v$, which can be associated with the 
states of the graviton in $D=10$ dimensions (the field-theoretic 
identification of the various states will be clarified in the
appendix). Therefore this supermultiplet will be called the 
{\em graviton multiplet.} Multiplication with ${\bf 8}_c$ or 
${\bf 8}_s$ goes in the same fashion, except that we will
associate the {\bf 8}$_c$ and {\bf 8}$_s$ representations with 
fermionic 
quantities (note that these are the representations to which the 
fermion
states of the Yang-Mills multiplet and the supersymmetry charges are 
assigned).
Consequently, we interchange the boson and fermion assignments in 
these 
products. Multiplication with ${\bf 8}_c$ then leads to ${\bf 
8}_c\times {\bf 8}_c$ bosonic and ${\bf 8}_c\times{\bf 8}_v$ 
fermionic states, whereas multiplication with ${\bf 8}_s$ gives 
${\bf 8}_s\times{\bf 8}_c$
bosonic and ${\bf 8}_s\times{\bf 8}_v$ fermionic states. These
supermultiplets contain fermions transforming according to the ${\bf 
56}_s$ and ${\bf 56}_c$ representations, respectively, which can 
be associated  with gravitino states (see the appendix for the 
helicity assignment of gravitino states), but no graviton states as 
those transform in the  ${\bf 35}_v$ representation. Therefore 
these two supermultiplets are called {\em  gravitino
multiplets}. We have thus established the existence of two 
inequivalent gravitino multiplets. The explicit SO(8) 
decompositions of the vector,
graviton and gravitino supermultiplets are shown in 
Table~\ref{D10N1-susy-mult}.
         
By combining a graviton and a gravitino multiplet it is possible to 
construct an $N=2$ supermultiplet of 128 + 128 bosonic and fermionic states. 
However, since there are two inequivalent gravitino multiplets, 
there will also be two inequivalent $N=2$ supermultiplets 
containing the states corresponding to a
graviton and two gravitini. According to the construction presented 
above, one $N=2$  
supermultiplet may be be viewed as the tensor product of two identical
supermultiplets (namely ${\bf 8}_v + {\bf 8}_c$). Such a multiplet 
follows if one starts from a supersymmetry algebra based on {\em 
two} Majorana-Weyl spinor charges $Q$ with the {\em same} 
chirality. The states of this multiplet decompose as follows:
\begin{eqnarray} \label{eq:IIB}
\mbox{\it Chiral $N=2$ supermultiplet {\rm (IIB)}}\!\!&& 
\nonumber\\[-.55cm]
({\bf 8}_v+ {\bf 8}_c)\times({\bf 8}_v+{\bf 8}_c)&\Longrightarrow&  
\left\{\begin{array}{l}
\mbox{\bf bosons}: \\
{\bf 1}+ {\bf 1} +  {\bf 28} +  {\bf 28}+ {\bf 35}_v + {\bf 35}_c \\[.5cm] 
\mbox{\bf fermions}: \\
{\bf 8}_s+ {\bf 8}_s +  {\bf 56}_s+ {\bf 56}_s 
\end{array}
\right. 
\end{eqnarray}
This is the multiplet corresponding to IIB supergravity 
\cite{SH}.
Because the supercharges have the same chirality, one can perform rotations
between these spinor charges which leave the supersymmetry algebra 
unaffected. Hence the automorphism group $H_{\rm R}$ is equal to 
SO(2). This feature reflects itself in the multiplet 
decomposition, where the $\bf 1$, ${\bf 8}_s$, $\bf 28$ and ${\bf 
56}_s$ representations are degenerate and constitute doublets 
under this SO(2) group. 

A second supermultiplet may be viewed as the tensor  
product of a (${\bf 8}_v + {\bf 8}_s$) supermultiplet with a 
second supermultiplet
(${\bf 8}_v + {\bf 8}_c$). In this case the supercharges 
constitute two Majorana-Weyl spinors of opposite chirality. Now the 
supermultiplet decomposes as follows: 
\begin{eqnarray} \label{eq:IIA}
\mbox{\it Nonchiral $N=2$ supermultiplet {\rm (IIA)}}\!\!&& 
\nonumber\\[-.55cm]
({\bf 8}_v+ {\bf 8}_s)\times({\bf 8}_v+{\bf 8}_c)&\Longrightarrow&  
\left\{\begin{array}{l}
\mbox{\bf bosons}: \\
{\bf 1} +  {\bf 8}_v+ {\bf 28} +{\bf 35}_v + {\bf 56}_v \\[.5cm] 
\mbox{\bf fermions}: \\
{\bf 8}_s+ {\bf 8}_c +  {\bf 56}_s+ {\bf 56}_c 
\end{array}
\right. 
\end{eqnarray}
This is the multiplet corresponding to IIA supergravity \cite{IIASG}.
It can be obtained by a straightforward reduction of $D=11$ 
supergravity. The 
latter follows from the fact that two $D=10$ Majorana-Weyl 
spinors with opposite 
chirality can be combined into a single $D=11$ Majorana spinor.
The formula below summarizes the massless states 
of IIA supergravity from an 11-dimensional perspective. The 
massless states of 11-dimensional supergravity transform according
to the $\bf 44$, $\bf 84$ and $\bf 128$ representation of the 
helicity  group SO(9).  They correspond to the degrees of freedom 
described by the metric, a 3-rank antisymmetric gauge field and 
the gravitino field, respectively. We also show how the 
10-dimensional states can subsequently be branched into 
9-dimensional states,  
characterized in terms of representations of the helicity group 
SO(7):  
\begin{eqnarray} \label{eq:11-IIA-9}
{\bf 44} &\Longrightarrow&  
\left\{\begin{array}{lcl}
{\bf 1}   & \longrightarrow &  {\bf 1} \\
{\bf 8}_v & \longrightarrow & {\bf 1} + {\bf 7}\\
{\bf 35}_v& \longrightarrow & {\bf 1} + {\bf 7} + {\bf 27}
\end{array}
\right. 
\nonumber\\
{\bf 84} &\Longrightarrow&  
\left\{\begin{array}{lcl}
{\bf 28}   & \longrightarrow &   {\bf 7}+{\bf 21}  \\
{\bf 56}_v & \longrightarrow & {\bf 21} + {\bf 35}
\end{array}
\right. 
\\
{\bf 128} &\Longrightarrow&  
\left\{\begin{array}{lcl}
{\bf 8}_s   & \longrightarrow &   {\bf 8}  \\
{\bf 8}_c   & \longrightarrow &  {\bf 8}   \\
{\bf 56}_s  & \longrightarrow &  {\bf 8} +{\bf 48} \\
{\bf 56}_c  & \longrightarrow &  {\bf 8}+{\bf 48}  \\
\end{array}
\right. 
\nonumber
\end{eqnarray}
Clearly, in $D=9$ we have a higher degeneracy of states, related 
to the automorphism group SO(2). We note the presence of graviton and 
gravitino states, transforming in the $\bf 27$ and $\bf 48$ 
representations. 

One could also take the states of the IIB supergravity and 
decompose them into $D=9$ massless states. This leads to 
precisely the same supermultiplet as the reduction of the states 
of IIA supergravity. Indeed, the reductions of IIA and IIB 
supergravity to 9 dimensions, yield the same theory 
\cite{DHS,d-brane1,berg}. To see this at the level of the Lagrangian 
requires certain duality transformations, which we discuss in 
section~3. Hence $Q=32$ 
supergravity is unique in all spacetime  dimensions, except for 
$D=10$. Maximal supergravity will be discussed in subsection~2.4. 
The field-content of the maximal $Q=32$ 
supergravity theories for dimensions $3\leq D\leq 11$ will be 
presented in  two tables (cf. Table~\ref{maximal-sg-bosons} and 
\ref{maximal-sg-fermions}).

 
%
 \subsubsection{$D=6$}
In 6 dimensions we have chiral spinors, which are not Majorana. 
The charge conjugated spinor has the same chirality, so that the 
chiral rotations of the spinors can be 
extended to the group   
USp$(2N_+)$, for $N_+$ chiral spinors. Likewise $N_-$  
negative-chirality spinors transform under USp$(2N_-)$. This is 
already incorporated in  
Table~\ref{susy-charges}. In principle we have $N_+$ positive- 
and $N_-$ negative-chirality charges, but almost all 
information follows from first considering the purely chiral 
case. In Table~\ref{D6-chiral-susy-mult} we present the  
decomposition of the various helicity representations of 
the smallest supermultiplets based on $N_+=1,2,3$ or 4 
supercharges. In $D=6$ dimensions the helicity group SO(4) 
decomposes into the product of two SU(2) groups: SO(4)$\cong ({\rm 
SU}_+(2) \times {\rm SU}_-(2))/{Z_2}$. When we have 
supercharges of only one chirality, the smallest supermultiplet will 
only transform under one SU(2) factor of the helicity group, as 
is shown in Table~\ref{D6-chiral-susy-mult}.\footnote{%
The content of this table also specifies the smallest 
{\em massive} supermultiplets in four dimensions. The SU(2) 
group is then associated with spin in three space dimensions. 
} 

\begin{table}
\begin{center}
\begin{tabular}{l l l l l}\hline
SU$_+$(2)&$N_+=1$&$N_+=2$& $N_+=3$ & $N_+=4$ \\\hline
{\bf 5} & ~      &    ~  &  ~      &   1    \\
{\bf 4} & ~      &    ~  &  1      &   8    \\
{\bf 3} & ~      &    1  &  6      &   27   \\
{\bf 2} & 1      &    4  &  14     &   48   \\
{\bf 1} & 2      &    5  &  14     &   42   \\\hline
~      & $(2+2)_{\bf C}$ & $(8+8)_{\bf R}$ & $(32+32)_{\bf C}$ & 
$(128+128)_{\bf R}$ \\ \hline
\end{tabular}
\end{center}
\caption{Shortest massless supermultiplets of $D=6$ $N_+$-extended 
chiral supersymmetry. The states transform both in the SU$_+$(2) 
helicity group and under a USp(2$N_+$) group. For odd values of 
$N_+$ the representations are complex, for even $N_+$ they can be chosen 
real. Of course, an identical table can be given for 
negative-chirality spinors. }\label{D6-chiral-susy-mult} 
\end{table}

Let us now consider specific supermultiplets. All these 
multiplets are summarized in Table~\ref{D6-susy-mult}. The 
helicity assignments of the states described by gravitons, 
gravitini, vector and tensor gauge fields, and spinor fields are 
presented in the appendix. The simplest case is $(N_+, N_-)=(1,
0)$, where the smallest supermultiplet is the (1,0) {\it 
hypermultiplet}, consisting of a complex doublet of spinless 
states and a chiral 
spinor. Taking the tensor product of the 
smallest supermultiplet with the $(2,1)$ helicity representation 
gives the (1,0) {\it tensor multiplet}, with a selfdual tensor, a 
spinless state and a doublet of chiral spinors. The tensor product with 
the $(1,2)$ helicity representation yields the (1,0) {\it vector 
multiplet}, with a vector state, a doublet of chiral spinors and a scalar. 
Multiplying the latter with the (2,3) helicity representation, one obtains 
the states of (1,0) {\it supergravity}. Observe that the selfdual 
tensor fields in the tensor and supergravity supermultiplet are 
of opposite selfduality phase. 

Next consider $(N_+,N_-)=(2,0)$ supersymmetry. The smallest 
multiplet, shown in Table~\ref{D6-chiral-susy-mult}, then 
corresponds to the (2,0) {\it tensor multiplet}, with the bosonic states 
decomposing into a selfdual tensor, and a five-plet of spinless 
states, and a four-plet of chiral fermions. Multiplication with 
the (1,3) helicity representation yields the (2,0) supergravity 
multiplet, consisting of the graviton, four chiral gravitini and 
five selfdual tensors  \cite{townsend}. Again, the selfdual 
tensors of the tensor   
and of the supergravity supermultiplet are of opposite selfduality  
phase.

\begin{table}
\begin{center}
\begin{tabular}{l l l l}\hline
multiplet          &\#& bosons        & fermions  \\ \hline
(1,0) hyper        &$4+4$  & $(1,1;2,1) + {\rm h.c.} $ & $(2,1;1,1)$ \\[1mm]
(1,0) tensor       &$4+4$  & $(3,1;1,1) + (1,1;1,1)$   & $(2,1;2,1)$ \\[1mm]
(1,0) vector       &$4+4$  & $(2,2;1,1)$               & $(1,2;2,1)$ \\[1mm]
(1,0) supergravity &$12+12$& $(3,3;1,1) + (1,3;1,1)$   & $(2,3;2,1)$ \\[1mm]
(2,0) tensor       &$8+8$  & $(3,1;1,1) + (1,1;5,1)$   & $(2,1;4,1)$ \\[1mm]
(2,0) supergravity &$24+24$& $(3,3;1,1) + (1,3;5,1)$   & $(2,3;4,1)$ \\[1mm]
(1,1) vector       &$8+8$  & $(2,2;1,1) + (1,1;2,2)$   & $(2,1;1,2) 
         + (1,2;2,1)$ \\[1mm]
(1,1) supergravity &$32+32$& $(3,3;1,1)$               & $(3,2;1,2) 
         + (2,3;2,1)$ \\
~                  &~ & $+(1,3;1,1) + (3,1; 1,1)$ & $+(1,2;1,2) 
         + (2,1;2,1)$ \\
~                  &~ & $+(1,1;1,1) + (2,2;2,2)$  & ~\\[1mm]
(2,2) supergravity &$128+128$& $(3,3;1,1)$              & $(3,2;4,1) 
         + (2,3;1,4)$ \\
~                  &~ & $+(3,1;1,5)+(1,3;5,1)$    & $+(2,1;4,5) 
         + (1,2;5,4)$ \\
~                  &~ & $+(2,2;4,4) + (1,1;5,5)$  & ~ \\ \hline
\end{tabular}
\end{center}
\caption{Some relevant $D=6$ supermultiplets with $(N_+,N_-)$ 
supersymmetry. The states $(n,m;\tilde n,\tilde m)$ are assigned 
to $(n,m)$ representations of  
the helicity group SU$_+(2)\times {\rm SU}_-(2)$ and $(\tilde n,
\tilde m)$ representations of  
USp$(2N_+)\times {\rm USp}(2N_-)$. The second column lists 
the number of bosonic + fermionic states for each 
multiplet. }\label{D6-susy-mult}
\end{table}

Of course, there exists also a nonchiral version with 16 supercharges, 
namely the one corresponding to $(N_+,N_-)=(1,1)$. The smallest 
multiplet is now given by the tensor product of the 
supermultiplets with (1,0) and (0,1) supersymmetry. This yields 
the vector multiplet, with the vector state and four scalars, 
the latter transforming with respect to the (2,2) representation of 
USp$(2) \times {\rm USp}(2)$. There are two doublets of chiral 
fermions with opposite chirality, each transforming as a doublet 
under the corresponding USp(2) group.
Taking the tensor product of the vector multiplet with the (2,2) 
representation of the helicity group yields the states of the (1,
1) {\it supergravity} multiplet. It consists of 32 bosonic states, 
corresponding to a graviton, a tensor, a 
scalar and four vector states, where the latter transform under the 
(2,2) representation of USp$(2) \times {\rm USp}(2)$. The 32 
fermionic states comprise two doublets of chiral gravitini and two 
chiral spinor doublets, transforming as doublets under the 
appropriate USp(2) group.

Finally, we turn to the case of $(N_+,N_-)=(2,2)$. The smallest 
supermultiplet is given by the tensor product of the smallest 
(2,0) and (0,2) supermultiplets. This yields the 128 + 128 states 
of the (2,2) {\it supergravity} multiplet. These states transform 
according to representations of USp$(4) \times {\rm USp}(4)$. 

In principle, one can continue and classify representations for 
other values of $(N_+,N_-)$. As is obvious from the construction 
that we have presented, this will inevitably lead to states 
transforming in higher-helicity representations. As we 
will discuss in subsection~2.4, the higher-spin gauge fields 
associated with these representations can not be coupled to 
gravity. Although the representations exist and can be described by 
appropriate free-field theories, they have no future as 
nontrivial quantum field theories.

 \subsection{Maximal supersymmetry: $Q\leq32$}

In the above we have restricted ourselves to (massless) supermultiplets 
based on $Q\leq32$ supercharge components. {From} the general analysis  
it is clear that increasing the number of supercharges leads to 
higher and higher helicity representations. Obviously some of 
these representations will also occur in lower-$Q$ supermultiplets, 
by multiplying shorter multiplets by suitable helicity 
representations. It is not so easy to 
indicate in arbitrary dimension what we mean by a higher helicity 
representation, but we have in mind those representations that are 
described by gauge fields that are {\it symmetric} Lorentz tensors. 
Symmetric tensor gauge fields for 
arbitrary helicity states can be constructed (in four dimensions, 
see, for instance,  \cite{fronsdal}). However, it turns out that 
symmetric gauge fields cannot consistently couple to themselves 
or to other fields. An exception is the graviton field, which 
can interact with itself as well as to low-spin matter, but not to 
other higher-spin gauge fields. By consistent, we mean that their 
respective gauge invariances of the higher-spin fields (or 
appropriate deformations thereof) cannot be  
preserved at the interacting level. 

There have been many efforts to circumvent this apparent no-go 
theorem. What seems clear, is that one needs a combination of the 
following ingredients in order to do this (for a recent review, 
see \cite{vasiliev}): (i) an  
infinite tower of higher-spin gauge fields; (ii) interactions 
that are inversely proportional to the cosmological constant; 
(iii) extensions of the super-Poincar\'e or the super-de~Sitter 
algebra with additional fermionic and bosonic charges. 

Conventional supergravity theories are not of 
this kind. This is the reason why we have avoided (i.e. in 
Table~\ref{D6-susy-mult}) to  
list supermultiplets with states transforming in higher-helicity 
representations. The bound $Q\leq32$ originates from the 
necessity of avoiding the higher-spin fields. It implies that 
supergravity does not exist for spacetime 
dimensions $D\geq11$ (at least, if one assumes a single time 
coordinate), because  Lorentz spinors have more than 32 
components beyond $D=11$ \cite{nahm}. 

Most of the search for interacting higher-spin fields was 
performed in four spacetime dimensions \cite{higher-spin}. When 
one increases the  
number of supercharges beyond $Q=32$, then a supermultiplet will 
contain several massless states of spin-2 and at least spin-5/2 
fermions. In higher spacetime dimensions, more than 32 
supercharges are excluded (in the absence of higher-spin gauge 
fields), because, upon dimensional reduction, 
these theories would give rise to theories that are inconsistent 
in $D=4$. There is also direct  
evidence in $D=3$, where graviton and gravitini fields do not 
describe dynamic degrees of freedom. Hence, one can write down 
supergravity theories based on a graviton field and an arbitrary 
number of gravitino fields, which are topological. However, when 
coupling matter to this theory, described by scalars and spinors, 
the theory supports not more than 32 supercharges. 
Beyond $Q=16$ there are four unique theories with $Q=18,20, 
24$~and~32 \cite{DWTN}.

\begin{table}
\begin{center}
\begin{tabular}{l l l l l l l l }\hline
$D$ &$H_{\rm R}$& graviton & $p=-1$ & $p=0$ & $p=1$ & $p=2$ & 
$p=3\!$  \\   \hline
11  & 1      & 1  & 0   & 0   & 0  & 1  & 0 \\[.5mm]
10A$\!\!\!$ & 1      & 1  & 1   & 1   & 1  & 1  & 0 \\[.5mm]
10B$\!\!\!$ & SO(2)  & 1  & 2   & 0   & 2  & 0  & $1^\ast$ \\[.5mm]
9   & SO(2)  & 1  &$2+1$&$2+1$& 2  & 1  & ~ \\[.5mm]
8   & U(2)   & 1  & $5+1+\bar 1\!$ & $3+\bar 3$ &3 &$[1]$&~\\[.5mm]
7   & USp(4) & 1  & 14  & 10  & 5  & ~  & ~ \\[.5mm]
6   & USp$(4)\!\times\! {\rm USp}(4)$  & 1 & (5,5) & (4,4) 
   & $(5,1)+(1,5)\!\!\!\!$ & ~ & ~ \\[.5mm]
5   & USp(8) & 1   & 42 & 27   & ~  & ~  & ~ \\[.5mm]
4   & U(8)   & 1   & $35+\overline{35}$ & $[28]$ & ~ & ~ & ~ \\[.5mm]
3   & SO(16) & 1   & 128 & ~   & ~  & ~  & ~ \\ \hline
\end{tabular}
\end{center}
\caption{Bosonic field content for maximal supergravities. The 
$p=3$ gauge field in $D=10$B has a self-dual field strength. The 
representations [1] and [28] (in $D=8,4$, respectively) are 
extended to U(1) and SU(8) representations through duality transformations 
on the field strengths. These transformations can not be 
represented on the vector potentials. In $D=3$ dimensions, the 
graviton does not describe propagating degrees of freedom.
}\label{maximal-sg-bosons} 
\end{table}

\subsection{Maximal supergravities} 
In this section we review the maximal supergravities in various 
dimensions. These theories have precisely $Q=32$ supercharge 
components. We restrict our discussion to $3\leq D\leq 11$. 

The bosonic fields always comprise the metric tensor for the 
graviton field and a number of antisymmetric gauge fields. For 
the antisymmetric gauge fields, it is a priori unclear whether to 
choose a $(p+1)$-rank gauge field or its dual $(D-3-p)$-rank  
partner, but it turns out that the interactions often prefer  
the rank of the gauge field to be as small as possible. 
Therefore, in Table~\ref{maximal-sg-bosons}, 
we restrict ourselves to $p\leq 3$, as in $D=11$ dimensions, 
$p=3$ and $p=4$ are each other's dual conjugates. This table presents 
all the field configurations for maximal supergravity in various 
dimensions. Obviously, the problematic higher-spin fields are 
avoided, because the only symmetric gauge field is the one 
describing  the graviton. In Table~\ref{maximal-sg-fermions} we also 
present the fermionic fields, always consisting of gravitini and 
simple spinors. All these fields are classified as 
representations of the automorphism group $H_{\rm R}$. In order 
to compare these tables to similar tables in the literature, one 
may need to use the (local) equivalences:  USp(4)$\sim$SO(5), 
USp(2)$\sim$SU(2) and SU(4)$\sim$SO(6).

\begin{table}
\begin{center}
\begin{tabular}{l l l l}\hline
$D$ &$H_{\rm R}$& gravitini & spinors   \\ \hline
11  & 1      &  1  & 0    \\[.5mm]
10A & 1      & 1+1 & 1+1    \\[.5mm]
10B & SO(2)  &  2  & 2    \\[.5mm]
9   & SO(2)  &  2  &$2+2$ \\[.5mm]
8   & U(2)   &  $2+\bar 2$  & $2+\bar 2+ 4 +\bar 4$  \\[.5mm]
7   & USp(4) &  4  & 16  \\[.5mm]
6   & USp$(4)\!\times\! {\rm USp}(4)$  & $(4,1)+(1,4)$ 
   & $(4,5) + (5,4)$  \\[.5mm]
5   & USp(8) & 8   & 42  \\[.5mm]
4   & U(8)   & $8+\bar 8$   & $56 +\overline{56}$ \\[.5mm]  
3   & SO(16) & 16  & 128 \\ \hline
\end{tabular}
\end{center}
\caption{Fermionic field content for maximal supergravities. For 
$D=5,6,7$ the fermion fields are counted as symplectic Majorana 
spinors. For $D=4,8$ we include both chiral and antichiral spinor 
components, which transform in conjugate representations of 
$H_{\rm R}$. In $D=3$ dimensions the gravitino does not correspond 
to propagating degrees of freedom.
}\label{maximal-sg-fermions} 
\end{table}

The supersymmetry algebra of the maximal supergravities comprises 
general coordinate transformations, local supersymmetry 
transformations and the gauge transformations associated with the 
antisymmetric gauge fields\footnote{%
   There may be 
   additional gauge transformations that are of interest to us. 
   As we discuss in subsection~2.5.1, it is possible to have 
   (part of the) the automorphism group $H_{\rm R}$ realized as a 
   local invariance. However, the corresponding gauge fields are 
   then composite  
   and do not give rise to physical states (at least, 
   not in perturbation theory).}. %
These gauge transformations usually appear in the 
anticommutator of two supercharges, and may be regarded as 
central charges. In perturbation theory, the theory 
does not contain charged fields, so these central charges simply vanish 
on physical states. However, at the nonperturbative level, there 
may be solitonic or other states that carry charges. An example 
are magnetic monopoles, dyons, or extremal black holes. On 
such states, some of the central charges may take finite values. 
Without further knowledge about the kind of states  
that may emerge at the nonperturbative level, we can  
generally classify the possible central charges, by considering a decomposition of the anticommutator. 
This anticommutator carries at least two spinor indices and two 
indices associated with the group $H_{\rm R}$. Hence we may write 
\be
\{Q_\a, Q_\b\} \propto \sum_r \;(\G^{\m_1\cdots \m_r} C)_{\a\b} \, 
Z_{\m_1\cdots \m_r}\,,
\ee
where $\G^{\m_1\cdots \m_r}$ is the antisymmetrized product of 
$r$ gamma matrices, $C$ is the charge-conjugation matrix and 
$Z_{\m_1\cdots \m_r}$ is the central charge, which transforms as 
an antisymmetric $r$-rank Lorentz tensor and depends on 
possible additional $H_{\rm R}$ indices attached to the 
supercharges. The central  
charge must be symmetric or antisymmetric in these indices, 
depending on whether the product of the gamma matrices with $C$ is 
symmetric or antisymmetric, so that the product is symmetric in 
the combined indices of the supercharges. For given spacetime 
dimension all 
possible central charges can be classified.\footnote{%
For a related discussion see for example
\cite{Mtheory,sergio} and references therein.}
 For the maximal 
supergravities in spacetime dimensions $3\leq D\leq11$ this 
classification is given in Table~\ref{maximal-central-extension}. 
Because we have 32 supercharge components, the sum of the 
independent momentum operators and the central charges must be 
equal to $(32\times 33)/2 = 528$. 

\begin{table}
\begin{center}
\begin{tabular}{l l l l l l l l}\hline
$D$ &$H_{\rm R}$& $r=0$ & $r=1$ & $r=2$ & $r=3$ & $r=4$ & 
$r=5$  \\ \hline
11  & 1      & ~  & ~           & 1   & ~  & ~  & 1 \\
~  & ~      & ~   & ~           & $[55]$  & ~  & ~  & 
$[462]$ \\[1mm]
10A$\!\!\!$ & 1      & 1  & $1$       & 1   & ~  & 1    & $1+1$ \\
~   & ~      & $[1]$  & $[10]$    &$[45]$   & ~  & $[210]$  &  
$[126]$ \\[1mm]
10B$\!\!\!$ & SO(2)  & ~  & $2$& ~   & 1   & ~  & $1+2$ \\
~   & ~      & ~  & $[10]$          & ~   & $[120]$ & ~  & 
$[126]$          \\[1mm]
9   & SO(2)  & $1+2$ & $2$ & 1 & 1  & $1+2$ & ~ \\
~   &        & $[1]$           & $[9]$          &$[36]$ &$[84]$  
& $[126]$         & ~ \\[1mm]
8   & U(2)   & $3+\bar 3$  & $3$      & $1+\bar 1$& $1+3$ 
&$3+\bar 3$ & ~ \\
~   &        & $[1]$           & $[8]$          & $[28]$        & 
$[56]$    &$[35]$ & ~ \\[1mm]
7   & USp(4) & 10  & $5$ & $1+5$      & 10  & ~  & ~ \\
~   &        & $[1]$   & $[7]$     & $[21]$         & $[35]$  & ~ 
 & ~ \\[1mm]
6   & USp$(4)\!\!\times\!\!{\rm USp}(4)\!\!$  & $(4,4)$ & $(1,1) + 
(5,1)\!$ & (4,4) & $(10,1)$ & ~ & ~ \\ 
~   & ~  &   ~   & $+(1,5)$& ~    & $+(1,10)$ &~&~\\
~   & ~  & $[1]$ & $[6]$ & $[15]$ & $[10]$ & ~ & ~ \\[1mm] 
5   & USp(8) & $1+27$   & $27$ & 36   & ~  & ~  & ~ \\
~   & ~      & $[1]$        & $[5]$      & $[10]$   & ~  & ~  & ~ 
\\[1mm]
4   & U(8) & $28+\overline{28}$ & $63$& $36+\overline{36}$ & ~ & ~ & ~ \\  
~   & ~    & $[1]$                 & $[4]$    & $[3]$      & ~ & 
~ & ~ \\[1mm] 
3   & SO(16) & 120   & $135$ & ~   & ~  & ~  & ~ \\
~   & ~      & $[1]$     & $[3]$   & ~   & ~  & ~  & ~ \\[1mm]
 \hline
\end{tabular}
\end{center}
\caption{Decomposition of the central extension in the 
supersymmetry algebra with $Q=32$ supercharge components in terms 
of $r$-rank Lorentz tensors. The second row specifies the number of 
independent components for each $r$-rank tensor charge. The total 
number of central charges is equal to $528-D$, because we have 
not listed the $D$ independent momentum operators}
\label{maximal-central-extension} 
\end{table}

%
 \subsubsection{$D=11$}
Supergravity in 11 spacetime dimensions is based on an ``elfbein" 
field $E_M^{\;A}$, a Majorana gravitino field $\Psi_M$  
and a 3-rank antisymmetric gauge field $C_{MNP}$. With chiral 
(2,0) supergravity in 6 dimensions, it is the 
only $Q\geq16$ supergravity theory without a scalar field. Its 
Lagrangian can be written as follows \cite{cjs},  
\bea\label{D11-lagrangian}
{\cal L}_{11}&\!\!=\!\!& {1\over \kappa_{11}^2} \bigg[ -\ft12 E\, R(E,
\Omega)  -\ft12 E\bar\Psi_M\G^{MNP}D_N(\Omega)\Psi_P -\ft1{48}E 
(F_{MNPQ})^2 \nonumber\\   
&& \hspace{8.5mm}- \ft1{3456} \sqrt{2} \, \varepsilon^{MNPQRSTUVWX} 
\,F_{MNPQ} \,F_{RSTU} \,C_{VWX}  \\
&&\hspace{8.5mm} - \ft1{192}\sqrt{2} E \Big(\bar\Psi_R \G^{MNPQRS} 
\Psi_S + 12 \,\bar\Psi^M  \G^{NP} \Psi^Q\Big) F_{MNPQ} + 
\cdots\bigg] \,, \nonumber
\eea
where the ellipses denote terms of order $\Psi^4$, $E= \det 
E_M^{\;A}$ and $\Omega_M{}^{\!AB}$ denotes the spin connection. 
The supersymmetry transformations are equal to 
\bea
\d E^{\,A}_M   &\!=\!& \ft12 \,\bar \e\,\G^A\Psi_M\,,\nonumber \\
\d C_{MNP} &\!=\!& -\ft18\sqrt{2} \,\bar \e\,\G_{[MN}\Psi_{P]} \,,
 \\
\d\Psi_M   &\!=\!&  D_M(\hat\Omega)\,\e + \ft1{288} \sqrt{2} \,
\Big(\G_M{}^{\!NPQR}- 8\, \d_M^{\,N}\, \G^{PQR} \Big)\,\e\, \hat 
F_{NPQR}\,.  \nonumber 
\eea
Here the covariant derivative is covariant with respect to local 
Lorentz transformation
\be
D_M(\Omega)\,\e = \Big(\pa_M -\ft14\Omega_M{}^{AB} \G_{AB}\Big) \e\, ,
\ee
and $\hat F_{MNPQ}$ is the supercovariant field strength
\be
\hat F_{MNPQ} = 24\, \pa_{[M}C_{NPQ]} + \ft32 \sqrt{2} \,\bar 
\Psi_{[M}\G_{NP}\Psi_{Q]} \,.
\ee
Note the presence in the Lagrangian of a Chern-Simons-like term 
$F\wedge F \wedge C$, so that the action is only invariant up to 
surface terms. We also wish to point out that the 
quartic-$\Psi$ terms can be included into the Lagrangian 
\eqn{D11-lagrangian} by replacing the spin-connection field 
$\Omega$ by $(\Omega+\hat \Omega)/2$ in the 
covariant derivative of the gravitino kinetic term and by 
replacing $F_{MNPQ}$ in the last line by $(\hat 
F_{MNPQ}+F_{MNPQ})/2$. These substitutions ensure that the field 
equations corresponding to \eqn{D11-lagrangian} are 
supercovariant.  The Lagrangian is derived in the context of the 
so-called ``1.5-order'' formalism, in which the spin connection is 
defined as a dependent field determined by its (algebraic) 
equation of motion, whereas its supersymmetry variation in the 
action is treated as if it were an independent field 
\cite{1.5-order}. The supercovariant spin connection is the 
solution of the following equation,
\be
D_{[M} (\hat\Omega) \,E_{N]}^{\,A} - \ft14 \bar \Psi_M \G^A\Psi_N 
=0\,.
\ee
The left-hand side is the supercovariant torsion tensor.  

We have the following bosonic field 
equations and Bianchi identities, 
\bea\label{D11-field-eq}
R_{MN} &=& \ft1{72}g_{MN} \, F_{PQRS}F^{PQRS} -\ft16 
F_{MPQR}\,F_N{}^{\!PQR}\,, \nonumber\\ 
\pa_{M}\Big(E\,  F^{MNPQ}\Big) &=& \ft1{1152} \sqrt{2}\, 
\varepsilon^{NPQRSTUVWXY} F_{RSTU}\,F_{VWXY}\,, \nonumber \\
\pa_{[M}F_{NPQR]}&=&0\,,
\eea
which no longer depend on the antisymmetric gauge field. 
An alternative form of the second equation is \cite{page}
\be\label{D11-field-eq2}
\pa_{[M}H_{NPQRSTU]} =0\,,
\ee
where $H_{MNPQRST}$ is the dual field strength,
\be
H_{MNPQRST} = {1\over 7!}E\, \varepsilon_{MNPQRSTUVWX} F^{UVWX} 
-\ft12 \sqrt{2}  \, F_{[MNPQ}\,C_{RST]}\,.
\ee
One could imagine that the third equation of \eqn{D11-field-eq} and 
\eqn{D11-field-eq2} receive contributions from charges that would 
give rise to source terms on the right-hand side of the equations. 
These charges are associated with the `flux'-integral of $H_{MNPQRST}$ 
and $F_{MNPQ}$ over the boundary of an 8- and a 5-dimensional 
spatial volume, respectively. This volume is transverse to a  
$p=2$ and $p=5$ brane configuration, and the corresponding charges are 2- and 
5-rank Lorentz tensors. These are just the charges that can appear as 
central charges in the supersymmetry algebra, as one can verify 
in Table~\ref{maximal-central-extension}. Solutions of 
11-dimensional supergravity that contribute to these charges were 
considered in \cite{duffstelle,gueven,solirev,Mtheory}. 

It is straightforward to evaluate the supersymmetry algebra on 
these fields. The commutator of two supersymmetry transformations 
yields a general-coordinate transformation, a local Lorentz 
transformation, a supersymmetry transformation and a gauge 
transformation associated with the tensor gauge field,
\be
[\d(\e_1),\d(\e_2) ] = \d_{\rm gct} (\xi^M) + \d_Q(\e_3)
+ \d_L(\l^{AB}) + \d_A(\xi_{MN}) \,.
\ee
The parameters of the transformations on the right-hand side 
are given by
\bea
\xi^M&\!=\!& \ft12 \,\bar\e_2\G^M\e_1 \,,\nonumber\\[1.1mm] 
\e_3&\!=\!& -\xi^M \Psi_M \,, \nonumber\\[.8mm]
\l^{AB} &\!=\!& -\xi^M \hat \Omega^{AB}_M + \ft1{288}\sqrt{2}\, 
\bar\e_2\Big[ \G^{ABCDEF}\hat F_{CDEF} + 24\,   
\G_{CD}\hat F^{ABCD}\Big]\e_1\,, \nonumber\\
\xi_{MN} &\!=\!& -\ft18\sqrt{2} \,\bar \e_2 \G_{MN}\e_1\,.
\eea
Note that the normalizations differ from the ones used in the 
supersymmetry algebra in previous subsections.  The tensor  
gauge field transforms under gauge transformations as $\d 
C_{MNP} = \pa_{[M}\xi_{NP]}$. 

Finally, the constant $1/\kappa_{11}^2$ in front of the Lagrangian 
\eqn{D11-lagrangian}, which has the dimension $[{\rm 
length}]^{-9}\sim [{\rm mass}]^9$, is 
undetermined and depends on  
fixing some length scale.
To see this consider a continuous  rescaling of the fields, 
\be  
E_M^{\;A}\to {\rm e}^{-\a} E_M^{\;A}\,, \qquad \Psi_M\to {\rm 
e}^{-\a/2}\Psi_M\,, \qquad C_{MNP} \to {\rm e}^{-3\a}C_{MNP} \,.  
\label{scale-11}
\ee
Under this rescaling the Lagrangian changes 
according to
\be
{\cal L}_{11} \to {\rm e}^{-9\a}{\cal L}_{11}\,.
\ee
This change can then be absorbed into a redefinition of 
$\kappa_{11}$,\footnote{%
  Note that the rescalings also leave the supersymmetry 
  transformation rules unchanged, provided the supersymmetry 
  parameter $\e$ is changed accordingly.} 
\be \label{scale2-11}
\kappa_{11}^2\to {\rm e}^{-9\a} \kappa_{11}^2\,.
\ee
The indetermination of $\kappa$ is not a special property of 
$D=11$ but occurs in any spacetime dimension
where the Einstein-Hilbert action displays a similar
scaling property, 
\be\label{scaleD}
g_{\m\n}^D \to {\rm e}^{-2\a}g_{\m\n}^D\, ,\qquad
{\cal L}_{D} \to {\rm e}^{(2-D)\a}{\cal L}_{D}\,,\qquad
\kappa_{D}^2\to {\rm e}^{(2-D)\a} \kappa_{D}^2\,.
\ee
Newton's constant, 
$(\kappa^2_D)^{\scriptscriptstyle \rm physical}$, 
does not necessarily coincide with the parameter 
$\kappa_{D}^2$ but
also depends on the precise value 
adopted for the (flat) metric in 
the ground state of the theory. Up to certain 
convention-dependent normalization factors
one defines 
\be
(\kappa^2_D)^{\scriptscriptstyle \rm physical} := \kappa_D^2\, 
\l^{(2-D)/2} \,, \label{Newtoncc}
\ee
where $g_{\m\n}^D$ is expanded about $\l\,\eta_{\m\n}^D$, 
with $\eta_{\m\n}^D$ equal to the Lorentz-invariant flat metric 
with diagonal elements equal to $\pm 1$. Note that 
$(\kappa^2_D)^{\scriptscriptstyle \rm physical}$ is invariant under
the scale transformations (\ref{scaleD})
and thus a physically meaningful scale.

When the Lagrangian contains additional terms, for instance, of 
higher order in the Riemann tensor, then the corresponding 
coupling constant will scale differently under \eqn{scaleD}. 
Its physical value will therefore depend in a different way on the 
parameter $\lambda$ that parametrizes the (flat) metric in the ground state. 
An even simpler example is a scalar massive field, added to the 
Einstein-Hilbert Lagrangian. Its physical mass is equal to 
$\lambda$ times the mass parameter in the Lagrangian. However, we 
should stress that the physics never depends {\it explicitly} on 
$\lambda$, provided one expresses all physical quantities in 
terms of physical parameters, all determined for the same value 
of $\lambda$. We return to the issue of frames and scales 
in section~2.5.1 and in appendix B.

   \subsection{Dimensional reduction and hidden symmetries}
The maximal supergravities in various dimensions are related by 
dimensional reduction. Here some of the spatial dimensions are 
compactified on a hyper-torus whose size is shrunk to zero. In 
this situation some of the gauge symmetries that are related to 
the compactified dimensions survive and take the form of internal 
symmetries. The aim of our discussion here is to elucidate a 
number of features related to these symmetries, mainly in the 
context of the reduction of $D=11$ supergravity to $D=10$ 
dimensions. 

We denote the compactified coordinate by $x^{11}$ which now 
parameterizes a circle of length $L$.\footnote{%
   Throughout these lectures we enumerate spacetime coordinates 
   by $0,1,\ldots, D-1$. Nevertheless, we denote the compactified  
   coordinate by $x^{11}$, to indicate that it is the eleventh 
   spacetime coordinate.} %
The fields are thus decomposed 
as periodic functions in $x^{11}$ on the interval $0\leq 
x^{11}\leq L$. This results  
in a spectrum of massless modes and an infinite tower of massive 
modes. The massless modes form the basis of the lower-dimensional 
supergravity theory. Because a toroidal background does not break 
supersymmetry, the resulting supergravity has  
the same number of supersymmetries as the original one. For 
compactifications on less trivial spaces than the hyper-torus 
(which we will discuss in section~3) this is not necessarily the case and the 
number of independent supersymmetries can be reduced. 
Actually, fully supersymmetric compactifications are rare. For 
instance, in 11-dimensional supergravity 7 coordinates can be 
compactified in precisely two ways such that all supersymmetries 
remain unaffected \cite{biran}. One is the compactification on a 
torus $T^7$, the other one the compactification of a sphere 
$S^7$. However, in the latter case the resulting 4-dimensional 
supergravity theory acquires a cosmological  
term. In the context of these lectures, such 
compactifications are less relevant and will not be discussed.   

In the formulation of the compactified theory, it is important to 
decompose the higher-dimensional fields in such a way that they 
transform covariantly under the lower-dimensional gauge 
symmetries, and in particular under diffeomorphisms of the 
lower-dimensional spacetime. This ensures that various  
complicated mixtures of massless modes with the tower of massive 
modes will be avoided. It is a key element in ensuring that 
solutions of the lower-dimensional theory remain solutions of the 
original higher-dimensional one. Another point of interest concerns the 
nature of the massive supermultiplets. Because these originate 
from supermultiplets that are massless in higher dimensions, 
these multiplets must be shortened by the presence of central 
charges. The central charge here originates from the momentum 
operator in the compactified dimension. We return to this issue 
shortly. 

The emergence of new internal symmetries in theories that 
originate from a higher-dimensional setting, is a standard feature of 
Kaluza-Klein theories \cite{KK}. Following the discussion in \cite{DWVVP} 
we distinguish between symmetries that have a direct explanation 
in terms of the symmetries in higher dimensions, and symmetries 
whose origin is obscure from a higher-dimen\-sional viewpoint. 
Let us start with the symmetries associated with the metric  
tensor. The 11-dimensional metric can be decomposed according 
to 
\be \label{KKmetric}
{\rm d}s^2 = g_{\m\n} \,{\rm d}x^\m{\rm d}x^\n  + {\rm 
e}^{4\phi/3} 
({\rm d}x^{11}+ V_\m{\rm d}x^\m)({\rm d}x^{11}+ V_\n{\rm 
d}x^\n)\,,
\ee
where the indices $\mu,\nu$ label the 10-dimensional coordinates 
and the factor multiplying $\phi$ is for convenience later. 
The massless modes correspond to the $x^{11}$-independent parts of the 
10-dimensional metric $g_{\m\n}$, the vector field $V_\m$ and 
the scalar $\phi$. 
Here the $x^{11}$-independent component of $V_\m$ acts as a gauge 
field associated with reparametrizations of the circle coordinate 
$x^{11}$ with an arbitrary function $\xi(x)$ of the 10 remaining 
spacetime coordinates $x^\mu$. Specifically, we have $x^{11} \to 
x^{11} - \xi(x)$ and $x^\m\to x^\m$, corresponding to 
\be 
V_\m(x)\to V_\m(x)+ \pa_\m \xi(x)\,.
\ee
The massive modes, which correspond to the Fourier modes in terms 
of $x^{11}$, couple to this gauge field with a charge that is a 
multiple of 
\be 
e_{\scriptscriptstyle\rm KK} = {2\pi\over L}\,. \label{KK-charge}
\ee

Another symmetry of the lower-dimensional theory is more subtle 
to identify.\footnote{%
  There are various discussions of this symmetry in the 
  literature. Its existence in 10-dimensional supergravity was 
  noted long ago (see, e.g. \cite{cremmer2,cremmer3}) and an extensive 
  discussion can be found in \cite{berg}. Our derivation here was 
  alluded to in \cite{DWVVP}, which deals with isometries in 
  $N=2$ supersymmetric Maxwell-Einstein theories in $D=5,4$ 
  and 3 dimensions.} %
In the previous subsection we identified certain 
scale transformations of the $D=11$ fields, which did not leave 
the theory invariant but could be used to adjust the coupling 
constant $\kappa_{11}$. In the compactified situation we can also 
involve the compactification length into the dimensional 
scaling. The integration over $x^{11}$ introduces an overall 
factor $L$ in the action (we do not incorporate any $L$-dependent 
normalizations in the Fourier sums, so that the 10-dimensional 
and the 11-dimensional fields are directly proportional). Therefore, 
the coupling constant that emerges in the 10-dimensional theory 
equals
\be\label{def-L}
{1\over \kappa_{10}^2}= {L\over \kappa_{11}^2}\,,
\ee
and has the dimension $[{\rm mass}]^8$. 
However, because of the invariance under diffeomorphisms, $L$ 
itself has no intrinsic meaning. It simply expresses the length of the 
periodicity interval of $x^{11}$, which itself is a coordinate 
without an intrinsic meaning. Stated differently, we can 
reparameterize $x^{11}$ by some diffeomorphism, as long as we 
change $L$ accordingly. In particular, we may rescale $L$ according to
\be
L \to {\rm e}^{-9\a} L \,,
\ee
corresponding to a reparametrization of the 11-th coordinate,
\be
x^{11}\to {\rm e}^{-9\a} x^{11} \,,\label{diff-11}
\ee 
so that $\kappa_{10}$ remains invariant. Consequently we are then
dealing with a symmetry of the Lagrangian.  

In the effective 10-dimensional theory, the scale transformations 
\eqn{scale-11} are thus suitably combined with the diffeomorphism 
\eqn{diff-11} to yield an invariance of the Lagrangian. For the 
fields corresponding to the 11-dimensional metric, these 
combined transformations are given by\footnote{%
   Note that this applies to all Fourier modes, as they depend on 
   $x^{11}/L$, which is insensitive to the scale transformation.}
\be \label{scale-10x}
e_\m^a \to  {\rm e}^{-\a}e_\m^a \,,\qquad \phi \to  \phi + 12\a 
\,,\qquad V_\m\to {\rm e}^{-9\a} V_\m \,.
\ee
The tensor gauge field $C_{MNP}$ decomposes into a 3- and a 
2-rank tensor in 10 dimensions, which transform according to
\be \label{scale-10y}
C_{\m\n\rho}\to {\rm e}^{-3\a} C_{\m\n\rho}\,, \qquad 
C_{11\m\n} \to  {\rm e}^{6\a}  C_{11\m\n}\,.
\ee

The presence of the above scale symmetry is confirmed by the 
resulting 10-dimensional Lagrangian for the massless  
(i.e., $x^{11}$-independent) modes. Its purely bosonic terms  read
\bea \label{D10-lagrangian}
{\cal L}_{10}&\!\!=\!\!& {1\over \kappa_{10}^2} \bigg[ -\ft12 e\,{\rm 
e}^{2\phi/3} R(e,\omega)  -\ft18 e\,{\rm e}^{2\phi}(\pa_\m 
V_\n-\pa_\n V_\m)^2 \\ 
&&\hspace{9.5mm} -\ft1{48}e\,{\rm e}^{2\phi/3}(F_{\m\n\rho\s})^2  
 -\ft34 e\, {\rm e}^{- 2\phi/3}(H_{\m\n\rho})^2 \nonumber  \\
&& \hspace{9.5mm}+ \ft1{1152} \sqrt{2}\,
\varepsilon^{\m_1
- \m_{10}} 
\,C_{11\m_1\m_2} \,F_{\m_3\m_4\m_5\m_6} \,
F_{\m_7\m_8\m_9\m_{10}}\bigg] \,, \nonumber 
\eea
where $H_{\m\n\rho}= 6\pa_{[\m}C_{\n\rho]11}$ is the field 
strength tensor belonging to the 2-rank tensor gauge field.

The above reduction allows us to discuss a number of 
characteristic features.  First of all, the metric tensor 
produces an extra vector and a scalar, when dimensionally 
reducing the dimension by one unit. The scalar is invariant under 
certain shift symmetries, as shown above, which act 
multiplicatively on the other fields. Secondly, 
tensor fields generate tensor fields of a rank that is one unit 
lower. When this lower-rank field is a scalar field (or 
equivalent to it by a duality transformation), it will be subject to 
shifts by a constant parameter, which is simply associated with a 
gauge transformation that is linearly proportional to the extra 
higher-dimensional coordinate. Because of this, these shifts must 
leave the Lagrangian 
invariant. This last feature is still missing in 
the above discussion, as the rank-3 tensor decomposes into a 
rank-3 and a rank-2 tensor. But when descending to lower 
dimensions than 10, additional scalars will emerge and this 
phenomenon will be present.  

When consecutively reducing the dimension, this pattern repeats 
itself.\footnote{%
  For instance, in \eqn{D10-lagrangian} one considers the scale 
  transformations (for the bosonic fields),
\be
\begin{array}{rcl}
 e_\m^{\;a} &\!\!\to\!\!& {\rm e}^{-\b}\, e_\m^{\;a}\,,\\
 V_\m       &\!\!\to\!\!& {\rm e}^{-\b}\, V_\m      \,,
\end{array} \qquad 
\begin{array}{rcl}
 C_{\m\n11}   &\!\!\to\!\!& {\rm e}^{-2\b}\, C_{\m\n11}\,,\\
 C_{\m\n\rho} &\!\!\to\!\!& {\rm e}^{-3\b} \, C_{\m\n\rho}\,,
\end{array} 
\ee 
  while $\phi$ remains invariant. These transformations change the 
  Lagrangian by an overall factor $\exp[-8\b]$, which can be 
  absorbed into $1/\kappa_{10}^2$. When compactifying one more 
  dimension to a circle, these scale transformations yield another  
  isometry in 9 spacetime dimensions, that commutes with the scale 
  transformations (\ref{scale-10x},\ref{scale-10y}). 
} %
In this way, for each scalar field that is generated by the 
dimensional reduction, there is also an extra symmetry.
The dimension of the isometry group is thus (at least) equal to the 
dimension of the manifold.  Furthermore it is easy to see that 
the symmetries indicated  
above act {\it transitively} on the manifold, so that 
this manifold is homogeneous. The corresponding algebra of these 
isometries is solvable and the rank of the algebra is equal to 
${\bf r}= 11-D$, where $D$ is the  
spacetime dimension to which we reduce. This is because its Cartan 
subalgebra is precisely associated with the scale symmetries 
connected with the scalars that originate from the metric.

The above isometries not only leave the scalar manifold invariant 
but the whole supergravity Lagrangian. In $D=4$, or 8, these 
symmetries do not leave the Lagrangian, but only the field 
equations invariant. The reason for this is that the isometries  
act by means of a duality transformation on the field strengths 
associated with the vector or 3-rank gauge field, respectively. 
They cannot be implemented directly on the gauge fields 
themselves. The presence of  
these duality invariances is a well-known feature of 
supergravity theories, which was first observed many years ago 
\cite{dual,cremmer,dWVP,deroo,DWVVP}. However, it is  
easy to see that the scalar manifold (as well as the rest of the 
theory) must possess additional symmetries, simply because the 
isometries corresponding to the solvable algebra do not yet contain 
the automorphism group $H_{\rm R}$ of the underlying 
supermultiplet. We expect  
that $H_{\rm R}$ is realized as a symmetry, because the maximal 
supergravity theories have no additional parameters, so there is 
nothing that can break this symmetry. So we expect an homogeneous 
space with an isometry group whose algebra is the sum of the 
solvable algebra and the one corresponding to (part of) $H_{\rm R}$. A 
counting argument (of the type first used in \cite{cremmer}) then 
usually reveals what the structure of the homogeneous space is. In 
Table~\ref{maximal-cosets} we list the isometry and isotropy groups of 
these scalar manifolds for maximal supergravity in dimensions $3\leq 
D\leq 11$. Earlier version of such tables can, for instance,  be 
found in \cite{cremmer2,cremmer3}. A more recent discussion of 
these isometry groups can be found in, for example, 
\cite{hull,various}. We return to this discussion and related 
issues in section~3.   

\begin{table}
\begin{center}
\begin{tabular}{l l l l}\hline
$D$ &G        & H   & ${\rm dim}\,[{\rm G}]-{\rm dim}\,[{\rm H}]$ \\ \hline
11  & 1       & 1   & $0-0=0$      \\
10A & SO$(1,1)/Z_2$   & 1   & $1-0=1$  \\
10B & SL(2)    & SO(2) &  $3-1=2$  \\
9   & GL(2)    & SO(2) &  $4-1=3$ \\
8   & E$_{3(+3)}\sim {\rm SL}(3)\! \times \!{\rm SL}(2)$    &  
U(2) & $11- 4=7$  \\
7   & E$_{4(+4)}\sim {\rm SL}(5)$  & USp(4) &$24-10=14$   \\
6   & E$_{5(+5)}\sim {\rm SO}(5,5)$ & 
    USp$(4)\!\times\! {\rm USp}(4)$ &$45-20=25$   \\ 
5   & E$_{6(+6)}$  & USp(8) & $78-36= 42$    \\
4   & E$_{7(+7)}$  &  SU(8) & $133-63= 70$    \\  
3   & E$_{8(+8)}$  & SO(16) & $248 - 120= 128$   \\ \hline
\end{tabular}
\end{center}
\caption{Homogeneous scalar manifolds G/H for maximal 
supergravities in various dimensions. The type-IIB theory 
cannot be obtained from reduction of 11-dimensional supergravity
and is included for completeness. The difference of the 
dimensions of G and H equals the number of scalar fields, 
listed in Table~6.
}\label{maximal-cosets} 
\end{table}

We should add that it is generally possible to realize the group 
$H_{\rm R}$ as a {\it local} symmetry of the Lagrangian. The 
corresponding connections are then composite connections, 
governed by the Cartan-Maurer equations. In such a formulation 
most fields (in particular, the fermions) do not transform under 
the duality group, but only under the local $H_{\rm R}$ group. 
The scalars transform linearly under both the rigid duality group as well 
as under the local $H_{\rm R}$ group. After fixing a gauge, the 
isometries become nonlinearly realized. The fields which 
initially transform only under the local $H_{\rm R}$ group, will 
now transform under the duality group through field-dependent 
$H_{\rm R}$ transformations. This phenomenon is also realized for 
the central charges, which transform under the group $H_{\rm R}$ 
as we have shown in Table~\ref{maximal-central-extension}.

 \subsubsection{Frames and scales}

The Lagrangian \eqn{D10-lagrangian} does not contain the 
standard Einstein-Hilbert term for gravity, while a standard 
kinetic term for the scalar field $\phi$ is lacking. This does 
not pose a serious problem. In this form the 
gravitational field and the scalar field are entangled and one 
has to deal with the scalar-graviton system as a whole. To 
separate the scalar and gravitational 
degrees of freedom, one may apply a so-called Weyl rescaling of 
the metric $g_{\m\n}$ by an appropriate function of $\phi$. In 
the case that we include the massive modes, this rescaling may 
depend on the extra coordinate $x^{11}$. In the context of 
Kaluza-Klein theory this factor is therefore known as the `warp 
factor'. For these lectures two different Weyl rescalings are 
particularly relevant, which lead  
to the so-called Einstein and the string frame, respectively. 
They are defined by
\be \label{weyl-rescaling}
e_\m^a= {\rm e}^{-\phi/12}\, [e_\m^a]^{\scriptscriptstyle\rm 
Einstein}\,,\qquad e_\m^a=  {\rm e}^{-\phi/3}\,
[e_\m^a]^{\scriptscriptstyle\rm string}\,.
\ee

After applying the first rescaling \eqn{weyl-rescaling} to the 
Lagrangian  
\eqn{D10-lagrangian} one obtains the Lagrangian in the Einstein 
frame. This frame is characterized by a standard Einstein-Hilbert 
term and by a graviton field that is invariant under the scale 
transformations (\ref{scale-10x},\ref{scale-10y}). The corresponding 
Lagrangian reads\footnote{%
  Note that under a local scale transformation  
  $e_\m^a\to {\rm  e}^\L e_\m^a$, the Ricci scalar in $D$ 
  dimensions changes according to  
$$
R\to {\rm e}^{-2\L} \Big[ R + 2 (D-1) D^\m\partial_\m \L + 
(D-1)(D-2) g^{\m\n} \,\partial_\m\L\, \partial_\n\L\Big]\,.
$$} 
\bea\label{D10-lagrangian-E}
{\cal L}_{10}^{\scriptscriptstyle\rm Einstein}&\!\!=\!\!&
 {1\over \kappa_{10}^2} \bigg[ e\,  
\Big[-\ft12 R(e,\omega)  -\ft14 (\pa_\m\phi)^2 \Big]  -\ft18 e\,{\rm 
e}^{3\phi/2}(\pa_\m  V_\n-\pa_\n V_\m)^2 \nonumber\\ 
&&\hspace{9.5mm}   
 -\ft34 e\, {\rm e}^{- \phi}(H_{\m\n\rho})^2 -\ft1{48}e\,{\rm 
e}^{\phi/2}(F_{\m\n\rho\s})^2 \nonumber \\ 
&& \hspace{9.5mm}+ \ft1{1152} \sqrt{2}\,
\varepsilon^{\m_1%
-\m_{10}} 
\,C_{11\m_1\m_2} \,F_{\m_3\m_4\m_5\m_6} \,
F_{\m_7\m_8\m_9\m_{10}} \bigg] \,.  
\eea
Supergravity theories are usually formulated in this frame, 
where the isometries of the scalar fields do not act on the 
graviton.  

The second rescaling \eqn{weyl-rescaling} leads to the Lagrangian 
in the string frame,
\bea\label{D10-lagrangian-S}
{\cal L}_{10}^{\scriptscriptstyle\rm string} &\!\!=\!\!& 
{1\over\kappa_{10}^2} \bigg[ e\,{\rm  
e}^{-2\phi} \Big[-\ft12 R(e,\omega)  +2(\pa_\m  \phi)^2 -\ft34 
(H_{\m\n\rho})^2\Big]\nonumber  \\ 
&&\hspace{9.5mm} -\ft18 e\,(\pa_\m 
V_\n-\pa_\n V_\m)^2 -\ft1{48}e\,(F_{\m\n\rho\s})^2   \nonumber \\
&& \hspace{9.5mm}+ \ft1{1152} \sqrt{2}\, 
\varepsilon^{\m_1%
-\m_{10}} 
\,C_{11\m_1\m_2} \,F_{\m_3\m_4\m_5\m_6} \,
F_{\m_7\m_8\m_9\m_{10}}\bigg] \,. 
\eea
This frame is characterized by the fact that 
$R$ and $(H_{\m\n\rho})^2$ have the same coupling 
to the scalar $\phi$, or, equivalently, that 
$g_{\m\n}$ and $C_{11\m\n}$ transform 
with equal weights under the scale transformations 
(\ref{scale-10x},\ref{scale-10y}).
In string theory $\phi$ coincides 
with the dilaton field that couples 
to the topology of the worldsheet and whose
vacuum-expectation value defines the string
coupling constant according to
$\gstring=\exp(\langle\phi\rangle)$. 
We shall return to this in 
section~3, but here we already indicate  
the significance of the dilaton factors in the Lagrangian above. 
The metric $g_{\mu\nu}$, the antisymmetric tensor
$C_{\mu\nu 11}$ and the dilaton $\phi$ 
always arise in the Neveu-Schwarz sector and  
couple universally to  ${\rm e}^{-2\phi}$.
On the other hand the vector $V_\mu$ and the 3-form 
$C_{\mu\nu\rho}$ describe 
Ramond-Ramond (R-R) states and the specific
form of their vertex operators forbids
any tree-level coupling to the dilaton \cite{various,berg}.
In particular the Kaluza-Klein gauge field 
$V_\m$ corresponds in the string context to the R-R  
gauge field of type-II string theory. The infinite tower of 
massive Kaluza-Klein states carry a  
charge quantized in units of $e_{\scriptscriptstyle\rm KK}$, 
defined in \eqn{KK-charge}. In the context of 10-dimensional 
supergravity,  states with a 
R-R charge are solitonic. In string theory, the 
R-R charges are carried by the D-brane states. 

As we already discussed in the previous section,
Newton's constant is only defined after 
a choice of the metric in the ground
state is made. Expanding the metric in the Einstein
frame around $\l\,\eta_{\mu\nu}$
one obtains from 
(\ref{scaleD},\ref{D10-lagrangian-E}) 
\be\label{kappaE}
(\kappa^2_{10})^{\scriptscriptstyle \rm physical} 
= \kappa_{10}^2\,\l^{-4}\, , 
\ee
while expanding the metric in the string frame
around $\l\,\eta_{\mu\nu}$ leads to
\be\label{kappaS}
(\kappa^2_{10})^{\scriptscriptstyle \rm physical} 
= \kappa_{10}^2\, \l^{-4}\,  {\rm e}^{2\langle\phi\rangle}\, . 
\ee
Note that one cannot expand  both metrics
simultaneously around $\l\,\eta_{\mu\nu}$.

For later purposes let us note that
the above discussion can be generalized to 
arbitrary spacetime dimensions.
The Einstein frame in any dimension is defined by a 
gravitational action that is just the 
Einstein-Hilbert action, whereas in the string frame 
the Ricci 
scalar is multiplied by a dilaton term $\exp (-2\phi)$, as in 
\eqn{D10-lagrangian-E} and \eqn{D10-lagrangian-S}, respectively.
The Weyl rescaling which connects the two frames 
is given by,
\be
[e_\m^a]^{\scriptscriptstyle\rm string}= {\rm e}^{2\phi/(D-2)} \,
[e_\m^a]^{\scriptscriptstyle\rm Einstein}\,.
\ee
In arbitrary dimensions (\ref{kappaE}) and (\ref{kappaS})
read 
\bea\label{kappaD}
{\rm Einstein\ frame}:&&
(\kappa^2_{D})^{\scriptscriptstyle \rm physical} 
= \kappa_{D}^2\,\l^{(2-D)/2}\, ,\nonumber\\
{\rm string\ frame}:&&
(\kappa^2_{D})^{\scriptscriptstyle \rm physical} 
= \kappa_{D}^2 \,\l^{(2-D)/2}\, {\rm e}^{2\langle\phi\rangle}\, . 
\eea
This frame dependence does not only apply to 
$\kappa_{D}^2$ but to any dimensionful quantity.
For example, a mass, when measured in the same flat 
metric but specified in the two frames, is related by
\be\label{frel}
M^{\scriptscriptstyle\rm string}= {\rm e}^{-2\langle 
\phi\rangle/(D-2)}\,  M^{\scriptscriptstyle\rm Einstein}\,.
\ee
Of course this is consistent with the relation
(\ref{kappaD}). The physical masses in the above relation depend 
again on the value of $\lambda$. In the remainder of this 
subsection we choose $\lambda=1$ for convenience. 

Let us now return to 11-dimensional supergravity with the 11-th 
coordinate compactified to a circle so that $0\leq x^{11}\leq L$. 
As we stressed  
already, $L$ itself has no intrinsic meaning and it is better to 
consider the geodesic radius of the 11-th dimension, which reads
\be\label{Rphi}
R_{11}= {L\over 2\pi} \, {\rm e}^{2\langle \phi\rangle/3}\, .
\ee
This result applies to the frame specified by the 11-dimensional 
theory\footnote{This is the frame specified by the 
   metric given in \refeq{KKmetric}, which leads to the 
   Lagrangian \refeq{D10-lagrangian}.}. %
In the string frame, the above result reads
\be
(R_{11})^{\scriptscriptstyle\rm string}= {L\over 2\pi} \, {\rm 
e}^{\langle \phi\rangle}\, . 
\ee
It shows that a small 11-th dimension corresponds to small 
values of $\exp\langle \phi\rangle$
which in turn 
corresponds to a weakly coupled string theory. We come back to this 
crucial observation in section~3. Observe that $L$ is fixed 
in terms of $\kappa_{10}$ and $\kappa_{11}$ 
(cf. (\ref{def-L})).
 
{}From the 11-dimensional expressions, 
\be
E_a^{\,M}\pa_M = e_a^{\,\m} (\pa_\m - V_\m \,\pa_{11})\,, \qquad 
E_{11}^{\,M}\pa_M = {\rm e}^{-2\phi/3}\, \pa_{11}\,, 
\ee
where $a$ and $\m$ refer to the 10-dimensional Lorentz and world 
indices, 
we infer that, in the frame specified by the 11-dimensional 
theory, the Kaluza-Klein masses are multiples of 
\be
M^{\scriptscriptstyle\rm KK} = {1\over R_{11}}\,.
\ee
Hence Kaluza-Klein states have a mass and Kaluza-Klein charge  
(cf. \eqn{KK-charge}) related by   
\be 
M^{\scriptscriptstyle\rm KK}= 
\vert e_{\scriptscriptstyle\rm KK} \vert\, {\rm e}^{-2\langle 
\phi\rangle/3} \,. 
\ee
In the string frame, this result becomes
\be \label{Mphi}
(M^{\scriptscriptstyle\rm KK})^{\scriptscriptstyle\rm string}= 
\vert e_{\scriptscriptstyle\rm KK} \vert\, {\rm e}^{-\langle 
\phi\rangle} \,. 
\ee
Massive Kaluza-Klein states are always BPS states, meaning that 
they are contained in supermultiplets that are `shorter' than the 
generic massive supermultiplets because of nontrivial central 
charges. The central charge here is just the 11-th component of 
the momentum, which is proportional to the Kaluza-Klein charge. 

The surprising insight that emerged in recent years, is that the 
Kaluza-Klein features of 11-dimensional supergravity have a 
precise counterpart in string theory 
\cite{hull,Townsend,various}. There one has  
nonperturbative (in the string coupling constant) states which 
carry R-R charges. We return to this  phenomenon in section~3.

 \subsection{Nonmaximal supersymmetry}
In previous subsections we discussed a large number of 
supermultiplets. Furthermore, in Fig.~\ref{fig:sugry} we presented 
an overview of all supergravity theories in spacetime dimensions  
$4\leq D\leq 11$ with $Q=32,16,8$ or 4 supercharges. In this 
section we summarize a number of  
results on nonmaximal supersymmetric theories with $Q=16$ 
supercharges, which are now restricted to dimensions $D\leq 10$.  

\begin{table}
\begin{center}
\begin{tabular}{l l l l l l l l}\hline
$D$ &$H_{\rm R}$& $A_\m$ & $\phi$ & $\chi$   \\ \hline
10 & 1       & 1   & 0     &  1    \\
9  & 1       & 1   & 1     &  1   \\
8  & U(1)    & 1   & $1+\bar1$ &  $1+\bar 1$    \\
7   & USp(2)  & 1  & $3$   &   2  \\
6  & USp$(2)\!\times\! {\rm USp}(2)$  & 1 & $(2,2)$ & $(2,1)+(1,2)$\\ 
5   & USp(4)& 1   & 5     &  4    \\
4   & U(4)  & 1   & $6^\ast$ &  $4+\bar 4$    \\  
3   & SO(8) & ~   &  8   &   8     \\ \hline
\end{tabular}
\end{center}
\caption{Field content for maximal super-Maxwell theories in 
various dimensions. All supermultiplets contain a gauge field 
$A_\m$, scalars $\phi$ and spinors $\chi$. In $D=3$ dimensions the 
vector field is dual to a  
scalar. The $6^\ast$ representation of SU(4) is a selfdual 
2-rank tensor. }
\label{maximal-Maxwell} 
\end{table}

For $Q=16$ the automorphism group $H_{\rm R}$ is smaller.  
Table~\ref{maximal-Maxwell} lists this group in various 
dimensions and shows the field representations for the vector 
multiplet in dimension $3\leq D\leq 10$. This multiplet comprises 
$8+8$ physical degrees of freedom.

We also consider the $Q=16$ supergravity theories. The Lagrangian 
can be obtained by truncation of \eqn{D10-lagrangian}. However, 
unlike in the case of maximal supergravity, we now have the 
option of introducing additional matter fields. For $Q=16$ the 
matter will be in the form of vector supermultiplets, possibly 
associated with some nonabelian gauge group. 
Table~\ref{non-maximal-supergravity} summarizes $Q=16$ supergravity 
for dimensions $3\leq D\leq 10$. In $D=10$ dimensions the bosonic 
terms of the supergravity Lagrangian take the form \cite{BDDV},
\bea \label{D10-Q16-lagrangian}
{\cal L}_{10}\!&=\!& {1\over \kappa_{10}^2} \bigg[ -\ft12 e\,{\rm 
e}^{2\phi/3} R(e,\omega)  \nonumber \\
&&\hspace{9mm} -\ft34 e\, {\rm e}^{- 
2\phi/3}(H_{\m\n\rho})^2  - \ft14 e (\pa_\m A_\n- \pa_\n A_\m)^2 
\bigg] \,,  
\eea
where, for convenience, we have included a single vector gauge 
field, representing an abelian vector supermultiplet. A 
feature that deserves to be mentioned, is that the field strength 
$H_{\m\n\rho}$ 
associated with the 2-rank gauge field acquires a Chern-Simons 
term $A_{[\m} \pa_\n A_{\rho]}$. Chern-Simons terms play an important 
role in the anomaly cancellations of this theory. Note also that 
the kinetic term for the Kaluza-Klein vector field in 
\eqn{D10-lagrangian}, depends on  
$\phi$, unlike the kinetic term for the matter vector field in 
the Lagrangian above.  This reflects itself in the extension of 
the symmetry transformations noted in (\ref{scale-10x},
\ref{scale-10y}),
\be \label{scale-10z}
e_\m^a \to  {\rm e}^{-\a}e_\m^a \,,\quad \phi \to  \phi + 12\a 
\,, \quad
C_{11\m\n} \to  {\rm e}^{6\a}  C_{11\m\n}\,, \quad A_{\m} 
\to  {\rm e}^{3\a}  A_{\m}\,. 
\ee
where $A_\m$ transforms differently from the Kaluza-Klein vector field 
$V_\m$. 

\begin{table}
\begin{center}
\begin{tabular}{l l l l l l l }\hline
$D$ &$H_{\rm R}$& graviton & $p=-1$ & $p=0$ & $p=1$  \\ \hline
10  & 1          & 1 & $1$ & ~    & 1     \\
9   & 1          & 1 & $1$ & 1    & 1     \\
8   & U(1)       & 1 & $1$ & $1+\bar1$& $1$ \\
7  & USp(2)      & 1 & $1$ &$3$   & 1    \\
6A$\!\!$ & USp$(2)\!\times\! {\rm USp}(2)$ & 1 &1&(2,2)&(1,1)\\ 
6B$\!\!$ & USp$(4)$ & 1 & ~&~ & $5^\ast$ \\ 
5   & USp(4)     & 1 & $1$ & 5    & ~    \\
4   & U(4)       & 1 & $1+\bar 1$ & $[6]$& ~ \\
3   & SO(8)      & 1 & $8k$       & ~    & ~ \\  \hline
\end{tabular}
\end{center}
\caption{Bosonic fields of nonmaximal supergravity with $Q=16$. 
In 6 dimensions type-A and type-B 
correspond to (1,1) and (2,0) supergravity. Note that, with the 
exception of the 6B and the 4-dimensional theory, all these 
theories contain precisely one scalar field. In $D=4$ dimensions, the 
SU(4) transformations cannot be implemented on the vector 
potentials, but act on the (abelian) field strengths by duality 
transformations. In $D=3$ dimensions 
supergravity is a topological theory and can be coupled to scalars and 
spinors. The scalars parameterize the coset space SO(8$,
k)/$SO$(8)\times {\rm SO}(k)$, where $k$ is an arbitrary 
integer. }
\label{non-maximal-supergravity} 
\end{table}

In this case there are three different Weyl rescalings that are 
relevant, namely 
\bea \label{weyl-rescaling2}
e_\m^a&=& {\rm e}^{-\phi/12}\, [e_\m^a]^{\scriptscriptstyle\rm 
Einstein}\,,\qquad
 e_\m^a=  {\rm e}^{-\phi/3}\,
[e_\m^a]^{\scriptscriptstyle\rm string}\,,\nonumber \\
 e_\m^a&=&  {\rm e}^{\phi/6}\,
[e_\m^a]^{\scriptscriptstyle\rm string'}\,.
\eea
It is straightforward to obtain the corresponding Lagrangians. In 
the Einstein frame, the graviton is again invariant under the 
isometries of the scalar field. The bosonic terms read
\bea \label{D10-Q16-lagrangian1}
{\cal L}_{10}^{\scriptscriptstyle\rm Einstein}\!&=\!& {1\over 
\kappa_{10}^2} \bigg[ -\ft12 e\,  
R(e,\omega)  -\ft14 e (\pa_\m\phi)^2 \nonumber\\
&& \hspace{9mm} -\ft34 e\, {\rm e}^{- 
\phi}(H_{\m\n\rho})^2  - \ft14 e \,{\rm e}^{-\phi/2} (\pa_\m A_\n- 
\pa_\n A_\m)^2   
\bigg] \,.  
\eea

The second Weyl rescaling leads to the following Lagrangian, 
\bea \label{D10-Q16-lagrangian2}
{\cal L}_{10}^{\scriptscriptstyle\rm string}\!&=\!& {1\over 
\kappa_{10}^2} {\rm e}^{-2\phi}\bigg[ -\ft12 e\, R(e,\omega) + 
2e(\pa_\m\phi)^2 \nonumber  \\
&& \hspace{15mm}  -\ft34 e\, 
(H_{\m\n\rho})^2   
- \ft14 e (\pa_\m A_\n- \pa_\n A_\m)^2  \bigg] \,,  
\eea
which shows a uniform coupling with the dilaton. This is the 
low-energy effective Lagrangian relevant to the heterotic string. 
Eventually the matter gauge field  
has to be part of an nonabelian gauge theory based on the group SO(32) 
or ${\rm E}_8\times{\rm E}_8$, in order to be anomaly-free. 

Finally, the third Weyl rescaling yields
\bea \label{D10-Q16-lagrangian3}
{\cal L}_{10}^{\scriptscriptstyle\rm string'}\!&=\!& {1\over 
\kappa_{10}^2} \bigg[e\,{\rm e}^{2\phi}  \Big[-\ft12 R(e,\omega) 
+2(\pa_\m\phi)^2 \Big] \nonumber \\ 
&& \hspace{9mm}  -\ft34 e\, 
(H_{\m\n\rho})^2  - \ft14 e\,{\rm  
e}^{\phi}(\pa_\m A_\n- \pa_\n A_\m)^2  \bigg] \,.
\eea
Here the dilaton seems to appear with the wrong sign. As it turns 
out, this is the low-energy effective action of the type-I 
string, where the type-I dilaton must be associated with $-\phi$. This 
will be further elucidated in section~3.

%
\section{String theories in various dimensions}

\subsection{perturbative string theories in $D=10$}
\begin{figure}[ht]
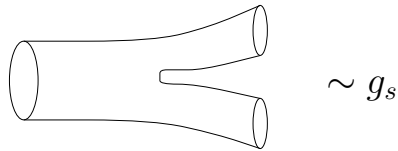
  
\begin{center}
\input fig1.pstex_t
\end{center}
\caption{The fundamental closed string vertex.}
\label{fig:fig1}
\end{figure}

\begin{figure}[thb]
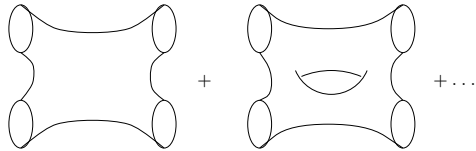
  
\begin{center}
\input fig2.pstex_t
\end{center}
\caption{The perturbative expansion of string scattering 
amplitudes. The order of $\gstring$ is governed by the number of 
holes in the world sheet. 
}
\label{fig:fig2}
\end{figure}

\paragraph{The perturbative expansion.}
In string theory the fundamental objects are one-dimensional strings
which, as they move in time, sweep out a 2-dimen\-sional worldsheet
$\Sigma$ \cite{tex1}.
Strings can be open or closed and their
worldsheet is embedded in some higher-dimensional target space
which is identified with a Minkowskian spacetime.
States in the target space appear as 
eigenmodes of the  string and their scattering amplitudes are 
generalized by appropriate scattering amplitudes of strings.
These scattering amplitudes  are built from
a fundamental vertex, which for closed strings is depicted 
in Fig.~\ref{fig:fig1}.
It represents the splitting of a string or 
the joining of two strings and 
the strength of this interaction is governed by 
a  dimensionless string coupling constant $\gstring$.
Out of the fundamental
vertex one composes all possible 
closed string scattering
amplitudes ${\cal A}$, for example the four-point amplitude
shown  in Fig.~\ref{fig:fig2}.
The expansion in the topology of the Riemann surface
(i.e.~the number of holes in the surface)
coincides with  a power series expansion in the string
coupling constant formally written as
\begin{equation} \label{eq:pert}
{\cal A}\, =\, \sum_{n=0}^\infty  \gstring^{-\chi}\, {\cal A}^{(n)}\ ,
\end{equation}
where ${\cal A}^{(n)}$ is the scattering amplitude
on a Riemann surface of genus $n$ and 
$\chi(\Sigma)$ is  the Euler characteristic
of the Riemann surface
\begin{equation}\label{euler}
\chi(\Sigma)=\frac{1}{4\pi}\int_\Sigma R^{(2)} = 2-2n-b\ .
\end{equation}
$R^{(2)}$ is the
curvature on $\Sigma$ and $b$ the number of 
boundaries of the Riemann surface
(for the four-point amplitude of Fig.~\ref{fig:fig2}
one has $b=4$).\footnote{%
  For open strings different diagrams contribute at the same 
  order of the string loop expansion. See \cite{tex1} for 
  further details.} %

In all string theories there is a massless scalar field
$\D$ called the dilaton which couples to 
$\sqrt{h} R^{(2)}$
and therefore its vacuum-expectation value 
determines the size of the string coupling; 
one finds \cite{AAT,tex1}
\begin{equation}\label{dilaton}
\gstring\, =\, e^{\langle\D\rangle}\ .
\end{equation}
$\gstring$ is a free parameter since $\phi$ is a flat 
direction (a modulus) of the effective potential.
Thus,
string perturbation theory is defined 
in that region of the parameter space
(which is also called the moduli space) where 
$\gstring < 1$ and the tree-level amplitude
(genus-$0$)
is the dominant contribution with higher-loop amplitudes 
suppressed by higher powers of $\gstring$.
Until three years ago this was the only regime
accessible in string theory.

\paragraph{The spacetime spectrum of the string.}
The propagation of a free string ($\gstring=0$) is governed
by the 2-dimensional action 
\be\label{stringaction}
S_{\rm free} = -{1\over 4\pi\alpha'} \int_\Sigma 
\partial_i X^\mu(\sigma,\tau)\,
\partial^i X^\nu(\sigma,\tau)\; \eta_{\mu\nu}\ ,
\ee
where $\pa_i$ denotes $\pa/\pa\sigma$ and $\pa/\pa\tau$. Here 
$\sigma$ parameterizes the spatial direction on $\Sigma$ while $\tau$
denotes the 2-dimensional time coordinate. The coordinates
of the $D$-dimensional target spacetime in which the string 
moves, are represented by $X^\mu$, with $\mu=0,\ldots,D-1$;
in terms of the 2-dimensional
field theory they appear as $D$ scalar fields. 
For $S$ to be dimensionless $\alpha'$ has dimension
[length]$^2\sim$ [mass]$^{-2}$;
it is the fundamental mass scale of string theory
which is also denoted by $\Mstr$ 
with the identification $\alpha'=\Mstr^{-2}$. 
The mass of all perturbative string states is a multiple
of $\Mstr$. 
Just as the coupling constant $\kappa$ in the supergravity 
Lagrangians  
in section~2, this scale has no intrinsic meaning and must be 
fixed by some independent criterion.
Demanding that string theory contains Einstein
gravity as its low-energy limit relates
the characteristic scales of the two theories.
By comparing for example physical graviton-graviton
scattering amplitudes in both theories
one finds in the following expression for Newton's constant in $D$
dimensions \cite{GS},\footnote{%
  The relation (\ref{Mrel}) holds in arbitrary dimensions 
  with ${\rm e}^{\langle \phi\rangle}$ being the string coupling
  constant for a string moving in $D$ spacetime dimensions.
  Later in these lectures we consider compactifications of string 
  theory and then there is a volume-dependent
  relation between the string couplings defined in different 
  dimensions. This relation is discussed in appendix B.} %
\be\label{Mrel}
(\kappa^2_D)^{\scriptscriptstyle \rm physical} = 
{\alpha^\prime}^{(D-2)/2} \, {\rm e}^{2\langle \phi\rangle}\, ,
\ee
where we dropped convention-dependent numerical 
proportionality factors.

The equations of motion of the action \refeq{stringaction}
are given by
\be
(\partial^2_\tau - \partial^2_\sigma) X^\mu =0\ ,
\ee
with the solutions 
\be
X^\mu = X_L^\mu (\sigma+\tau) + X_R^\mu (\sigma-\tau)\ .
\ee
A closed string satisfies the boundary condition
$X^\mu (\sigma) = X^\mu (\sigma + 2\pi)$,
which does not mix  $X_L^\mu$  and $X_R^\mu$ 
and leaves them as independent solutions.
This splitting into left ($L$) and right ($R$) moving
fields has the consequence that upon quantizing the 2-dimensional
field theory, also the Hilbert space splits into a direct product
${\cal H}={\cal H}_L\otimes {\cal H}_R$
where ${\cal H}_L ({\cal H}_R)$ contains states built from
oscillator modes of $X_L (X_R)$.
These states also carry a representation of the 
$D$-dimensional target space
Lorentz group and thus can be identified as perturbative
states in spacetime of a given spin and mass.\footnote{These
are perturbative states since the quantization procedure
is a perturbation theory around the free string theory
with $\gstring=0$.}

In open string theory one has a choice to
impose at the end of the open string
either Neumann (N) boundary conditions, $\partial_\sigma X^\mu 
=0$, 
or Dirichlet (D) boundary conditions, $X^\mu =$ constant.
The boundary conditions mix left- and right-movers
and the product structure of the closed string is not
maintained. As a consequence a perturbative
spectrum of states is built from a single Hilbert space.
Neumann boundary conditions leave the $D$-dimensional Lorentz 
invariance unaffected.

Dirichlet boundary conditions, on the other hand, lead to very 
different types of objects and a very different
set of states (D-branes) in spacetime \cite{d-brane4}. 
In this case the end of 
an open string is constrained to only move in a
fixed spatial hyper-plane.  This plane must be regarded 
as a dynamical object with degrees of freedom
induced by the attached open string.
A careful analysis shows that the 
corresponding states in spacetime
are not part of the perturbative spectrum but
rather correspond to nonperturbative solitonic
type excitations\footnote{They are nonperturbative 
in that their mass (or rather their tension for higher-dimensional
D-branes) goes to infinity in the weak coupling limit
$\gstring\to 0$.}.
It is precisely these
states which dramatically affect
the properties of string theory 
in its nonperturbative regime.
These aspects will be subject of section~3.3.

So far we discussed the free string governed  by the action
\refeq{stringaction}; its
interactions are incorporated by 
promoting $S_{\rm free}$ to a nonlinear
2-dimensional $\sigma$-model. The amplitude 
${\cal A}$ can be interpreted 
as a unitary scattering amplitude 
in the target space
whenever this 2-dimensional field theory is
conformally invariant. The action is found to be
\bea\label{sigmaaction}
S&=& -{1\over 4\pi\alpha'} \int_\Sigma 
\partial_i X^\mu(\sigma,\tau)\,
\partial^i X^\nu(\sigma,\tau)\; g_{\mu\nu} (X(\sigma,\tau)) 
\nonumber \\
&&-{1\over 4\pi\alpha'} \int_\Sigma 
\,\varepsilon^{ij}\,\partial_i X^\mu(\sigma,\tau)\,
\partial_j X^\nu(\sigma,\tau)\; b_{\mu\nu} (X(\sigma,\tau)) 
 \\
&& + \frac{1}{4\pi} \int_\Sigma  R^{(2)}\, \D(X(\sigma,\tau))
+\cdots\ ,\nonumber
\eea
where $g_{\mu\nu} (X)$ is the metric\footnote{As we mentioned
  already in subsection~2.5, the metric $g_{\mu\nu} (X)$
  in (\ref{sigmaaction}) is the metric in the string 
  frame.} %
of the target space, $b_{\mu\nu} (X)$ is the antisymmetric
target-space tensor and $R^{(2)}$
is the curvature scalar of the 2-dimensional worldsheet
$\Sigma$. The target-space field $\D(X)$ represents a scalar coupling
and corresponds to the dilaton, since the coefficient
of its constant
vacuum-expectation value $\langle\D\rangle$ 
is the Euler number
$\chi(\Sigma) = \frac{1}{4\pi} \int_\Sigma  R^{(2)}$.
The ellipses  denote  further terms 
depending on the type of string theory
and the number of spacetime dimensions. 

The spacetime properties of a string theory significantly
change once one introduces supersymmetry on the worldsheet.
In two dimensions the irreducible supercharges 
are Majorana-Weyl spinors (see Table~\ref{susy-charges}). In addition
there are independent 
left- and right-moving supercharges $Q_L$, $Q_R$, so that
in general one can have $p$ supercharges 
$Q_L$ and $q$ supercharges $Q_R$; this  is also termed
$(p,q)$ supersymmetry.
A supersymmetric version of the action \refeq{sigmaaction}
requires the presence of 
Majorana-Weyl worldsheet fermions $\chi^\mu$
with appropriate couplings;
for example a scalar supermultiplet of 
$(1,0)$ supersymmetry contains the fields 
$\big(X_L(\sigma+\tau), \chi_L (\sigma+\tau)\big)$.
Depending on the amount of worldsheet supersymmetry
one defines the different {\it closed} string theories:
the bosonic string, the superstring and the
heterotic string (see Table~\ref{table-string}).

\begin{table}
\begin{center}
\begin{tabular}{lcc}\hline
closed           &worldsheet      & $D_{\rm max}$ \\
string theories  &supersymmetry   & {}  \\\hline
bosonic string   & $\quad (0,0)$ & $26$\\
superstring      & $\quad (1,1)$ & $10$\\
heterotic string & $\quad (0,1)$ & $10$\\ \hline
\end{tabular}
\end{center}
\caption{The closed-string theories, their worldsheet 
supersymmetry and the maximal possible spacetime 
dimension.}\label{table-string} 
\end{table}

For {\it open} string theories the left- and right-moving
worldsheet supercharges are not independent.
One can either have a bosonic open string
(with no worldsheet supersymmetry) or an open superstring
with one supercharge
which is a linear combination of $Q_L$ and $Q_R$.
The latter string theory is called type-I.
It contains (unoriented) open and closed strings with SO(32)
Chan-Paton factors coupling to the ends of the open string.

The bosonic string (open or closed) is tachyonic and cannot 
accommodate spacetime fermions;
for these reasons we omit
it from our subsequent discussion.
The superstring, the heterotic string and the type-I
string can all be
tachyon-free and do have spacetime fermions
in the massless spectrum. 
In addition, in most cases they are 
also spacetime supersymmetric and
contain (at least) a massless gravitino. There are also
tachyon-free non-supersymmetric string theories
\cite{DH}
but they have a dilaton tadpole at one-loop
and thus do not seem to correspond to stable vacuum 
configurations.\footnote{For a recent discussion of non-supersymmetric
string theories, see \cite{Dienes}.}
For this reason we solely focus on supersymmetric string theories 
henceforth.

The worldsheet fermions $\chi^\mu$ can have two distinct 
type of boundary conditions when transported around the closed 
string, 
\be
\chi^\mu (\sigma) =\left\{ \begin{array}{ll}
                  +\chi^\mu (\sigma+2\pi)& \mbox{\rm Ramond\ (R)}\,,
\\[2mm]
     -\chi^\mu (\sigma+2\pi)& \mbox{\rm Neveu-Schwarz\ (NS)\,.}
                    \end{array}
                     \right.
\ee
Consequently the states of the closed string Hilbert space can
arise in four different
sectors of fermion boundary conditions:
\bea
\left.
\begin{array}{llll}
{\rm NS}_L&\otimes&{\rm NS}_R\\
{\rm R}_L&\otimes&{\rm R}_R
\end{array}\right\} & {\rm spacetime\ bosons} 
\nonumber\\
\left.
\begin{array}{lll}
{\rm NS}_L&\otimes&{\rm R}_R\\
{\rm R}_L&\otimes&{\rm NS}_R
\end{array}\right\} & {\rm spacetime\ fermions} \ .\nonumber
\eea
The first two sectors contain the spacetime bosons, while the last two
sectors generate spacetime fermions.
The bosons from the R-R sector are built from bi-spinors
and thus the representation theory 
of the Lorentz group constrains these bosons
to always be antisymmetric Lorentz tensors of varying rank.
Furthermore, in the effective action and in all scattering
processes these tensors can only appear via their
field strength and thus there are no states in
perturbative string theory which  carry any charge
under the antisymmetric tensors in the R-R sector.
However, it turns out that this is an artifact
of perturbation theory and states carrying R-R charge 
do appear in the nonperturbative spectrum;
they are precisely the states generated
by appropriate D-brane configurations \cite{d-brane4}.

Conformal invariance on the worldsheet
(or equivalently unitarity in spacetime)
imposes a restriction on the maximal number
of spacetime dimensions and the spacetime spectrum.
All supersymmetric string theories
necessarily have $D\le 10$ and they
are particularly simple in 
their maximal possible dimension $D=10$.\footnote{%
  For closed strings an additional constraint arises from the 
  requirement of  
  modular invariance of one-loop amplitudes which results in an 
  anomaly-free spectrum of the corresponding low-energy effective 
  theory \cite{ASNW}. For open strings anomaly cancellation is a 
  consequence of the the absence of tadpole diagrams 
  \cite{tex1}.} %

In $D=10$ there are only five consistent spacetime supersymmetric 
string theories: type-IIA, type-IIB, heterotic 
${\rm E}_8\times {\rm E}_8$ (HE8), heterotic SO(32) (HSO)
and the type-I SO(32) string.
The first two have $Q=32$ supercharges and thus 
there is a unique massless multiplet in each case 
with a field content given in Table~\ref{maximal-sg-bosons}.
As we already indicated perturbative string theory
distinguishes between states arising in the NS-NS sector
from states of the R-R sector in that the coupling 
to the dilaton is different.
In the type-IIA theory one finds the graviton
$g_{\mu\nu}$, an antisymmetric tensor $b_{\mu\nu}$
and the dilaton $\phi$ in the NS-NS sector while
an abelian vector $V_\mu$ and a 3-form $C_{\mu\nu\rho}$
appear in the R-R sector.
The corresponding low-energy effective Lagrangian
was already given in section 2.5.1.
In type-IIB one has exactly the same states in the 
NS-NS sector, but in the R-R sector one has a 2-form 
$b'_{\mu\nu}$, an additional scalar $\phi'$ and a 4-form 
$c^*_{\mu\nu\rho\sigma}$ whose field strength is selfdual. Its 
field equations can be found in \cite{SH}.

The other three string theories all have $Q=16$ 
supercharges. In 
this case, the supersymmetric 
representation theory alone does not completely determine the spectrum.
The gravitational multiplet is unique 
(see Table~\ref{non-maximal-supergravity}), but 
the gauge group representation of the 
vector multiplets (see Table~\ref{maximal-Maxwell})
is only fixed if also anomaly cancellation is imposed.
The low-energy effective Lagrangian for the two heterotic 
theories is displayed in \refeq{D10-Q16-lagrangian1} and 
\refeq{D10-Q16-lagrangian2}, with the abelian
vector appropriately promoted to vector fields of
${\rm E}_8\times {\rm E}_8$ or SO(32), respectively.
The type-I string has the same supersymmetry but 
$b_{\mu\nu}$ arises in the R-R sector and thus has
different (perturbative) couplings to the dilaton.
The corresponding low-energy effective Lagrangian
is given by \refeq{D10-Q16-lagrangian3}
with $\phi$ replaced by $-\phi$.
In
Table~\ref{tab:tab2} we summarize the bosonic massless spectra for 
the five string theories,  which is in direct correspondence with 
some of the material collected in the 
Tables~\ref{maximal-sg-bosons}, \ref{maximal-Maxwell} and 
\ref{non-maximal-supergravity}, presented in section~2.

\begin{table}[t]
\begin{center}\begin{tabular}{lllll}
\hline 
{}type   & $Q$&$\;$& \multicolumn{2}{c}{bosonic spectrum}\\ 
\hline
IIA & 32 & & NS-NS & $g_{\mu\nu}$, $b_{\mu\nu}$, $\D$ \\[.5mm] 
\cline{4-5} 
 &&  
& R-R & $V_{\mu}$, $C_{\mu\nu\rho}$
\\ \hline
IIB & 32 &&  NS-NS &  $g_{\mu\nu}$, $b_{\mu\nu}$, $\D$ \\[.5mm] 
\cline{4-5}
 &&  
& R-R & 
$c^{*}_{\mu\nu\rho\sigma}$, $b_{\mu\nu}^{\prime}$, $\D^{\prime}$ \\
\hline
HE8& 16  & &\multicolumn{2}{c}{$g_{\mu\nu}$, $b_{\mu\nu}$, 
$\D$}\\[.5mm]
 & &  &\multicolumn{2}{c}{$A_{\mu}$ in
 adjoint of ${\rm E}_8\times {\rm E}_8$ }
\\ \hline
HSO & 16 & &\multicolumn{2}{c}{$g_{\mu\nu}$, $b_{\mu\nu}$, 
$\D$}\\[.5mm]
 & & &\multicolumn{2}{c}{$A_{\mu}$ in adjoint of SO(32) }
\\ \hline
I & 16 & &NS-NS  & $g_{\mu\nu}$,  $\D$\\[.5mm] \cline{4-5}
& & & R-R & $b_{\mu\nu}$    \\[.5mm]  \cline{4-5}
 &  & &open string &$A_{\mu}$ in
adjoint of SO(32)
\\ \hline\end{tabular}
\end{center} 
\caption{Supersymmetric string theories in $D=10$ and their 
fields describing the bosonic massless spectrum.} \label{tab:tab2}
\end{table}

%
\subsection{Calabi--Yau compactifications and perturbative 
dualities}
So far we discussed the various string theories in 10 
spacetime dimensions. Lower-dimensional theories
can be obtained by compactifying the $D=10$ theories
on an  internal, `curled up' compact manifold 
$Y$.\footnote{%
  There are also string vacua which cannot be viewed as a 
  compactification of the 10-dimensional string theories. 
  Their duality properties have been much 
  less investigated and for a lack of space we
  neglect them in our discussion here
  and solely focus on string vacua with a geometrical
  interpretation.} %
Unitarity in spacetime requires  $Y$ to be a Calabi-Yau 
manifold \cite{tex1}.\footnote{By a slight abuse
of terminology we include in our discussion here
also the circle $T^1$, which strictly speaking is not a
Calabi-Yau manifold but does give unitary S-matrices.}
Calabi-Yau  manifolds are  Ricci-flat K\"ahler manifolds
of vanishing first Chern class ($c_1(Y)=0$) with 
holonomy group SU($\M$) where $\M$ is
the complex dimension of $Y$. 
A  one dimensional (complex) Calabi-Yau manifold is  
topologically always
a torus $T^2$ and  toroidal compactifications 
leave all supercharges intact.
For $\M=2$ all Calabi-Yau manifolds are topologically
equivalent to the 4-dimensional 
$K3$ surface \cite{aspinwallrev} and  
as a consequence of the nontrivial SU(2) holonomy
half of the supercharges are broken.
For $\M=3$ there exist many topologically distinct 
Calabi--Yau threefolds $Y_3$ and they all
break $\frac34$ of the supercharges \cite{mirrorrev}.
We summarize this situation in the following table:

$$
\begin{array}{lll}
\M=1:  & T^2  &   {\rm breaks\ no\ supercharges}\\
\M=2:   & K3  &   {\rm breaks}\ 1/2\ {\rm of\ the\  supercharges}\\
\M=3:   & Y_3  &   {\rm breaks}\ 3/4\ {\rm of\ the\ supercharges}\\
\end{array}
$$

\begin{figure}  
\epsfig{figure=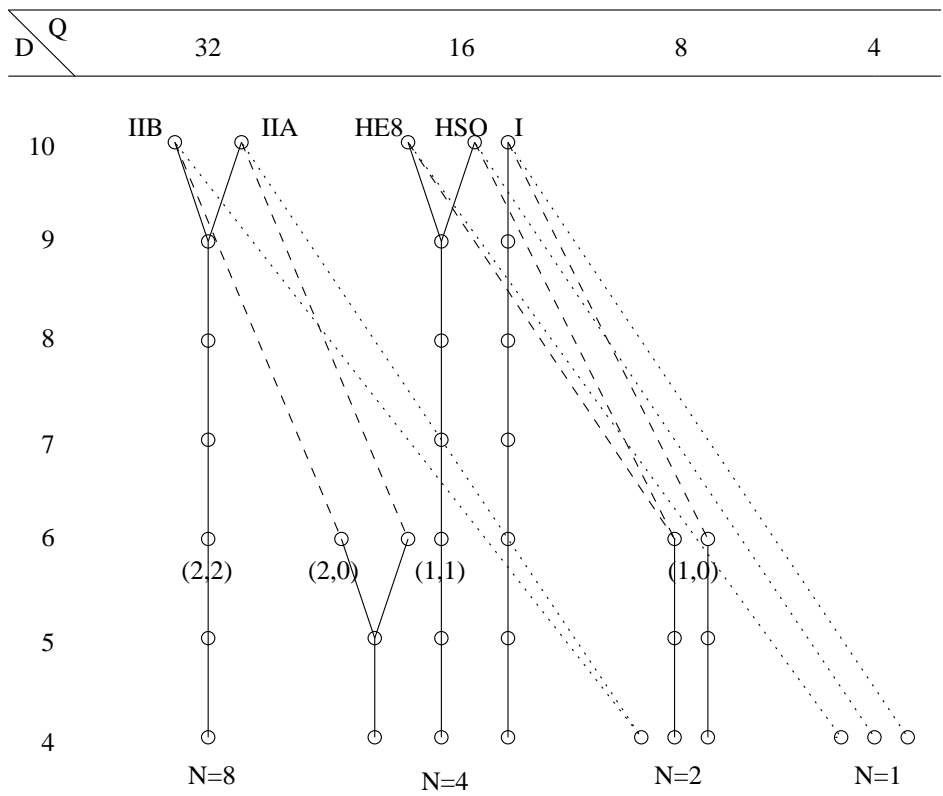,width=\textwidth}
\caption{Calabi-Yau compactifications of the 10-dimensional
string theories. The solid line ($-$) denotes toroidal
compactification, the dashed line ($--$) denotes 
$K3$ compactifications and the dotted line ($\cdots$)
denotes $Y_3$ compactifications. Whenever two 
compactifications (two lines) terminate in the same point,
the two string theories are related by a perturbative
duality. (A line crossing a circle is purely
accidental and has no physical significance.)}
\label{fig:sugra}
\end{figure}

The different theories obtained by compactifying
on such Calabi--Yau manifolds
are depicted in Fig.~\ref{fig:sugra}.
Note that different compactifications can have
the exact same representation of supersymmetry. 

%
\subsubsection{Toroidal compactifications and $T$-duality.}
Compactifying the 10-dimensional string theories on an $n$-dimensional
torus $T^n$ leads to string theories with $D=10-n$. 
Supersymmetry remains unbroken and thus 
one moves down vertically in the $D$-$Q$ plane
of Fig.~\ref{fig:sugra}. 
The massless spectrum 
can be obtained by
dimensional reduction of the appropriate 10-dimensional theories.
For simplicity we start by considering closed string theories with 
one compact dimension which we take to be a circle
$T^1$.
In this case there are
nine spacetime coordinates $X^\mu$
satisfying the boundary conditions\footnote{%
  Throughout this section the indices $\mu,\nu$ always denote 
  the uncompactified spacetime directions.}
\begin{equation}
X^\mu \left(\sigma = 2\pi , \tau\right) 
= X^\mu\left(\sigma = 0 , \tau \right),
\end{equation}
and one internal coordinate $X^{10}$,  
which can wrap  $m$ times around the $T^1$ of radius $R$,
\begin{equation} \label{eq:compact}
X^{10}\left( \sigma = 2\pi , \tau \right) 
= X^{10}\left( \sigma = 0, \tau\right) +2\pi m R\ .
\end{equation}
The massless 
spectrum of the 9-dimensional theory  
includes the two abelian Kaluza-Klein gauge bosons
$g_{\mu\,10}$ and $b_{\mu\, 10}$ as well as
a massless 
scalar field $g_{10\, 10}$ 
which is a flat direction of the effective potential
and whose vacuum-expectation value parameterizes
the size of the internal $T^1$ (cf. section~2.5)
The appearance of flat directions is a 
generic feature of string compactifications
and such scalar fields are called moduli. 
For the case at hand the moduli space is 
one-dimensional and hence there is a 
one-parameter family of inequivalent 
string vacua.\footnote{The different solutions
of a given string theory are often referred to as 
the vacuum states of that string theory
or simply as the string vacua.}
The boundary condition 
(\ref{eq:compact}) leads to a quantization of the 
internal  momentum component $p^{10}=k/R$
and a whole tower of massive Kaluza--Klein
states labelled by the integer $k$.
In addition there are also massive winding modes 
labelled by $m$ and altogether one finds a mass spectrum 
\begin{equation}\label{eq:mass}
 M^2 =  \ft12 (P_L^2 + P_R^2)+\, \cdots \ , \qquad 
P_{L,R}=  \frac{k}{R} \pm \frac{m R}{\alpha'} \ .
\end{equation}  
The ellipses  stand for $R$-independent 
contributions of the oscillator modes.
The mass spectrum is invariant under the exchange of 
$R\leftrightarrow \alpha'/R$ if simultaneously one also exchanges
$m\leftrightarrow  k$.
This $Z_2$ invariance of the spectrum is the first 
example of a {\sl T--duality} \cite{j-gang}.

These considerations can be generalized to higher-dimensional 
toroidal compactifications on a torus $T^n$ \cite{narain}.
The boundary conditions on the $n$ compact coordinates
are
\begin{equation} \label{eq:compactn}
X^i\left( \sigma = 2\pi , \tau \right) 
= X^i\left( \sigma = 0, \tau\right) +2\pi R^i\ ,
\end{equation}
where $R^i=\sum_I m_I e_I^i$ 
are vectors on an $n$-dimensional lattice 
with basis $e_I^i$.\footnote{
The index $i$ runs over the internal dimensions,
i.e.\ $i=1,\ldots,n=10-D$.} 
The momenta $p^i$
live on the dual lattice and one finds
\begin{equation}\label{eq:massn}
M^2 =  \ft12 (P_L^2 + P_R^2)+ \cdots \ , \qquad 
P_{L,R}^i=  p^i \pm \frac{R^i}{\alpha'} \ .
\end{equation}  
Modular invariance constrains the lattice to be 
an even, selfdual Lorentzian lattice.
The inequivalent lattices of this type can be labelled
by points in the coset space 
\cite{narain}
\begin{equation}\label{eq:Mtoroidal}
{\cal M}=\left. \frac{{\rm SO}(n,n)}{{\rm SO(n)}\times {\rm 
SO}(n)}\right/\Gamma_T \ , 
\end{equation}
where 
\begin{equation}\label{eq:Ttoroidal}
\Gamma_T = {\rm SO}\left(n,n, Z\right)
\end{equation}
is the T--duality group identifying equivalent 
lattices or in other
words equivalent toroidal compactifications.
The mass formula \refeq{eq:massn} (as well as the entire
partition function)
shares the invariance under $\Gamma_T$.

\paragraph{Type-II compactified on $T^n$.}
Toroidal compactifications of type-II string theory
all have the
maximal number of  supercharges $Q=32$.
The associated supergravities have been discussed in section 2.4
and for each of these cases there is a unique
gravitational multiplet containing
all massless fields.\footnote{We briefly mentioned
in section 2.2.2  that below $D=10$ the type-IIA
and type-IIB supergravities are equivalent.
A careful analysis in $D=9$ reveals 
a flip of the chiralities of the space-time
fermions in  the limits $R\to 0$ versus
$R\to\infty$ 
and thus the non-chiral type-IIA and the chiral
type-IIB theory can be viewed as two distinct
limits of one and the same type-II theory
in $D=9$ \cite{DHS,d-brane1}.
One also often refers to this fact
by stating that in $D=9$ type-IIA and type-IIB
are T-dual to each other in that 
type-IIA at a large compactification radius 
is equivalent to
type-IIB at a small compactification radius 
and vice versa.}
However in perturbative string theory there is a 
clear distinction between states arising from 
the NS-NS versus the R-R sector.
In the NS-NS sector one finds 
the Kaluza--Klein gauge bosons
$g_{\mu i}$ and $b_{\mu i}$ of a gauge
group $G=U(1)^{2n}$, the dilaton $\D$  and the  moduli
$g_{ij}$, $b_{ij}$;  the latter are 
precisely the $n^2$ coordinates 
of the toroidal  moduli space ${\cal M}$ 
given in \refeq{eq:Mtoroidal}.
The dilaton and the scalars of the R-R sector are not part 
of this ${\cal M}$.
However, all scalar fields -- NS-NS and R-R -- 
reside in the same gravitational multiplet and so 
the supergravity considerations discussed in 
section~2.5  suggest that they 
combine into a larger moduli space with nontrivial mixings.
For a long time this state of affairs seemed incompatible
with perturbative string theory since the specific form
of the vertex operator of the R-R scalars implies that they 
can have no nontrivial couplings to the 
NS-NS scalars \cite{d-brane4}.
However, nonperturbative corrections alter this conclusion
and by now it is believed that taking perturbative
and nonperturbative contributions together exactly
reproduces the geometrical structures suggested
by supergravity. 
We will return to this point in more detail in section~3.3.

\paragraph{Heterotic string compactified on $T^n$.}
The heterotic string compactified on $T^n$ 
has 16 supercharges and
there are $n$ additional scalars $A_{i}^a$
transforming in the adjoint representation
of  ${\rm E}_8\times {\rm E}_8$ or $SO(32)$. 
However, only the $16 n$ 
scalars in the Cartan subalgebra 
are flat directions and their (generic) vacuum expectation values
break the non-abelian gauge symmetry
to U(1)$^{16}$. Together with the $2n$ Kaluza--Klein
gauge bosons $g_{\mu i}$ and $b_{\mu i}$ they 
form\footnote{$n+16$ of these
vectors reside in vector multiplets while the
remaining $n$ vectors are part of the gravitational
multiplet (see Tables~\ref{maximal-Maxwell}
and \ref{non-maximal-supergravity}).}
the total gauge group ${\rm G}={\rm U}(1)^{2n+16}$.
On special subspaces of the moduli space 
there can be nonabelian enhancement of the 
${\rm U}(1)^{16}$, at most up to 
the original E$_8 \times {\rm E}_8$ or SO(32) (at least 
perturbatively).

The $16 n$ scalars in the Cartan subalgebra 
parametrize together with the toroidal 
moduli $g_{ij}$, $b_{ij}$ and the dilaton $\D$
the $n(n+16)+1$ dimensional moduli space \cite{narain}
\begin{equation}\label{eq:Mhettoroidal}
{\cal M}\ =\ \left. R^+\ \times\
\frac{{\rm SO}(n,n+16)}{{\rm SO}(n)\times {\rm SO}(n+16)}\right/\Gamma_T ,
\end{equation}
with the T--duality group
\begin{equation}\label{eq:Thettoroidal}
\Gamma_T = {\rm SO}\left(n,n+16, Z\right)\ .
\end{equation}
The toroidal moduli all reside in $n+16$
(abelian) vector multiplets and 
the heterotic dilaton is the unique scalar 
in the gravitational multiplet 
(see Table~\ref{non-maximal-supergravity}). 
Supergravity implies that there can be no mixing
between the dilaton and the other $n(n+16)$ scalars and 
thus the dilaton spans the $R^+$ component of 
${\cal M}$.\footnote{%
  In $D=4$ the antisymmetric tensor 
  in the gravitational multiplet is dual
  to a pseudoscalar and thus can be combined
  with the dilaton into one complex scalar field.
  Since this scalar still resides in the gravitational
  multiplet it does not mix with the other toroidal
  moduli and  again spans a separate component 
  of the moduli space which is found to be locally equivalent to 
  the SU$(1,1)/{\rm U}(1)$ coset space, 
  which replaces $R^+$ in \refeq{eq:Mhettoroidal}.}
Locally, the moduli space is  uniquely
determined by supersymmetry \cite{deroo} and so  already 
from this point of view the moduli space
\refeq{eq:Mhettoroidal} is likely to be exact. 

It has also been shown that below ten dimensions
the heterotic ${\rm E}_8\times {\rm E}_8$ theory and
the heterotic SO(32) theory are continuously connected
in the moduli space. That is, the two theories 
sit at different points of the
same moduli space of one and the same heterotic 
string theory \cite{gins}.
The continuous path which connects the two theories in $D=9$
involves a transformation $R\to \alpha'/R$ and hence
they are also called T--duals of each other.

\paragraph{Type-I compactified on $T^n$.}

Toroidal compactifications of the 
type-I theory are slightly more involved.
Locally the moduli space is dictated by 
supersymmetry to be the coset
$R^+\times {\rm SO}(n,n+16)/{\rm SO}(n)\times {\rm SO}(n+16)$
but perturbatively there is no T--duality
symmetry and thus the global moduli space
does not coincide with \refeq{eq:Mhettoroidal}.
However, once D-branes are included 
as possible open string configurations, 
type-I theories also have  T--duality and the moduli space 
is given by \refeq{eq:Mhettoroidal} and \refeq{eq:Thettoroidal} 
\cite{d-brane1,d-brane2,d-brane4}. 
In 
fact, establishing T--duality was a guiding 
motive in the original
discovery of D-branes \cite{d-brane1}. 
{}From the open string point of view 
T--duality is not a perturbative symmetry
since it necessarily involves the presence
of solitonic-like excitations.

%
\subsubsection{$K3$ compactifications}

So far we discussed toroidal compactifications
which leave all supercharges intact.
Compactifications on a  4-dimensional $K3$ surface
break half of the supercharges
and hence one moves one column to the right 
and four rows down in the $D$-$Q$ plane 
of Fig.~\ref{fig:sugra}; the resulting string theories
therefore have $D=6$ and either 16 or 8 supercharges.
The massless modes of $K3$ compactifications
arise from nontrivial deformations of the metric 
and from nontrivial harmonic forms on the $K3$ 
surface \cite{aspinwallrev}.
The moduli space of nontrivial metric deformations 
is known to be 58-dimensional and given by the coset space 
\begin{equation}\label{MKthree}
{\cal M}\ =\  R^+\ \times\
\frac{{\rm SO}(3,19)}{{\rm SO}(3)\times {\rm SO}(19)}\ ,
\end{equation}
where the second factor is the Teichm\"uller space 
for Einstein metrics of unit volume and 
the first factor is associated with the
size of the $K3$.

On any (complex) K\"ahler manifold, the differential forms 
can be decomposed into $(p,q)$-forms with $p$ holomorphic
and $q$ antiholomorphic differentials.
The harmonic $(p,q)$-forms form the cohomology groups 
$H^{p,q}$ of dimension $h^{p,q}$ and for $K3$ one has
a Hodge diamond
\begin{equation} \label{eq:hodge}
\begin{array}{c c c c c}
            &              &  h^{0,0} &            &            \\
            & h^{1,0}\! &              &\! h^{0,1}\!&            \\
  h^{2,0}\! & \!             & h^{1,1} & \!            &\! h^{0,2}\\
            & h^{2,1} &              &h^{1,2}&             \\
             &              &h^{2,2}&              & 
\end{array} =
\begin{array}{c c c c c}
  &   & 1   &  & \\
  & 0\! &      &\! 0& \\
1 &\!   & 20 & \! & 1\\
  & 0\! &      &\! 0&  \\
  &   &1    &  & 
\end{array}\  .
\end{equation}
Thus there are 22 harmonic $p+q=2$--forms which represent the 
nontrivial deformations of an antisymmetric tensor 
$b_{ij}$. On a 4-dimensional manifold
an antisymmetric tensor can be constrained
to a selfdual or an anti-selfdual tensor and 
on $K3$ one finds 
that the 22 2-forms decompose into 3 selfdual
and 19 anti-selfdual 2-forms  
\cite{aspinwallrev}.
For later reference we also record
that the Euler number of $K3$ is found to be 
\begin{equation}\label{Euler}
\chi(K3) = \sum_{p,q} (-)^{p+q}\, h^{p,q} = 24.
\end{equation}
%
\paragraph{Type-IIA compactified on $K3$}
Compactifying the type-IIA string on $K3$ breaks half 
of the 32 (non-chiral) supercharges in $D=10$ and thus
results in a $(1,1)$ supergravity 
in $D=6$ coupled to vector multiplets
which we already discussed in section~2.2.3.
The massless bosonic modes of such compactifications are 
given by $g_{\mu\nu}$, $b_{\mu\nu}$,  $\D$, 
$g_{ij}$, $b_{ij}$ all of which arise 
from the NS-NS sector.
$g_{ij}$ denotes the 58 zero modes of the metric on $K3$
and $b_{ij}$ are the 22 harmonic 2-forms.\footnote{%
  Contrary to toroidal compactifications there are no massless 
  vectors $g_{\mu i}$ or $b_{\mu i}$ since there are no one-forms 
  on $K3$  (cf. \refeq{eq:hodge}).} %
In the R-R sector one finds 
$V_\mu, C_{\mu\nu\rho}$ and 22 vectors $C_{\mu ij}$. 
In $D=6$ a 3-form  is dual to a
vector field so that altogether there are 
24 vector fields in the R-R sector and 81 scalars 
in the NS-NS sector.
The multiplets of $(1,1)$ supergravity are discussed in 
section~2.2.3. and one infers that the bosonic states
of the $K3$ compactification fill out 
one gravity multiplet and 20 vector multiplets.

All 81 scalars arise in the NS-NS 
sector. The deformations of the metric   
span the moduli space given in \refeq{MKthree}.
Together with the 22 harmonic 2-forms 
they combine into
the 81-dimensional moduli space  \cite{NS,aspinwall}
\begin{equation}  \label{eq:2aonk3}
{\cal M}\ =\ R^+\ \times \ 
 \left. \frac{{\rm SO}(4,20)}{{\rm SO}(4)\times {\rm 
SO}(20)}\right/ \Gamma_T\ , 
\end{equation}
where the factor $R^+$ is again parameterized by
the single scalar in the
gravitational multiplet which can be identified
with the 6-dimensional dilaton. 
The second coset factor is spanned by the scalar fields
of the vector multiplets.
Similar to toroidal compactifications one finds 
perturbative identifications of the parameter space
which are directly related to the properties 
of the underlying 2-dimensional conformal field theory.
Such equivalences are also termed $T$-duality and for
the case at hand 
the  T--duality group is found to 
be \cite{NS,aspinwall}
\be\label{gammatk3}
\Gamma_T = {\rm SO}(4,20,Z) \ .
\ee

%
\paragraph{Type-IIB compactified on $K3$.}

Compactifying the type-IIB string on $K3$ breaks half 
of the 32 chiral supercharges in $D=10$ and thus
results in a $(2,0)$ supergravity in $D=6$ 
coupled to tensor multiplets
which we already discussed in section~2.2.3.
The massless bosonic modes of such compactifications are 
given by $g_{\mu\nu}, b_{\mu\nu}, \D, g_{ij}, b_{ij}$
in the NS-NS sector and 
$b_{\mu\nu}', \D^\prime, b_{ij}^\prime, 
C_{\mu\nu\rho\sigma},C_{\mu\nu ij}$
in the R-R sector; 
both  $b_{ij}$ and  $b_{ij}'$ are 22 harmonic 2-forms on $K3$.
In $D=6$ a 4-form tensor $C_{\m\n\rho\s}$ describes only one physical
degree of freedom and is dual to a real scalar.
Furthermore, there are 22 spacetime 
tensors $C_{\mu\nu ij}$, proportional to the 22
harmonic forms on $K3$. 
Since $C$ is chosen selfdual in $D=10$, the tensor fields
$C_{\m\n ij}$  are
selfdual or anti-selfdual and their selfduality phase is
correlated with the (anti)selfduality of the corresponding  $K3$
harmonic forms. Hence the $C_{\mu\nu ij}$ decompose  into
3 selfdual and 19 anti-selfdual $D=6$ antisymmetric tensors.
Altogether 
there are thus 81 NS-NS and 24 R-R scalars and 
5 selfdual and 21 anti-selfdual R-R antisymmetric tensors.
The corresponding supermultiplets 
were already discussed in section~2.2.3.;
we immediately infer that  the massless modes arising from
the $K3$ compactification combine into one gravitational 
and 21 tensor multiplets of $(2,0)$ supersymmetry.
This theory being chiral is potentially anomalous;
however, it was shown in \cite{townsend} that 
precisely this  combination of multiplets 
is anomaly free.

The 81 scalars of NS-NS sector span the same 
moduli space as in \refeq{eq:2aonk3} and 
similar to toroidal compactifications 
the scalars from the R-R cannot have any nontrivial
mixing with the NS-NS scalars  at
the perturbative level. However, 
the $(2,0)$ gravitational multiplet contains no
scalar at all but rather all scalars appear
in the 21 tensor multiplets.
On the basis of supersymmetry it was conjectured 
\cite{romans} that all
105 scalars locally parameterize the  moduli space 
\begin{equation}\label{Mtwob}
{\cal M} =  \frac{{\rm SO}(5,21)}{{\rm SO}(5)\times {\rm SO}(21)}\ .
\end{equation}
Indeed, once nonperturbative 
corrections of string theory are taken into account
this moduli space is generated; we return to this point
in section~3.3.4.

%
\paragraph{The heterotic string compactified on $K3$}
The heterotic string compactified on $K3$ has
8 unbroken supercharges
or $(1,0)$ supergravity in $D=6$ coupled to 
vector-, tensor- and hypermultiplets.
Contrary to the previously discussed type-II
compactifications this heterotic string theory does not 
uniquely fix the massless spectrum 
but instead one finds distinct families of 
string vacua with different contents of massless 
states.\footnote{Of course, it is a general
property of supersymmetry and string theory that fewer unbroken
supercharges lead to a much richer variety
of low-energy spectra.} 
However, $(1,0)$
supersymmetry is chiral and thus gauge and gravitational
anomaly cancellation do impose some 
constraints on the allowed spectra of supermultiplets.
One finds the condition \cite{gswest}
\begin{equation}\label{eq:Rfourconstraint}
 n_H - n_V + 29\, n_T -273 = 0,
\end{equation}
where $n_H$, $n_V$ and $n_T$ are the numbers of hyper, 
vector and tensor multiplets, respectively.
(The specific field content of these multiplets
is displayed in Table~\ref{D6-susy-mult}.)
In the perturbative spectrum of the 
heterotic string there is only one tensor multiplet 
which contains the selfdual part of
$b_{\mu\nu}$ (the anti-selfdual piece resides in the
gravitational multiplet) and the dilaton 
and hence anomaly cancellation in any perturbative
heterotic string vacua demands
$n_H-n_V=244$.
In addition the Green--Schwarz mechanism
requires a modified field strength $H$
for the antisymmetric tensor
$H= db +\omega^L - \sum_a v_a\, \omega_a^{YM}$ where 
$\omega^L$ is a Lorentz--Chern--Simons term and $\omega_a^{YM}$ 
are the Yang--Mills Chern--Simons terms \cite{gswest}.
The index $a$ labels the factors $G_a$ of the gauge group
$G= \otimes_a G_a$ and $v_a$ are some computable
constants which depend on the specific massless spectrum.
In order to ensure a globally well-defined $H$ on the compact 
$K3$ the integral $\int_{K3} {\rm d}H$ has to vanish. 
This implies
\begin{equation}\label{eq:instanton}
\sum_a n_a\, \equiv\, \sum_a \int_{K3}\left( {\rm tr}\,F^2\right)_a
\,=\, \int_{K3}{\rm tr}\, R^2\, =\, 24,
\end{equation}
where the last equation used the fact that 
24 is the Euler number of $K3$.
{}From (\ref{eq:instanton}) we learn that 
in any compactification of the heterotic string on $K3$ 
the original 10-dimensional gauge symmetry 
(${\rm E}_8\times {\rm E}_8$ or SO(32)) is necessarily broken
since consistency requires
a non-vanishing instanton number $n_a$.

As before we can also ask for perturbative equivalences
on the space of heterotic $K3$ compactifications.
It has been shown that the $K3$ compactifications of 
the 10-dimensional heterotic string with
gauge group SO(32) lie in the same moduli space
as (particular) $K3$ compactifications of 
the 10-dimensional heterotic string with
gauge group ${\rm E}_8\times {\rm E}_8$ \cite{morva1,bele}.
More precisely, 
the gauge group is really Spin$(32)/Z_2$ and 
one has to distinguish 
two different types of instantons which 
are characterized by the second Stieffel--Whitney
class \cite{bele}.
The corresponding distinct
compactifications
of the Spin$(32)/Z_2$ heterotic string
are also called compactifications
with and without vector structure.
It has been shown that
compactifications with vector structure have a common 
moduli space with ${\rm E}_8\times {\rm E}_8$ compactifications
of instanton numbers $n_1=8, n_2=16$ \cite{morva1}
while compactifications without vector structure 
have a common 
moduli space with ${\rm E}_8\times {\rm E}_8$ compactifications
of instanton numbers $n_1= n_2=12$ \cite{bele}.
The continuous path which connects the two pairs
of theories involves a transformation 
$R\to \alpha'/R$
and hence they are also called T--dual.
Furthermore, the ${\rm E}_8\times {\rm E}_8$ compactifications
with instanton numbers
$n_1=n_2=12$ are part of the same moduli space
as the compactifications
with instanton numbers
$n_1=10, n_2=14$ \cite{aldazabal,morva1}.

\paragraph{Type-I compactified on $K3$.}
Also this compactification leads to $(1,0)$ 
supersymmetry and thus anomaly cancellation
imposes the same constraint \refeq{eq:Rfourconstraint}
on the massless spectrum. However, in type-I 
compactification there can be more than one tensor multiplet
and as a consequence also a generalized Green-Schwarz
mechanism can be employed \cite{sagnotti}.
The resulting spectra are much less investigated 
and we refer the reader to the literature for further
details \cite{sagnotti,gimon}.

\subsubsection{Calabi-Yau threefolds compactifications}
Compactifications on a 6-dimensional Calabi-Yau threefold
$Y_3$ break $3/4$  of the supercharges present in $D=10$
and hence one moves two columns to the right 
and six rows down in the $D$-$Q$ plane 
of Fig.~\ref{fig:sugra}; the resulting string theories
therefore have $D=4$ and either 8 or 4 supercharges.
The massless modes of such compactification
arise from the nontrivial harmonic forms on $Y_3$ 
which again are most conveniently summarized by
the Hodge diamond 
\begin{equation} \label{eq:hodge-3}
\begin{array}{c c c c c c c}
           &  &              &  1 &   &  &    \\
      & &      0\! &  & \! 0 & &        \\
  & 0\! & &h^{1,1} & &\! 0 & \\
  1 & & h^{1,2}\! & & \! h^{1,2}& &  1\\
   & 0\! & & h^{1,1} & & \! 0& \\
 & & 0\! & &\! 0 & & \\
 & & & 1 & & &  
\end{array}\ ,
\end{equation}
where $h^{1,1}$ and $h^{1,2}$ are arbitrary integers  
\cite{tex1,mirrorrev}.
The corresponding $(1,1)$ and $(1,2)$ forms are the 
deformations of the Calabi-Yau metric 
and the complex structure, respectively.
Using the definition \refeq{Euler} one finds the Euler number 
of a threefold to be
$\chi(Y_3) = 2(h^{1,1}-h^{1,2})$.
It is believed that most (if not all) 
Calabi--Yau threefolds $Y_3$ have an associated 
mirror manifold $\tilde{Y}_3$ with the property
that its Hodge numbers are exactly reversed, 
i.e.~$h^{1,1}(\tilde{Y_3})=h^{1,2}(Y_3)$ 
and
$h^{1,2}(\tilde{Y_3})=h^{1,1}(Y_3)$, so that
$\chi(Y_3)=-\chi(\tilde Y_3)$ \cite{mirror}.

The moduli space of Calabi--Yau threefolds is 
locally a direct product space 
\begin{equation}
{\cal M}_{Y_3} 
= {\cal M}_{(1,1)} \otimes {\cal M}_{(1,2)}\  ,
\end{equation}
where ${\cal M}_{(1,1)}$ (${\cal M}_{(1,2)}$)
 is the  moduli space 
parameterized by the $(1,1)$-forms ($(1,2)$-forms).
Both factors are constrained to be
special K\"ahler manifolds \cite{dWVP,CdlO,strom,vPrev}.
The metric $G_{i \bar{j}}$ of a K\"ahler manifold
is determined by a real scalar function,
the K\"ahler potential $K$ 
\begin{equation}
G_{i \bar{j}} = \frac{\partial}{\partial \Z^i}
\frac{\partial}{\partial \bar{\Z}^{\bar{j}}}\,
K\left( \Z, \bar{\Z}\right),
\end{equation}
where $\Z^i$ are the (complex) coordinates on the moduli space.
For a  special K\"ahler manifold  the K\"ahler potential 
satisfies the additional constraint
\begin{equation}      \label{eq:kahlerpot}
K = -\log \left[ 2({\cal F}(\Z)+\bar{{\cal F}}(\bar\Z)) -
(\Z^i -\bar{\Z}^{\bar{i}})
({\cal F}_i(\Z) - \bar{\cal F}_{\bar{i}}(\bar\Z))\right]\ , 
\quad {\cal F}_i \equiv \frac{\partial{\cal F} }{\partial \Z^i}\ .
\end{equation}
That is, $K$ is determined by a single 
holomorphic function ${\cal F}(\Z)$.

\paragraph{Type-II compactified on $Y_3$.}
Such compactifications have  8 unbroken supercharges
which is also called  $N=2$ supersymmetry in $D=4$.
The multiplets are the 
gravitational multiplet, the vector multiplet and
the hypermultiplet.

Compactifications of the type-IIA string 
results in the massless modes $g_{\mu\nu}, b_{\mu\nu}, \phi,
g_{ij}, b_{ij}$ from the NS-NS sector and 
$A_\mu, C_{\mu ij}, C_{ijk}$ from the R-R sector.
{}From $g_{ij}, b_{ij}$ one obtains $h^{1,1}+h^{1,2}$
complex massless scalar fields in the NS-NS
sector. $C_{\mu ij}$ leads to 
$h^{1,1}$ abelian vectors  
while  $h^{1,2}$
complex scalars arise from $C_{ijk}$ 
(all in the R-R sector)
\cite{NS,CFG}.
These states (together with their fermionic
partners) combine into 
$h^{1,1}$ vector multiplets
and $h^{1,2}$ hypermultiplets.
Furthermore, in $D=4$ an antisymmetric tensor is dual to a 
scalar and hence $\D, b_{\mu\nu}$ and two R-R scalars from
$C_{ijk}$ form an additional hypermultiplet.
Thus the total number of vector multiplets is 
$n_V=h^{1,1}$ while the number of hypermultiplets
is given by $n_H=h^{2,1}+1$.

For type-IIB vacua 
one also has $h^{1,1}+h^{1,2}$
complex massless scalar fields in the NS-NS
sector but now 
$h^{1,2}$ abelian vectors together 
with $h^{1,1}$
complex scalars in the R-R sector (the universal
hypermultiplet containing the dilaton is again present)
\cite{NS,CFG}.
Hence,
$n_V=h^{2,1}$ and  $n_H=h^{1,1}+1$ holds
for the type-IIB theory.

The gauge group is always abelian and given
by $(h^{1,1}+1)$  U(1) factors in type-IIA
and $(h^{1,2}+1)$ $U(1)$ factors in type-IIB
(the extra U(1) is the graviphoton of the
gravitational multiplet).
We summarize the spectrum of type-II vacua in
Table~\ref{typeII}.

\begin{table}[t]
\begin{center}
\begin{tabular}{lcll} \hline
{}&$\;$&IIA&IIB \\ \hline
$n_H$&&$h^{1,2}+1$&$h^{1,1}+1 $\\
$n_V$&&$h^{1,1}$&$h^{1,2}$\\
G&&U$(1)^{h^{1,1}+1}$& U$(1)^{h^{1,2}+1}$\\ \hline
\end{tabular}
\end{center}
\caption{Massless spectra of type-II vacua.}\label{typeII}
\end{table}

As we see the role of $h^{1,1}$ and $h^{1,2}$
is exactly interchanged between type-IIA 
and type-IIB. Therefore,
compactification of type-IIA on 
a Calabi--Yau threefold $Y_3$ is equivalent to 
compactification of type-IIB on the mirror
Calabi--Yau $\tilde Y_3$.
This is another example of a perturbative 
equivalence of two entire classes of string vacua.

\paragraph{Heterotic string compactified on $Y_3$.}Such 
compactification have 4 unbroken supercharges 
which corresponds to $N=1$ supersymmetry in $D=4$.
Now the $(1,1)$ and $(1,2)$ forms both correspond to massless
chiral multiplets. Generically, there are many distinct
families of string vacua with varying gauge groups and 
matter content and relatively little can be said
in general about their properties.

As in $K3$ compactifications of the heterotic string
the field strength $H$ of the antisymmetric tensor
has to be modified by appropriate Lorentz-- and Yang--Mills
Chern--Simons terms. However, the requirement for a globally
defined $H$ on $Y_3$ is slightly more involved since
the compact manifold is now 6-dimensional and the integral
over ${\rm d}H$ is no longer a topological invariant.

A special class of consistent 
compactifications is obtained
by embedding the
spin connection in the gauge connection
of  ${\rm E}_8\times {\rm E}_8$ or SO(32) \cite{CHSW}.
In the first case one obtains a gauge group ${\rm E}_8\times {\rm E}_6$
with $h^{1,1}$ chiral multiplets in the ${\bf 27}$
and $h^{1,2}$ chiral multiplets in the ${\bf \overline{27}}$
representation of ${\rm E}_6$. In addition there are also
$h^{1,1}+ h^{1,2}$ gauge neutral moduli multiplets.

Similarly,
compactifications of the SO(32) heterotic string
lead to a gauge group ${\rm SO}(26)\times {\rm U}(1)$
with $h^{1,1}$ chiral multiplets in the 
${\bf 26}_1 \oplus {\bf 1}_{-2}$
and $h^{1,2}$ chiral multiplets in the 
${\bf 26}_{-1} \oplus {\bf 1}_{2}$
representation of ${\rm SO}(26)\times {\rm U}(1)$. In addition, there are
$h^{1,1}+ h^{1,2}$ gauge neutral moduli multiplets.

Again we encounter a perturbative equivalence:
A heterotic string compactified on a given $Y_3$ 
leads to the exact same string vacuum as a compactification
on the mirror manifold $\tilde Y_3$. This can be seen 
from the above assignment of the massless spectrum 
(it is only convention what is called ${\bf 27}$
versus ${\bf \overline{27}}$)
but also has been shown more generally for the full 
string theory \cite{mirror}.

There are many compactifications 
with different embeddings than the one discussed above
and they  can lead to very
different spectra.
The space of heterotic $Y_3$ compactifications
displays a much bigger variety of spectra than 
any of the
compactifications discussed so far and 
many of the properties can only be discussed 
on a case-by-case basis \cite{DQ}.

\paragraph{Type-I string compactified on $Y_3$.}
This class of string compactification also has
4 supercharges ($N=1$) but has been much less 
investigated as the heterotic string. 
There are also many 
distinct families of string vacua with varying spectra and couplings \cite{bianchi}.

\subsection{Duality in string theory}

The concept of duality is very common in physics.
Generically it means that there are two (or more) 
different descriptions of the same physical
system. 
Frequently the different descriptions
are only valid in specific domains of the 
parameter space and only together they
can be used to cover the entire parameter space
of the physical system.
The past few years have shown \cite{dualrev,Mtheory,d-brane4} that also 
the various string theories of Fig.~\ref{fig:sugra}
are interrelated by a complicated `web' of duality
relations; they 
are not at all independent but instead are 
different regions of a common parameter space.
In fact, it seems that a given representation of the
supersymmetry algebra (with a given number of 
supersymmetries and spacetime dimensions)
lead only to one
distinct quantum theory with a parameter space that can
encompass various perturbatively distinct 
string theories. In Fig.~\ref{fig:sugra} we plotted the different
perturbative compactifications, some of which 
share the same representation of supersymmetry.
As we will see in this section they all turn out 
to be different regions in a common parameter space.

One distinguishes 
perturbative and nonperturbative dualities.
Perturbative dualities already hold at weak string
coupling and the map which identifies the perturbative
theories 
does not involve the dilaton.
On the other hand nonperturbative dualities identify regions
of the parameter space which are not simultaneously
at weak coupling and
the duality map involves the dilaton in a nontrivial way.
Such nonperturbative dualities
are of utmost importance since they 
map the strong-coupling region
of a given (string) theory to the weak-coupling region
of a dual theory where perturbative methods
are applicable and hence the strong-coupling limit
gets (at least partially) under quantitative control.

The perturbative dualities are well established and 
we have already seen them in the previous section.
The nonperturbative dualities are more difficult to 
deal with and they cannot be proven at present.
Rather their validity has only been checked for quantities
or couplings which do not receive quantum corrections.
Such quantities or couplings  exist
in supersymmetric (string) theories;
they are the BPS states of the theory as well as
the holomorphic couplings 
(such as the prepotential ${\cal F}(z)$
of $N=2$ supergravity in $D=4$) of the effective action.
It is precisely for this reason that supersymmetry has played
such an important (technical) role in 
establishing nonperturbative dualities.

Let us first briefly discuss the perturbative dualities
from a common point of view. Then we focus  on
the  nonperturbative dualities
and  discuss the various relations  in turn.

\subsubsection{T--dualities}

All perturbative dualities are now called
T--dualities but one can divide them into two classes.
In toroidal compactifications (which we discussed in section~3.2.1.)
different points in the parameter space 
of the compactification correspond to equivalent theories
with the exact same spectrum and interactions.
As a consequence there is a discrete symmetry $\Gamma_T$
acting on the space of toroidal compactifications.
The same situation is encountered
in $K3$ compactifications of 
type-IIA  string theories where also a discrete
symmetry $\Gamma_T$ has been identified (cf. \refeq{gammatk3}).

A different situation occurs
in $T^1$ or $K3$ compactifications of the heterotic string.
What was thought are two distinct perturbative
string theories -- the  ${\rm E}_8\times {\rm E}_8$ heterotic
string compactified on $T^1$ or $K3$ and the 
SO(32) string compactified on $T^1$ or $K3$ --
turn out to be merely different regions of a common
parameter space. In other words
there is a continuous path which connects the two
theories and thus also their respective parameter spaces
are continuously connected. 
Finally, the equivalence 
of type-IIA compactified on $Y_3$ with 
type-IIB compactified on the mirror $\tilde Y_3$
identifies compactifications on
geometrically distinct manifolds as identical
and hence maps the parameter space of type-IIA
compactifications onto the parameter space of type-IIB
compactifications. 
The common feature
of all of these examples is a perturbative equivalence
between string compactifications.
Let us now turn to nonperturbative equivalences.

\subsubsection{S--dualities}

Let $A$ and $B$ be two perturbatively distinct 
string theories each with its own string coupling
$g_A$ and $g_B$, respectively.
However, it is possible that once all quantum corrections 
(including the nonperturbative corrections) 
are taken into account
$A$ and $B$ are equivalent as quantum theories
and one has
\be
A\ \equiv\  B   \ .
\ee
This situation can occur in two different
ways:
\begin{itemize}
\item[(a)]
The strong-coupling limit of $A$ is mapped to the weak 
coupling limit of $B$
\be\label{couplingrel}
\lim_{g_A\to\infty} A\ \rightarrow\  \lim_{g_B\to 0} B \ ,
\ee
or in other words $g_A\sim g_B^{-1}$.
Using \refeq{dilaton} one finds  
in terms of the corresponding dilatons 
the identification
\be \label{dilatonrel}
\D_A = - \D_B\ .
\ee
Along with this strong-weak coupling relation
goes a map of the elementary excitations of
theory $A$ to the nonperturbative,
solitonic excitations of theory $B$ and vice versa. 
The theories $A$ and $B$ are called S--dual 
and one also refers to this situation as
a `string--string duality'.

Examples of S--dual string theories are:
\begin{itemize}
\item[{\bf $\bullet$}] 
The heterotic SO(32) string and 
the type-I string are S--dual in $D=10$.
The evidence for this duality is the
agreement of the low-energy effective actions\footnote{%
In (\ref{D10-Q16-lagrangian1})--(\ref{D10-Q16-lagrangian3})
we displayed the heterotic Lagrangian in different frames.
The last frame (\ref{D10-Q16-lagrangian3})
shows the equivalence with type-I.}
once one identifies $\D_{\rm HSO} = - \D_{\rm I}$ 
\cite{various,type1a,type1b}
and the fact that the perturbative
heterotic SO(32) string
has been identified as the D--string of the type-I
theory \cite{polch-witt}. 
In the limit of strong coupling in the type-I
theory ($g_{\rm I}\to\infty$) the heterotic SO(32) 
string becomes the `lightest' and thus perturbative
object. 

\item[{\bf $\bullet$}] 
The type-IIA string compactified on $K3$ and
the heterotic string compactified on $T^4$ are S--dual
\cite{dufflu,hull,various,sen-he,hav-stro}.
Both theories have $(1,1)$ supersymmetry in $D=6$ with exactly
the same massless spectrum.
Furthermore, from (\ref{eq:Mhettoroidal}, \ref{eq:2aonk3})
one learns that also the moduli spaces of the two
string compactifications coincide.
The effective actions of the two perturbative theories
agree if one  identifies \cite{various}
\bea \label{eq:strowe}
\D_{\rm H} &=& - \D_{\rm IIA}\ ,\nonumber \\
H_{\rm H} &=& e^{-2\D_{\rm IIA}} \,{}^\ast H_{\rm IIA}\ ,  \\
(g_{\mu\nu})_{\rm H} &=& e^{-2\D_{\rm IIA}} (g_{\mu\nu})_{\rm IIA}\ ,
\nonumber 
\eea
where $H={\rm d}b$ is the field strength of the antisymmetric
tensor and ${}^\ast H$ is its Poincare dual.
The first equation in (\ref{eq:strowe})
again implies a strong-weak coupling relation while
the second  is the equivalent of an 
electric-magnetic duality.
Further evidence for this S-duality 
arises from the observation that the zero modes 
in a solitonic string background
of the type-IIA theory compactified on $K3$ 
have the same structure
as the Kaluza--Klein modes 
of the heterotic string compactified 
on $T^4$ \cite{sen-he,hav-stro}.
\end{itemize}

\item[(b)]
There is a variant of the above situation
where the dilaton of theory $A$ is not mapped to
the dilaton of theory $B$ as in \refeq{dilatonrel}, 
but rather to any of the other perturbative moduli 
$R_B$ of theory $B$.
In this case one has the identifications
\be 
\D_A \sim R_B\ ,\qquad \D_B \sim R_A \  ,
\ee
or in other words the strong-coupling limit of 
$A$ is independent of $g_B$ 
\be 
\lim_{g_A\to\infty} A\ =\  {\rm independent\ of}\ g_B \ .
\ee
As in case (a) also here the strong-coupling limit
of $A$ is controlled by the perturbative regime
of theory $B$ and thus accessible in perturbation
theory (at least in principle).

This situation is found in the following 
examples:
\begin{itemize}
\item[{\bf $\bullet$}] The type-II string compactified on $Y_3$
and the heterotic string
compactified on $K3\times T^2$ are  S--dual in the sense 
just defined \cite{kachru,fhsv}.
The heterotic dilaton is a member of a vector multiplet
and mapped to one of the geometrical moduli 
of the Calabi-Yau
threefold $Y_3$.\footnote{More precisely, $Y_3$ has to be
a $K3$-fibration and the heterotic dilaton is mapped
to the modulus parameterizing the size of the ${\bf P^1}$
base of the fibration \cite{mayr,VW,luy}.}
Conversely, the type-II dilaton is part of
a hypermultiplet and mapped to one of the geometrical
moduli of the $K3$.
The validity of this duality has been checked
in a variety of ways for quite a number 
of dual string vacua \cite{mayr,VW,luy,IIfour}.
In particular it has been shown that the 
prepotential ${\cal F}(z)$ 
appearing in \refeq{eq:kahlerpot} 
agrees for dual pairs of string vacua.

\item[{\bf $\bullet$}] In the same sense the heterotic string 
compactified on $K3$ 
and the type-I string compactified on $K3$ 
are S--dual \cite{bele,APT}.
\end{itemize}
\end{itemize}

Let us summarize the known S--dualities in the following table
$$
\begin{array}{lllcl}
D=10\!\!&:& {\rm HSO}&\stackrel{S}{\longleftrightarrow}& {\rm I}\\
D=6\!\!&:&  {\rm IIA}/K3&\stackrel{S}{\longleftrightarrow} & {\rm H}/T^4\\
   & &    {\rm H}/K3&\stackrel{S}{\longleftrightarrow} &  {\rm I}/K3\\
D=4\!\!&:&  {\rm II}/Y_3&\stackrel{S}{\longleftrightarrow} & {\rm 
H}/K3\times T^2\\ 
\end{array}
$$
In Fig.~\ref{fig:dual} these S-dualities are denoted
by a horizontal bar $(\protect\rule[0.5mm]{3mm}{0.5mm})$.

\subsubsection{Self-duality and $U$-duality}

Another situation is encountered when 
the strong-coupling limit of a theory $A$ is controlled
not by a distinct theory $B$, but rather 
by a different perturbative region of the same theory $A$. 
That is, the strong-coupling regime of $A$
has an alternative weakly-coupled description
within the same theory $A$ but in terms of 
a different set of elementary degrees of freedom.
The new perturbation theory
is often called the magnetic theory and its 
perturbative degrees of freedom are called magnetic 
degrees of freedom. This stems from the fact
that the first duality (in $D=4$) 
put forward by Montonen and Olive \cite{mo-ol}
suggested that an electric U(1) gauge theory is dual
to a magnetic U(1) gauge theory with a magnetic photon
and magnetic monopoles as perturbative degrees 
of freedom.
This situation is more general and can appear also
for extended objects. However, for such
a self-duality to hold the theory 
$A$ has to have a nontrivial  (discrete) symmetry
which maps the strong-coupling region to a region
of weak coupling and simultaneously 
the different elementary excitations onto each other.
One has to make a further subdivision of
this case:

\begin{itemize}

\item[(a)]
The symmetry group is $\Gamma_S={\rm SL}(2,Z)$ which acts on a 
single complex scalar field containing the
dilaton as its real (or imaginary) part.  
Unfortunately this situation is also called  S--duality and the  
associated symmetry group $\Gamma_S$ is called  
the S--duality group.
(We prefer to call it a special case of a  U--duality.)\\
Examples of this case are:

\begin{itemize}
\item[{\bf $\bullet$}]
The type-IIB string in $D=10$ is conjectured to
have $\Gamma_S={\rm SL}(2,Z)$ \cite{hull,various,type1b}.
The corresponding supergravity theory has a ${\rm SL}(2,R)$ 
as a symmetry group \cite{SH}
(see also Table~\ref{maximal-cosets})
but quantum corrections break this continuous symmetry
to its discrete version ${\rm SL}(2,Z)$.
This exact symmetry predicts an infinite number
of equivalent weakly coupled type-IIB strings 
which carry R-R charge;
such strings have indeed been identified as 
appropriate D-strings \cite{JSchwarz,wittenbs}.

\item[{\bf $\bullet$}]
A second example is the heterotic string compactified
on $T^6$ which has $D=4$ and also $\Gamma_S={\rm SL}(2,Z)$.
In toroidal compactifications of the heterotic
string the dilaton is the unique scalar in the
gravitational multiplet and parameterizes the
$R^+$ component of the heterotic moduli
space \refeq{eq:Mhettoroidal}.
However, in $D=4$ an antisymmetric tensor
of rank 2 is on-shell equivalent to a 
pseudoscalar and thus can be combined together
with the dilaton 
into one complex scalar of the gravitational
multiplet spanning the SU(1,1)/U(1) component
of the moduli space. It is this complex scalar
on which $\Gamma_S$ acts, leaving all other
toroidal moduli invariant.\footnote{%
  Note that this is rather different than the previous case 
  where $\Gamma_S$ acted on the two scalars
  of the IIB string.} %
The vacuum-expectation value of the pseudoscalar
plays the role of the $\theta$-angle and so
this duality is nothing but the string theoretical version of
the original electric--magnetic Montonen--Olive duality
which in many respects started the subject of 
string dualities \cite{mo-ol,harveyrev,FILQ,Sen}.
\end{itemize}

In Fig.~\ref{fig:dual} we mark the theories with
$\Gamma_S={\rm SL}(2,Z)$ by an `S' next to it.

There also is a nontrivial  generalization of this 
case:\\

\item[(b)]
The product of $\Gamma_S$ and the $T$-duality
group $\Gamma_T$ is contained as a maximal subgroup 
in a bigger group $\Gamma_U$, called the U--duality 
group \cite{hull}.
This situation is encountered in toroidal $T^n$
compactifications of the type-II string.
$\Gamma_S$ is `inherited' from the type-IIB string
in $D=10$ and $\Gamma_T$ 
already exists at the perturbative level 
(cf. \refeq{eq:Ttoroidal}).
As we already discussed extensively in section~2.5, the 
corresponding supergravities 
do have a much larger continuous symmetry group
which transforms all scalar fields
into each other irrespective of their
NS-NS or R-R origin 
(cf.\ Table~\ref{maximal-cosets}). 
This is a consequence of the
fact that the supergravities have a unique 
gravitational multiplet, which contains all scalar
fields on an equal footing.
Furthermore, they are constructed as 
toroidal compactifications of the 11-dimensional
supergravity  while the string vacua arise
as compactifications of 10-dimensional string theories.
Within the perturbative regime there can never be 
a symmetry which mixes NS-NS scalars with their R-R
`colleagues'  due to their rather different dilaton couplings.
However, nonperturbatively, when 
also D-brane configurations are taken
into account, such a symmetry is no longer forbidden
and evidence in its favour has been 
accumulated \cite{hull,various}.
The necessary states carrying R-R charge have been 
identified and the nonperturbative BPS-spectrum
assembles in representations of $\Gamma_U$.
The U--duality groups in arbitrary dimensions
are summarized in Table~\ref{tab:Uduality} \cite{hull};
they are just the discrete version of the group
G in Table~\ref{maximal-cosets}. 
In Fig.~\ref{fig:dual} these theories are marked with 
a `U'.
\end{itemize}
%

\begin{table}[t]
\begin{center}\begin{tabular}{cccc}
\hline 
D&$\;$ & $\Gamma_T$ & $\Gamma_U$ 
\\ \hline
10A &&1& $1$  \\  
10B &&1& {\rm SL}(2,Z)\\  
9 && $Z_2$ & ${\rm SL}(2,Z)\!\times\! Z_2$\\ 
8 && ${\rm SO}(2,2,Z)$ & ${\rm SL}(3,Z)\!\times\! {\rm SL}(2,Z)$\\ 
7 && ${\rm SO}(3,3,Z)$ & ${\rm SL}(5,Z)$\\ 
6 && ${\rm SO}(4,4,Z)$ & ${\rm SO}(5,5,Z)$\\ 
5 && ${\rm SO}(5,5,Z)$ & ${\rm E}_{6(+6)}(Z)$\\ 
4 && ${\rm SO}(6,6,Z)$ & ${\rm E}_{7(+7)}(Z)$\\ \hline 
\end{tabular}\end{center}
\caption{U--duality groups for type-II strings.}\label{tab:Uduality}
\end{table}

\subsubsection{M-theory}
The various dualities discussed so far relate different
perturbative string theories. In these cases the 
strong-coupling limit of a given string theory is 
controlled by another (or the same) perturbative string theory.
However, not all strong-coupling limits are of this type.
Instead it is possible that the 
strong-coupling limit of a given theory 
is something entirely new, not
any of the other string theories \cite{various}.
Only limited amount of information is so far
known about this new theory which is called  $M$-theory.
Examples of this situation are:

\begin{itemize}

\item[{\bf $\bullet$}] The strong-coupling
limit of the type-IIA theory in $D=10$. 
The low-energy effective action was discussed in section
2.5  where we also indicated how it can be constructed
as a $T^1$ compactification of 11-dimensional
supergravity. This implied a relation between the radius
$R_{11}$ of the 11-th dimension and
the string coupling constant $\gstring =e^{\langle\phi\rangle}$
\cite{various} (cf. \refeq{Rphi})
\be\label{rgrel}
R_{11} = \frac{L}{2\pi}\, \gstring^{\frac23} \ ,
\ee
where $L$ is the length of the 11-th dimension introduced in section~2.5.
Moreover, the Kaluza-Klein spectrum of this theory
obeys (in the string frame) 
\be
M^{\scriptscriptstyle\rm KK} = \frac{2\pi |n|}{\gstring L}\ ,
\ee
where $n$ is an arbitrary integer (cf. \refeq{Mphi}). These KK-states are not
part of the perturbative type-IIA spectrum since
they become heavy 
in the weak-coupling limit $\gstring\to 0$.
However, in the strong-coupling limit $\gstring\to\infty$
they become light and can no longer be neglected
in the effective theory.  This  infinite number of light
states (which can be identified with D-particles
of type-IIA string theory, or extremal black holes of IIA 
supergravity) signals that the theory effectively decompactifies, 
which  can also be seen from \refeq{rgrel}.
Supersymmetry is unbroken in this limit
and hence the KK-states assemble in supermultiplets
of the 11-dimensional supergravity.
Since there is no string theory which has 
11-dimensional supergravity as the low-energy limit,
the strong-coupling limit of type-IIA string theory
has to be a new theory, called M-theory, which cannot be a theory 
of (only) strings. M-theory is supposed to capture all degrees of 
freedom of all known string theories, both at the perturbative 
and the nonperturbative 
level \cite{Townsend,various,horava1,Mtheory}. There exists a 
conjecture according to which the degrees of freedom of M-theory 
are captured in U($N$) supersymmetric matrix models in the $N\to 
\infty$ limit \cite{BFSS}. These matrix models have been known 
for some time \cite{CH} and were also known to describe 
supermembranes \cite{membranes} in the lightcone gauge 
\cite{DWHN}. The same quantum-mechanical models describe the 
short-distance dynamics of $N$ D-particles, caused by the exchange 
of open strings \cite{wittenbs,d-brane4}. 
A review of these developments is beyond the scope of these
lectures and we refer the reader to the 
literature \cite{Matrix}.

\item[{\bf $\bullet$}] 
A second and maybe even more surprising result
shows that also the strong-coupling limit
of the heterotic ${\rm E}_8\times {\rm E}_8$ string is
captured by M-theory. In this case, 11-dimensional 
supergravity is not compactified on a circle but
rather on  a $Z_2$ orbifold of the circle \cite{horava1}.
The space coordinate 
$x^{11}$ is odd under the action of $Z_2$
and hence the three-form $C_{\mu\nu\rho}$
as well as $g_{\mu 11}$
are also odd. The $Z_2$-invariant spectrum 
in $D=10$ consists of the metric $g_{\mu \nu}$,
the antisymmetric tensor $C_{\mu\nu11}$
and the scalar $g_{11\, 11}$. 
Up to the gauge degrees of freedom
this is precisely the massless spectrum of the
10-dimensional heterotic string. 
The ${\rm E}_8\times {\rm E}_8$ 
Yang-Mills fields have to arise in the twisted sector 
of the orbifold.
One way to see this is to note that the 
$Z_2$ truncation of 11-dimensional supergravity 
is inconsistent
in that it gives rise to gravitational anomalies
\cite{gaume}. In order to cancel such 
anomalies non-abelian gauge fields have to be present
with appropriate couplings to the antisymmetric
tensor such that a Green-Schwarz mechanism can be employed
\cite{green-schwarz}.
Such additional states can only appear in the 
twisted sectors of the orbifold theory which 
are located at the orbifold fixed points 
$x^{11}=0$ and $x^{11}=L/2$. 
However, due to the $Z_2$ symmetry, these two 
10-dimensional hyperplanes
have to contribute equally to the anomaly.  
This can only be achieved for a gauge group 
which is a product
of two factors and thus ${\rm E}_8\times {\rm E}_8$ with one 
${\rm E}_8$ factor on each hyperplane is the only consistent candidate
for such a theory \cite{horava1}.
Just as in the type-IIA case 
one has $R_{11} = g_{\rm H}^{3/2} L/2\pi$ and thus 
weak coupling corresponds to  
small $R_{11}$ and the two 10-dimensional
hyperplanes sit close to each other;
in the strong-coupling limit the two 10-dimensional
hyperplanes move far apart (to the end of the world).
Using the previous terminology the heterotic 
${\rm E}_8\times {\rm E}_8$
string theory can be viewed as M-theory compactified 
on $T^1/Z_2$.

\item[{\bf $\bullet$}] 
There is an immediate corollary of the dualities discussed
so far. The strong-coupling limit of the type-IIA string compactified
on $K3$ is simultaneously governed by M-theory
compactified on $K3\times T^1$ and the heterotic string
compactified on $T^4$. Since there is a $T^1$ in both
theories one concludes that the strong-coupling limit
of the heterotic string
compactified on $T^3$ is governed by M-theory
compactified on $K3$ \cite{various}.
{}From \refeq{eq:Mhettoroidal} and \refeq{MKthree}
we learn that indeed the moduli spaces
of both theories agree if the heterotic
dilaton is related to  the overall size of the $K3$. 
A detailed comparison of the respective
effective actions reveals that the strong-coupling limit
on the heterotic side corresponds to the large-radius 
limit of the $K3$ on the M-theory side \cite{various}.

\item[{\bf $\bullet$}] 
The exact same argument can be repeated in $D=5$.
The strong-coupling limit of the type-IIA string  compactified
on (a $K3$-fibred) $Y_3$ 
is simultaneously governed by M-theory
compactified on $Y_3\times T^1$ and the heterotic string
compactified on $K3\times T^2$.
By the same argument used above one concludes that
the strong-coupling limit
of the heterotic string
compactified on $K3\times T^1$ is governed by M-theory
compactified on (a $K3$-fibred) $Y_3$ \cite{AFT}.
In this case the heterotic dilaton is directly related
to the size of the ${\bf P^1}$ base of the $K3$-fibration.

\item[{\bf $\bullet$}] 
A slightly more involved analysis is necessary
to conclude that the strong-coupling limit
of the IIB string compactified on $K3$ is governed by M-theory
compactified on $T^5/Z_2$ \cite{dasgupta}.
Compactifying 11-dimensional supergravity
on the  orbifold $T^5/Z_2$ one obtains the chiral $(2,0)$ 
supergravity with one gravity multiplet and 
five tensor multiplets  from the untwisted sector. 
The twisted sector is again inferred by 
anomaly cancellation and provides 16 further 
tensor multiplets. The weakly-coupled
type-IIB theory on $K3$ corresponds to a `smashed' 
$T^5/Z_2$ where the
32 fixed points degenerate into 16 pairs 
and the 16 tensor multiplets are
equally distributed among those pairs.
The 81 scalars from the NS-NS sector combine with
the 24 scalars from the R-R sector to form the
moduli space \cite{romans,aspe,hul}
\begin{equation}
{\cal M} = \left. \frac{{\rm SO}(5,21)}{{\rm SO(5)}\times {\rm 
SO}(21)}\right/ \Gamma_U,
\end{equation}
with a  U--duality group $\Gamma_U = {\rm SL}(5,21,Z)$.
The local structure of this moduli space is
already fixed by supergravity (cf. \refeq{Mtwob})
while the global structure follows from M-theory.
\end{itemize}

\noindent
Let us summarize the nontrivial compactifications
of M-theory:
$$
\begin{array}{lcl}
{\rm M}/T^1    &\to & {\rm IIA}\\
{\rm M}/T^1/Z_2&\to & {\rm HE8}\\
{\rm M}/K3     &\to & {\rm H}/T^3\\
{\rm M}/T^5/Z_2&\to & {\rm IIB}/K3\\
{\rm M}/Y_3    &\to & {\rm H}/K3\times T^1\\
\end{array}
$$
Theories whose strong-coupling limit is governed
by M-theory are marked with an `M' in Fig.~\ref{fig:dual}.

\subsubsection{$F$-theory}
As we discussed previously 
the type-IIB theory in 10 spacetime dimensions
is believed to have an exact ${\rm SL}(2,Z)$ quantum symmetry
which acts on the complex scalar 
$\tau = e^{-2\D}  + i \D^\prime$, 
where $\D$ and  $\D^\prime$ are the two scalar fields of
type-IIB theory (c.f.~Table~\ref{tab:tab2}).
This fact led Vafa to propose that the type-IIB string 
could be viewed as the toroidal compactification
of a twelve-dimensional theory, called F-theory,
where $\tau$ is the 
complex structure modulus of a  two-torus 
$T^2$ and the K\"ahler-class
modulus is frozen \cite{vafas-f}.
Apart from having a geometrical interpretation of 
the ${\rm SL}(2,Z)$ symmetry 
this proposal led to the construction
of new, nonperturbative string vacua in 
lower space-time dimensions. In order to preserve the 
${\rm SL}(2,Z)$ quantum symmetry the 
compactification manifold cannot be arbitrary
but has to be what is called an elliptic fibration.
That is, the manifold is locally a fibre bundle
with a two-torus $T^2$ over some base $B$ but there are 
a finite number of singular points where the 
torus degenerates.
As a consequence nontrivial closed loops 
on $B$ can induce a  ${\rm SL}(2,Z)$
transformation of the fibre.
This implies that the dilaton is not constant 
on the compactification manifold, but can 
have  ${\rm SL}(2,Z)$ monodromy \cite{stringy-cosmic}.
It is precisely this fact which results in
nontrivial (nonperturbative) string vacua
inaccessible in string perturbation theory. 

F-theory can be compactified on elliptic Calabi--Yau
manifolds and each of such compactifications is
conjectured to capture the nonperturbative physics
of an appropriate string vacua. One finds:
\begin{itemize}

\item[{\bf $\bullet$}] The IIB string 
in $D=10$ can be viewed as F-theory compactified
on $T^2$ with a frozen K\"ahler modulus.

\item[{\bf $\bullet$}]
F-theory compactified on an elliptic
$K3$ yields
an 8-dimensional vacuum with 16 supercharges
which is quantum equivalent to the heterotic string
compactified on $T^2$ \cite{vafas-f,sen-f}.

\item[{\bf $\bullet$}] 
F-theory compactified on an elliptic
Calabi--Yau threefold has 8 unbroken supercharges and
is quantum equivalent to
the heterotic string compactified on $K3$ \cite{morva1}.
In fact there is a beautiful correspondence
between the heterotic vacua labelled by the 
instanton numbers $(n_1,n_2)$ and 
elliptically fibred Calabi-Yau manifolds 
with the base being the
Hirzebruch surfaces $I\!\! F_{n_2-12}$
(we have chosen $n_2\ge n_1$) \cite{morva1}.

\item[{\bf $\bullet$}]
Finally, the heterotic string compactified on a
Calabi--Yau threefold $Y_3$ is quantum equivalent
to F-theory compactified on an elliptic
Calabi--Yau fourfold \cite{cyfour}.
Calabi-Yau fourfolds are Calabi-Yau manifolds of 
complex dimension
four and holonomy group SU(4).
\end{itemize}
Let us summarize the nontrivial compactifications
of F-theory:
$$
\begin{array}{lcl}
{\rm F}/T^2    &\to & {\rm IIB}\\
{\rm F}/K3     &\to &{\rm H}/T^2\\
{\rm F}/Y_3    &\to & {\rm H}/K3\\
{\rm F}/Y_4    &\to & {\rm H}/Y_3\\
\end{array}
$$
The theories governed by F-theory are marked with an `F'
in Fig.~\ref{fig:dual}.

\subsubsection{Summary of all strong-coupling limits.}
\begin{figure}  
\epsfig{figure=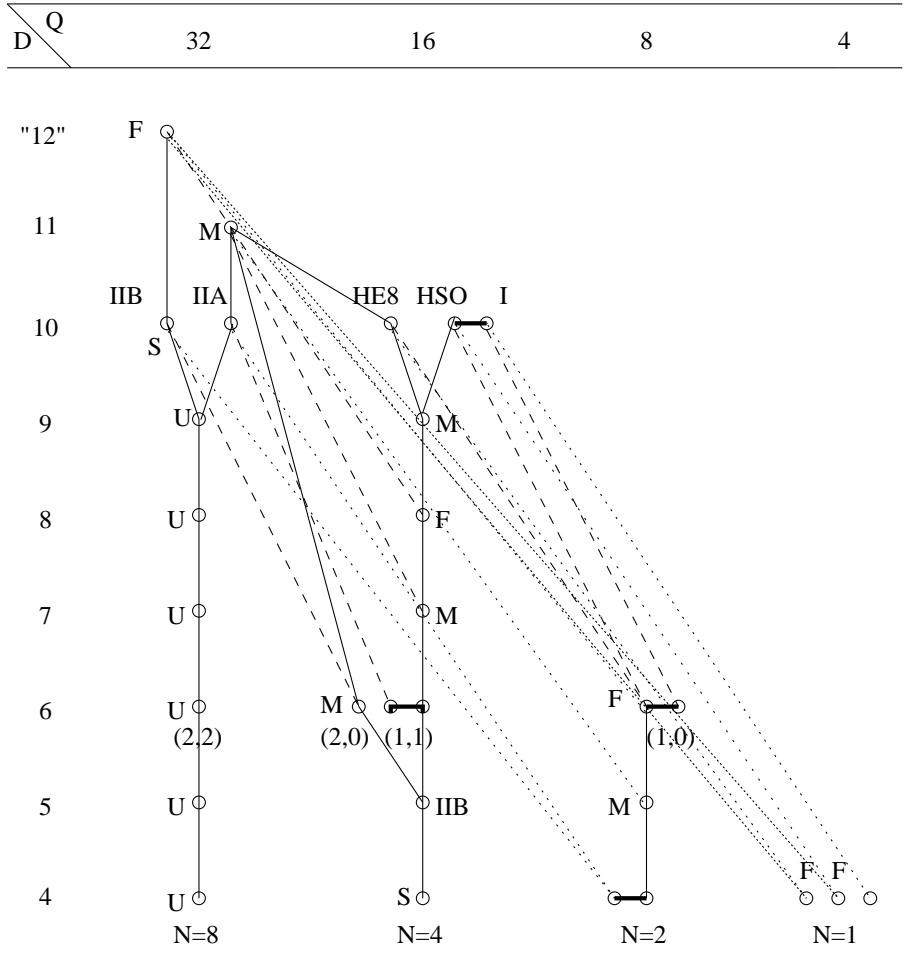}
\caption{The distinct string theory and their
strong-coupling limit. As in Fig.~4 
the solid line ($-$) denotes toroidal
compactification, the dashed line ($--$) denotes 
$K3$ compactifications and the dotted line ($\cdots$)
denotes $Y_3$ compactifications.
The fine-dotted
line ($\cdot\!\! \cdot\!\! \cdot$)
denotes $Y_4$ compactifications 
while the horizontal bar 
$(\protect\rule[0.5mm]{3mm}{0.5mm})$
indicates a string-string duality. 
The theories marked with a `U' (`S')
have a U--duality (S--duality); 
the strong-coupling limit of the theories marked
by `M' (`F') are controlled by M-theory (F-theory). With a slight 
abuse of convention, 
we also denote the two orbifold compactification
${\rm M}/T^1\!/Z_2$ and ${\rm M}/T^5\!/Z_2$
by a solid line.}
\label{fig:dual}
\end{figure}

So far we tried to systematically discuss
the different  possible strong-cou\-pling
limits of string theories along with the relevant
examples. 
In this final section 
let us one more time summarize all 
strong-coupling relations  but now organized by
the number of supercharges.
The following discussion is visualized in
Fig.~\ref{fig:dual}.

\paragraph{$Q=32$.}
Theories with $Q=32$ have a unique massless
multiplet which contains all scalar fields
on an equal footing and does not single out 
a string coupling constant.
As a consequence there is a discrete 
symmetry group $\Gamma_U$ 
(listed in Table~\ref{tab:Uduality})
which leads 
to global identifications in the moduli space 
and in any given region a different scalar
plays the role of the perturbative expansion
parameter.
A special situation occurs in $D=10$ where
the type-IIB string has $\Gamma_U=\Gamma_S={\rm SL}(2,Z)$ while 
the strong-coupling limit of the type-IIA string 
cannot be a string theory but is something new
-- M-theory -- whose low-energy limit is 
11-dimensional supergravity.

\paragraph{$Q=16$.}
Theories with $Q=16$ contain, besides the
gravitational supermultiplet, also a set 
of vector supermultiplets. The gravitational multiplet
always contains one scalar (cf. 
Table~\ref{non-maximal-supergravity}), which can be uniquely 
identified to play the role
of the coupling constant.
In $D=10$ the heterotic SO(32) string
and the type-I string are S-dual while
the strong-coupling limit of the ${\rm E}_8\times {\rm E}_8$
string is governed by M-theory compactified
on an orbifold $T^1/Z_2$.
In $D=9$ the two heterotic theories are
perturbatively equivalent and their strong
coupling limit
is governed both by M-theory and type-I.
In $D=8$ the strong
coupling limit
is governed both by F-theory and type-I;
in $D=7$ the strong-coupling limit 
is governed by M-theory compactified 
on $K3$.
In $D=6$ the heterotic string compactified
on $T^4$ and the type-IIA string compactified
on $K3$ are S-dual and the strong-coupling limit
of type-IIB compactified on $K3$ is captured
by M-theory compactified on $T^5/Z_2$.
In $D=5$ the strong-coupling limit 
of the heterotic string compactified
on $T^5$ is governed by type-IIB compactified
on $K3\times T^1$ \cite{various}.
In $D=4$ the antisymmetric tensor
and the dilaton can be combined into one 
complex scalar with an appropriate action of 
$\Gamma_U=\Gamma_S={\rm SL}(2,Z)$.

\paragraph{$Q=8$.}
Theories with $Q=8$ supercharges exist in $D=6$ and below.
In $D=6$ the heterotic ${\rm E}_8\times {\rm E}_8$ and the
heterotic SO(32) string are perturbatively
equivalent and their nonperturbative regime is 
governed by F-theory compactified on elliptic
Calabi--Yau threefolds. Furthermore, there also
exists an S-duality with type-I compactified
on $K3$.
In $D=5$ the strong-coupling limit is controlled
by M-theory while in $D=4$ the heterotic string 
compactified on $K3\times T^2$ is S-dual 
to the type-II string compactified on $Y_3$.

\paragraph{$Q=4$.}
Finally, theories with $Q=4$ only exist in $D=4$.
Both heterotic string theories are non-perturbatively
equivalent to F-theory compactified
on an elliptic Calabi--Yau fourfold $Y_4$, while
the strong-coupling limit of the type-I theory
is not yet completely understood.
Some of the type-I models seem to be S-dual to
the heterotic vacua \cite{kakus}. 
It might well be that all three theories are
part of a larger moduli space.

\vspace{8mm}

\noindent
{\bf Acknowledgement}\\
We would like to thank 
P.\ Aspinwall, G.L. Cardoso, V.\ Kaplunovsky, D.\ L\"ust, T. 
Mohaupt, H. Nicolai, J.\ Sonnenschein,
S.\ Theisen, A.\ Van Proeyen, S.\ Yankielowicz 
for the most fruitful collaboration on the 
subjects presented in these lectures.
S.\ F\"orste, C.M.\ Hofman, B.\ Kleijn, P.\ Mayr, E.\ Rabinovici, 
M.A.\ Vasiliev, G.\ Zwart also helped us with useful comments on 
the manuscript. 

We would also like to thank the organizers 
of this school
for providing such a stimulating environment.

\vspace{8mm}

\noindent
{\bf Appendix}
\appendix
 \section{Field representations} 

In section~2  we have outlined the derivation of various 
supermultiplets of states. At the noninteracting level, these 
states can easily be described in terms of local fields. The 
purpose of this appendix is to present suitable field 
representations for the relevant states. With 
the help of these field representations one can then write 
down free massless supersymmetric field theories. Interactions 
can be introduced separately, for instance, by iteration or 
by some more systematic procedure. We should stress that there 
are sometimes ambiguities, because different field representation 
can describe the same massless free states. At the interacting level, 
these ambiguities will usually disappear. So the proper choice of 
the field representation may be subtle.
Our strategy is to discuss a number of standard field 
representations, in $D$ spacetime dimensions, with their 
corresponding free wave equations and exhibit the behaviour of 
the corresponding states under helicity rotations. The 
supermultiplets discussed previously can then be converted into  
supersymmetric actions, quadratic in the  
fields. For selfdual tensor fields, the action must augmented by 
a duality constraint on the corresponding field strength.

 \subsection{Graviton fields}
The linearized Einstein equation for $g_{\mu\nu} = \eta_{\mu\nu} +
\kappa h_{\mu\nu}$ implies that (for $D\geq3$)
\begin{equation}\label{Einstein}
R_{\mu\nu} \propto \partial^2 h_{\mu\nu} + \partial_\mu 
\partial_\nu  h -
\partial_\mu \partial^\rho h_{\rho\nu} - \partial_\nu \partial^\rho 
h_{\rho\m}  = 0 \,,
\end{equation}
where $h \equiv h_{\mu\mu}$ and $R_{\mu\nu}$ is the Ricci tensor. 
To analyze the number of states implied by this equation, one may 
count the number of plane-wave solutions with given momentum 
$q^\m$. It then turns out that there are $D$ arbitrary solutions, 
corresponding to the linearized gauge invariance, $h_{\m\n} \to  
h_{\m\n} + \pa_\m\xi_\n +\pa_\n\xi_\m$, which can be discarded. 
Many other components vanish and the only nonvanishing ones 
require the momentum to be lightlike. These reside in the fields 
$h_{ij}$, where the components $i,j$ are in the  
transverse ($D-2$)-dimensional subspace. In addition, the trace 
of $h_{ij}$ must be zero. Hence, the relevant plane-wave 
solutions are massless and have polarizations (helicities) 
characterized by a symmetric traceless 2-rank tensor. This 
tensor comprises $\ft12D(D-3)$, which transform irreducibly under the 
SO$(D-2)$ helicity group of transverse rotations. 
For the special case of $D=6$ spacetime dimensions, the helicity 
group is SO(4), which factorizes into two SU(2) groups. The 
symmetric traceless  
representation then transforms as a doublet under each of the 
SU(2) factors and it is thus denoted by (2,2). 

As is well known, for $D=3$ there are obviously no
dynamic degrees of freedom associated with the gravitational 
field. When $D=2$ there are again no 
dynamic degrees of freedom, but here \eqn{Einstein} 
should be replaced by $R_{\mu\nu} = \frac{1}{2} g_{\mu\nu}R$. 

 \subsection{Antisymmetric tensor gauge fields}
Antisymmetric tensor gauge fields have field-strength tensors which 
are antisymmetric and of rank $p+2$. They satisfy field equations and 
Bianchi identities, generalizations of the Maxwell equations, 
which read 
\begin{equation}\label{p-gaugefield}
\partial_{[\mu_1}  F_{\mu_2 \cdots \mu_{p+3}]} = 0 \,, \qquad
\partial_\mu F^{\mu\nu_1 \cdots \mu_{p+1}} = 0 \,.
\end{equation}
A trivial example is the case $p=-1$, which describes an ordinary 
scalar field. For $p=-1$ the solution of the first equation of 
\eqn{p-gaugefield} yields $F_\mu =  
\partial_\mu \phi$, so that the second equation yields the 
Klein-Gordon equation for $\phi$. Another example is the case of a 
vector gauge field, which corresponds to $p=0$, where 
\eqn{p-gaugefield} are just the Maxwell equations.

There are two ways of dealing with \eqn{p-gaugefield}. One is to 
solve the first equation in terms of an antisymmetric 
tensor gauge field $A_{\mu_2 \ldots \mu_{p+1}}$ of rank $p+1$,
\begin{equation}
F_{\mu_1 \cdots \mu_{p+2}} = (p+2) \; \partial_{[\mu_1} A_{\mu_2 
\cdots \mu_{p+2}]} \,, 
\end{equation}
and then impose the second equation. 
The alternative is to first solve the second equation 
in terms of an antisymmetric gauge field $B_{\nu_1 \cdots 
\nu_{D-p+3}}$ of rank $D-p+3$,
\begin{equation}
F_{\mu_1 \cdots \mu_{p+2}} = \frac{1}{(D-p-1)!} \;
\varepsilon_{\mu_1 \cdots \mu_{p+2} \rho\nu_1 \cdots 
\nu_{D-p-3}}\; 
\partial^\rho B^{\nu_1\cdots \nu_{D-p-3}} \,,
\end{equation}
after which one imposes the first equation. 
The second procedure coincides with the first one, but it is based 
on the dual field strength defined by
\begin{equation}
F_{\mu_1\cdots \mu_{p+2}} = \frac{1}{(D-p-2)!} \;
\varepsilon_{\mu_1 \cdots \mu_{p+2}\nu_1 \cdots \nu_{D-p-2}}\;
\tilde{F}^{\nu_1 \cdots \nu_{D-p-2}} \,,
\end{equation}
which can be written as
\begin{equation}
\tilde{F}_{\nu_1 \cdots \nu_{D-p-2}} = (D-p-2) \;
\partial_{[\n_1} B_{\n_2 \cdots \nu_{D-p-3}]} \,.
\end{equation}
For $\tilde{F}$ the two equations \eqn{p-gaugefield} are 
interchanged and the solution in terms of $B_{\nu_1 \ldots 
\nu_{p^\prime +1}}$, with $p^\prime + p = D-4$ is the dual  
formulation of the one in terms of $A_{\mu_1\ldots \mu_{p+1}}$.  
As is well known, in a so-called ``first-order'' formulation it 
is possible to have a Lagrangian that encompasses both 
descriptions.  

Let us now examine the plane-wave solutions for the equations 
\eqn{p-gaugefield}. We will be somewhat more explicit here and  
start from a decomposition of $F_{\mu_1\cdots \mu_{p+2}}$ in  
the momentum representation, with a fixed momentum vector $q^\m$. 
Introducing $D-2$ transverse polarization vectors $\e^i_\m$, with 
$i= 1,2,\ldots,D-2$, and an additional vector $\bar q^\m= (-q^0,
\vec q)$, we decompose the field strength according to 
\begin{eqnarray}
F_{\mu_1 \ldots \mu_{p+2}}(q) & \propto & a_{i_1 \cdots 
i_{p+2}}(q)  \;
\epsilon^{i_1}_{[\mu_1} \cdots \epsilon^{i_{p+2}}_{\mu_{p+2}]} 
\nonumber \\ 
&& +  \left(b_{i_1 \cdots i_{p+1}}(q)\; \bar{q}_{[\mu_{1} }+ 
c_{i_1 \cdots i_{p+1}}(q)\;q_{[\mu_1} \right) 
\epsilon^{i_1}_{\mu_2 }\cdots  \epsilon^{i_{p+1}}_{\mu_{p+2}]} \nonumber \\
&& + \,d_{i_1 \ldots i_{p}} (q) \, \epsilon^{i_1}_{[\mu_1} \cdots 
\epsilon^{i_{p}}_{\mu_{p}} \,\bar{q}_{\mu_{p+1}} \,q_{\mu_{p+2} 
]} \;.
\end{eqnarray}
Imposing \eqn{p-gaugefield} yields
\begin{equation}
a_{i_1 \cdots i_{p+2}} (q) = b_{i_1 \cdots i_{p+1}} (q) =d_{i_1 
\cdots i_{p}} (q) = 0 \,, \qquad q^2 \,c_{i_1 \cdots i_{p+1}} (q) 
= 0 \,,
\end{equation}
so that the dynamic degrees of freedom are massless and reside in 
the antisymmetric $(p+1)$-th rank tensors $c_{i_1 \cdots i_{p+1}} 
(q)$ living in the transverse $(D-2)$-dimensional space. Hence the
number of degrees of freedom is equal to $(D-2)!/ [(p+1)! 
(D-p-3)!]$, which is, as expected, invariant under $p\to p^\prime 
= D-4-p$.

If $D=2$~mod~4 and $p+1=\frac{1}{2}(D-2)$, it is possible to restrict the 
tensor $F_{\mu_1 \ldots \mu_{p+2}}$ to be selfdual or 
antiselfdual, viz. 
\begin{equation}
F_{\mu_1 \ldots \mu_{p+2}} = \pm \frac{1}{(p+2)!} \,
\varepsilon_{\mu_1 \cdots \mu_{2p+4}} \; F^{\mu_{p+3} \ldots 
\mu_{2p+4}} \,.
\end{equation}
For such tensors the two equations \eqn{p-gaugefield} are no 
longer independent. The above duality condition on the field 
strength induces a corresponding $(D-2)$-dimensional duality 
condition (but now in the space of 
transverse momenta, which is Euclidean) on the coefficients 
$c_{i_1 \cdots i_{p+1}}(q)$, 
\begin{equation}
c_{i_1\cdots i_{p+1}}(q)  = \mp \frac{1}{(p+1)!}\, \varepsilon_{i_1 
\cdots i_{2p+2}} \; c^{i_{p+1} \cdots i_{2p+2}}(q) \,. 
\end{equation}
Consequently the number of independent solutions associated with the
antisymmetric tensor is reduced by a factor 2. For $D=6$ where 
the helicity group factorizes, the representation of the 
(anti-)selfdual tensor gauge fields, correspond to (3,1) and  
(1,3). In $D=10$ the (anti-)selfdual tensors correspond 
to the ${\bf 35}_s$ and ${\bf 35}_c$ representations. 

\subsection{Spinor fields}
Consider a spinor $u(q)$ in $D$ space-time dimensions, satisfying 
the massless Dirac equation (in momentum space),
\be
\rlap/q u(q) = 0 , 
\end{equation}
The Dirac equation implies that $q^2 = 0$. Using the same 
manipulations as those leading to \eqn{susy-algebra2}, we rewrite 
the Dirac equation as 
\be
q^0 \Big({\bf 1} - \tilde \G_D \tilde \G_\perp \Big)u(q) =0  \,,
\ee
where $\tilde \G_D$ and $\tilde \G_\perp$ were defined in 
section~2.2.

In odd dimensions $\tilde \G_D$ is proportional to the unit 
matrix, so that the above condition determines that the spinors 
are reduced to a subspace where $\tilde \G_\perp=\pm{\bf 1}$. For 
even dimensions the states constitute a spinor representation of 
the helicity group whose chirality is related to the 
$D$-dimen\-sional chirality of the spinor field. For instance, for 
$D=6$ dimensions a chiral spinor will transform under only one of the 
SU(2) groups of the helicity group, so we have either (2,1) of 
(1,2). For Majorana-Weyl spinors in $D=10$, the states transform 
as {\bf 8}$_c$ or {\bf 8}$_s$, depending on the chirality of the 
spinor field.

\subsection{Gravitino fields}
The gravitino field is a vector-spinor $\psi_\m$ and acts as the 
gauge field of local supersymmetry transformations. Free 
gravitini satisfy the Rarita-Schwinger equation
\begin{equation}\label{RS-eq}
\Gamma^\mu (\partial_\mu \psi_\nu - \partial_\nu \psi_\mu ) = 0 
\,.
\end{equation}
To examine the nature of plane-wave solutions, we again consider 
the momentum representation and decompose $\psi(q)$ as
\begin{equation}
\psi_\mu (q) = u_i (q) \,\epsilon^{i}_{\mu} + v(q) \,\bar{q}_\mu + 
w(q) \,q_\mu \,,
\end{equation}
where the coefficient functions $u_i(q)$, $v(q)$ and $w(q)$ are 
spinors. The field equation \eqn{RS-eq} takes the form
\begin{equation}
\rlap/ q  u_i (q) \,\epsilon^{i}_{\nu} - [ \rlap/\epsilon^i   
u_i (q) - \rlap/\bar{q}  v(q)]\, q_\nu - \rlap/q   v(q) \,
\bar{q}_\nu = 0 \,,
\end{equation}
where $\rlap/\epsilon^i = \epsilon^{i}_{\mu}\, \Gamma^\mu$. The 
spinor $w(q)$, which is subject to gauge transformations 
$\delta\psi_\mu =\partial_\mu \epsilon$, is not determined by the 
gauge invariant field 
equation \eqn{RS-eq} and can be discarded. The remaining 
spinors $u_i(q)$ and $v(q)$ satisfy
\begin{equation}
\rlap/q  u_i (q) =  \rlap/q  v(q) = 0\,,\qquad \rlap/\bar{q}  
v(q) = \rlap/ \epsilon^i  u_i(q) \,.
\end{equation}
Multiplying the last equation with $\rlap/q$ and using the first 
two equations and $\epsilon^i \cdot q = 0$, one derives 
$q\cdot\bar{q} v(q) = 0$. Hence $v(q) = 0$, so that we are left 
with two equations for $u_i(q)$,
\begin{equation} \label{u-equation}
\rlap/q   u_i (q) =  \rlap/\epsilon^i u_i (q) = 0 \,.
\end{equation}
Hence the gravitino states transform under transverse rotations 
according to the highest helicity representation contained in the 
product of a vector and a spinor representation. 

For instance, in $D=6$ dimensions $u_i(q)$ transforms as a chiral 
vector-spinor, which is a product of (2,2) with (2,1) or (1,2). 
This product decomposes into $(3,2)+(1,2)$, or $(2,3)+(2,1)$, 
respectively. The second representation is again suppressed by 
virtue of the second condition in \eqn{u-equation}, so that we are 
left with (3,2) or (2,3). 
    
In $D=10$ spacetime dimensions a chiral gravitino  
field  $u_i(q)$ transforms as 
a chiral vector-spinor, which constitutes a tensor product ${\bf 
8}_v \times {\bf 8}_c$ ( or ${\bf 8}_v\times {\bf 8}_s$, 
depending on the chirality).  According to the multiplication 
rules \eqn{so8-multiplication}, this product decomposes into 
${\bf 8}_s + {\bf 56}_s$ (or, ${\bf 8}_c +{\bf 56}_c$). However, 
the second equation in \eqn{u-equation}, which is SO(8) 
covariant, imposes eight conditions thus suppressing the ${\bf 
8}_s$  or ${\bf 8}_c$ representation. Consequently chiral 
gravitini transform according to the ${\bf 56}_s$ or ${\bf 56}_c$ 
representations of SO(8). 

\section{Coupling constants of low-energy effective field 
theories}  
In section~2 
we discussed various field 
theories that play a role as effective low-energy field theories 
for superstrings. The
effective field theories can be rigorously
derived from the underlying string theory
and in this process the free parameters of 
the field theories are expressed in terms of the 
parameters of the string theory 
itself. The purpose of this appendix is to briefly recall
the various possibilities of deriving the low-energy 
effective action. 

One method to obtain the low-energy effective action is known as 
the `$S$-matrix approach',  
which was pioneered in \cite{GS}.
Here one computes physical scattering amplitudes  
in both string theory and the low-energy field theory
and demands their equality in the limit
$p\ll \Mstr$ where $p$ is the characteristic momentum
of the scattering process.
This method is carried out most conveniently in the 
Einstein frame. 
Alternatively one can use 
the `$\sigma$-model approach' which was
pioneered in \cite{Callan}. 
One imposes conformal invariance on the 2-dimensional 
$\sigma$-model specified by the action
(\ref{sigmaaction}). This requirement leads to
field equations in spacetime which coincide with the field 
equations obtained from an action in the string frame.
(The $\sigma$-model approach
is not applicable to all string compactifications.)

Let us first outline how the relation
(\ref{Mrel}) emerges in the $S$-matrix approach.
{}From (\ref{euler}) and (\ref{sigmaaction}) we know 
that the dilaton couples to the topology of the world 
sheet, so that in leading order (genus-$0$), the 
$N$-particle $S$-matrix elements 
are proportional to 
$\gstring^{N-2}$ multiplied by an 
appropriate power of  
$\a^\prime$, in accordance with dimensional counting. 
On the other hand, the corresponding $S$-matrix elements when 
calculated from the effective field theory, are expressed in 
terms of Newton's constant. 
Comparing the $S$-matrix elements
one obtains (suppressing numerical factors)
the relation (\ref{Mrel}),
\be
(\kappa^2_D)^{\scriptscriptstyle \rm physical} = \kappa_D^2 \,
\l^{(2-D)/2} =  {\alpha^\prime}^{(D-2)/2} \, {\rm e}^{2\langle 
\phi\rangle}\, , \label{Newton-D}
\ee
with $\l$ given in the Einstein frame.

Note that the parameter $\kappa^2_D$ is not determined 
by \eqn{Newton-D}
in agreement with our previous arguments that
it is intrinsically undetermined. 
There are basically two 
ways to proceed:
First one may 
{\it choose} the constant $\kappa^2_D$ to be Newton's constant. 
This implies that one has to expand the  
metric around $g_{\m\n}=\eta_{\m\n}$ in the Einstein 
frame, so that 
$\l=1$. This is a convenient setting, which is most commonly 
used (see, e.g. \cite{GS}) and which leads to 
$\kappa^2_D=(\a^\prime)^{(D-2)/2}\gstring^2$.  
However, this choice implies that 
a coupling constant ($\kappa^2_D$) in the 
effective Lagrangian 
depends on a parameter ($\gstring$) 
that arises as the vacuum-expectation value of the 
dilaton. 
Alternatively one could insist 
that any  dependence on $\gstring$ only arises 
as a result of the explicit couplings of the dilaton field in the effective Lagrangian. 
Or in other words, no
{\it parameters} of the effective action
are chosen to explicitly depend on 
$\gstring$. In the Einstein frame this requires  
to expand the metric around $g_{\m\n}= \l \,\eta_{\m\n}$ with $\l^{(D-2)/4} = \gstring$,
while in the string 
frame an expansion around $g_{\m\n}= 
\eta_{\m\n}$ is necessary.
In both frames one obtains  
$\kappa_D^2 ={\alpha^\prime}^{(D-2)/2}$ 
with no explicit 
dependence on $\gstring$. 
In the string frame, this effective action  
coincide with the one obtained by the $\sigma$-model approach.\footnote{Observe that the 
$\sigma$-model approach does not insist on a particular 
ground-state value for the metric. Since it derives the 
effective action by integration of the field equations, it 
determines the Lagrangian only up to an overall constant.}
In the Einstein frame this choice is somewhat
awkward and rarely used.

As a further illustration of the two different
parameter choices let us consider higher order
gravitational interactions
which generically arise in string theory.
For example,
the string calculations (still to leading order in 
$\gstring$) of the $S$-matrix of 
graviton-graviton scattering give rise  
to contributions that require an effective interaction 
quartic in the Riemann tensor
\be
{\cal L}_{\rm eff} = {1\over \kappa_D^2}  \sqrt{-g}\,{\rm 
e}^{-2\phi}  \Big[ R + A\, \alpha^{\prime\,3}\, 
\Big(R_{abcd} R^{cdef} R_{efgh}  
R^{ghab}+\cdots\Big) \Big]\,. \label{R4-term}
\ee
We have displayed ${\cal L}_{\rm eff}$ is the 
string frame and the 
higher-order terms depend on a dimensionless constant $A$ which is independent of $\gstring$. 
In the 
Einstein frame the dilaton factor in front of the Ricci scalar is removed by a Weyl transformation which 
also changes the coupling  in front of
the $R^4$-terms into a factor  
$\exp(-12\phi/(D-2))$. Expanding the metric in 
the Einstein frame 
around $g_{\m\n}=\eta_{\m\n}$ and comparing the relevant 
$S$-matrix elements to the string calculation,  one finds that 
$\kappa^2_D = (\a^\prime)^{(D-2)/2}\,\gstring^2$ while $A= 
\gstring^{12/(D-2)}$  
\cite{GS}.\footnote{%
  Note that the relation with the dilaton field used in the second reference of \cite{GS} 
  is given by $\phi-\langle\phi\rangle = \sqrt{2} \kappa D$ (in 10 spacetime dimensions).} %
Again, the dependence on $\gstring$  
cannot be tied to the presence of a dilaton interaction in neither 
one of the two terms. 
However, if  one expands the Einstein 
metric around 
$g_{\m\n}=  \gstring^{-4/(D-2)}\,\eta_{\m\n}$ 
(or 
equivalently, the metric in the string frame 
around $g_{\m\n}=\eta_{\m\n}$)
one finds $\kappa^2_D=(\a^\prime)^{(D-2)/2}$ and $A=\rm constant$ -- both couplings
independent of $\gstring$. 
This form of the parameters is also obtained
in the $\sigma$-model approach where 
the $R^4$-term arises as a 4-loop counterterm
\cite{grisaru}.

The final point of this appendix 
concerns 
the dilaton in arbitrary spacetime dimensions.
It can be 
defined as the field in the $\sigma$-model
action (\ref{sigmaaction}) taken in $D$ 
dimensions. The corresponding vertex operator
is composed only out of 
operators of the spacetime sector of the 
conformal field theory (CFT) and no operators 
in the 
`internal CFT'.
Let us denote the dilaton defined  
in this manner by $\phi^{(D)}$. 
This definition has the virtue that 
$\phi^{(D)}$ is invariant under 
T-duality transformations of the $D$-dimensional
theory which originates from the 
existence of equivalence classes of the internal CFT.
The same is true for the graviton
and the antisymmetric tensor, whose
vertex operators are similarly 
composed solely out of the spacetime sector 
of the CFT.


Compactification of the low-energy effective 
actions relates the dilatons 
of different dimensions by a volume-dependent
factor of the compactification manifold $Y$ and the metric 
associated with the compactified dimensions. 
More precisely, starting
in $D=10$ one finds (in the string frame)
\be\label{phiC}
{1\over \kappa^{2}_{10}}\,  {\rm e}^{-2\phi^{(10)}}\, V_n\, 
\sqrt{\det g_n} 
={1\over \kappa^{2}_{10-n}}\,  {\rm e}^{-2\phi^{(10-n)}} \ ,
\ee
where $n$ is the dimension of $Y$, $V_n$ is the volume of the 
$n$-dimensional compactified coordinates and $g_n$ is the metric 
associated with the compactified 
dimensions.\footnote{To be precise, $V_n$ is the 
(higher-dimensional) analog of
the length $L$ introduced in section 2.5.
The geodesic volume which generalizes
$R_{11}$ in (\ref{Rphi}) is instead
proportional to $V_n\sqrt{\det g_n}$.}
 The latter is 
directly related to the vacuum-expectation value of certain 
moduli fields. 
Furthermore, the
 space-time part of the metric (in the string frame) 
is left unchanged in the parametrization
used in (\ref{phiC}) and 
the $D$-dimensional quantities are defined by 
\be\label{phiDrel}
{1\over \kappa^{2}_{10-n}} =  {V_n \over 
\kappa^{2}_{10}} \, ,
\qquad
\phi^{(10-n)} =
\phi^{(10)} - \ft14\, \log \det g_n \ .
\ee
The $D$-dimensional string metric, the dilaton 
and the antisymmetric tensor are invariant
under T-duality. For the 
perturbative dualities one can 
demonstrate this fact by performing a dimensional reduction on 
the 10-dimensional supergravity field theory  (for instance on the 
Lagrangian \eqn{D10-lagrangian-S}) from 
10 to $10-n$ dimensions. Using the arguments of section~2.5 one 
establishes the existence of a rank-$n$ group of invariances that 
leaves the string metric, the dilaton (defined according to 
\eqn{phiDrel}) and the antisymmetric tensor field invariant. 
However, the original  
10-dimensional dilaton $\phi^{(10)}$ transforms 
under these symmetries.


\end{document}